\documentclass[5p, authoryear, twocolumn]{elsarticle}   	% use "amsart" instead of "article" for AMSLaTeX format, 5p or review

\usepackage{geometry}                		% See geometry.pdf to learn the layout options. There are lots.
\usepackage{graphicx}				% Use pdf, png, jpg, or eps� with pdflatex; use eps in DVI mode
\usepackage{amssymb}

\usepackage{subfigure}

\usepackage{wrapfig}
\usepackage{lscape}
\usepackage{rotating}
\usepackage{epstopdf}
\usepackage{afterpage}
\usepackage{flafter}

 \usepackage{mathtools}
 \usepackage{bm}
 
\usepackage{csvsimple}
\usepackage{longtable}
\usepackage{booktabs}
\usepackage{array}

\usepackage{pdflscape}
 
 \usepackage{subfig}
 
 \usepackage{chapterbib}
 
 \usepackage[switch]{lineno}
\newcommand*\patchAmsMathEnvironmentForLineno[1]{%
  \expandafter\let\csname old#1\expandafter\endcsname\csname #1\endcsname
  \expandafter\let\csname oldend#1\expandafter\endcsname\csname end#1\endcsname
  \renewenvironment{#1}%
     {\linenomath\csname old#1\endcsname}%
     {\csname oldend#1\endcsname\endlinenomath}}% 
\newcommand*\patchBothAmsMathEnvironmentsForLineno[1]{%
  \patchAmsMathEnvironmentForLineno{#1}%
  \patchAmsMathEnvironmentForLineno{#1*}}%
\AtBeginDocument{%
\patchBothAmsMathEnvironmentsForLineno{equation}%
\patchBothAmsMathEnvironmentsForLineno{align}%
\patchBothAmsMathEnvironmentsForLineno{flalign}%
\patchBothAmsMathEnvironmentsForLineno{alignat}%
\patchBothAmsMathEnvironmentsForLineno{gather}%
\patchBothAmsMathEnvironmentsForLineno{multline}%
}
 
\usepackage{fancyhdr}
\begin{document}\sloppy

%\linenumbers
\clearpage
\clearpage

\begin{frontmatter}

\author[harv]{Simon J. Lock \corref{cor}}
\author[dav]{Sarah T. Stewart}
\author[harv]{Michail I. Petaev}
\author[bris]{Zo\"{e} M. Leinhardt}
\author[bris]{Mia T. Mace}
\author[harv]{Stein B. Jacobsen}
\author[seti]{Matija \'{C}uk}
\cortext[cor]{Corresponding author: slock@fas.harvard.edu}
\address[harv]{Department of Earth and Planetary Sciences, Harvard University, 20 Oxford Street, Cambridge, MA 02138, U.S.A.}
\address[dav]{Department of Earth and Planetary Sciences, U. California Davis, One Shields Avenue, Davis, CA 95616, U.S.A.}
\address[bris]{School of Physics, University of Bristol, Tyndall Avenue, Bristol, BS8 1TL, UK}
\address[seti]{SETI Institute, 189 Bernardo Avenue, Mountain View, CA 94043, USA}

\title{The origin of the Moon within a terrestrial synestia}

%\begin{keypoints}
%\item New model for lunar origin within an impact-generated synestia could explain the lunar composition
%\item Synestias are highly pressure supported and have different dynamics than condensate-dominated disks 
%\item Moon equilibrates with bulk silicate Earth vapor at tens of bars and a temperature buffered by silicate vaporization
%\end{keypoints}

%xxxxxxxxxxxxxxxxxxxxxxxxxxxxxxxxxxxxxxxxxxxxxxxxxxxxxxxxxxxxxxxxxxxxxxxxxxxxxxxxxxxxxxxxxxxxxxxxxxxxxxxxxxxxxxxxxxxxxxxxxxxxxxxxxxxxxxxx
%xxxxxxxxxxxxxxxxxxxxxxxxxxxxxxxxxxxxxxxxxxxxxxxxxxxxxxxxxxxxxxxxxxxxxxxxxxxxxxxxxxxxxxxxxxxxxxxxxxxxxxxxxxxxxxxxxxxxxxxxxxxxxxxxxxxxxxxx
%xxxxxxxxxxxxxxxxxxxxxxxxxxxxxxxxxxxxxxxxxxxxxxxxxxxxxxxxxxxxxxxxxxxxxxxxxxxxxxxxxxxxxxxxxxxxxxxxxxxxxxxxxxxxxxxxxxxxxxxxxxxxxxxxxxxxxxxx
\begin{abstract}
The giant impact hypothesis remains the leading theory for lunar origin. However, current models struggle to explain the Moon's composition and isotopic similarity with Earth. Here we present a new lunar origin model. High-energy, high-angular momentum giant impacts can create a post-impact structure that exceeds the corotation limit (CoRoL), which defines the hottest thermal state and angular momentum possible for a corotating body. In a typical super-CoRoL body, traditional definitions of mantle, atmosphere and disk are not appropriate, and the body forms a new type of planetary structure, named a synestia. Using simulations of cooling synestias combined with dynamic, thermodynamic and geochemical calculations, we show that satellite formation from a synestia can produce the main features of our Moon. We find that cooling drives mixing of the structure, and condensation generates moonlets that orbit within the synestia, surrounded by tens of bars of bulk silicate Earth (BSE) vapor. The moonlets and growing moon are heated by the vapor until the first major element (Si) begins to vaporize and buffer the temperature. Moonlets equilibrate with BSE vapor at the temperature of silicate vaporization and the pressure of the structure, establishing the lunar isotopic composition and pattern of moderately volatile elements. Eventually, the cooling synestia recedes within the lunar orbit, terminating the main stage of lunar accretion. Our model shifts the paradigm for lunar origin from specifying a certain impact scenario to achieving a Moon-forming synestia. Giant impacts that produce potential Moon-forming synestias were common at the end of terrestrial planet formation.

\end{abstract}

\end{frontmatter}

%xxxxxxxxxxxxxxxxxxxxxxxxxxxxxxxxxxxxxxxxxxxxxxxxxxxxxxxxxxxxxxxxxxxxxxxxxxxxxxxxxxxxxxxxxxxxxxxxxxxxxxxxxxxxxxxxxxxxxxxxxxxxxxxxxxxxxxxx
%xxxxxxxxxxxxxxxxxxxxxxxxxxxxxxxxxxxxxxxxxxxxxxxxxxxxxxxxxxxxxxxxxxxxxxxxxxxxxxxxxxxxxxxxxxxxxxxxxxxxxxxxxxxxxxxxxxxxxxxxxxxxxxxxxxxxxxxx
%xxxxxxxxxxxxxxxxxxxxxxxxxxxxxxxxxxxxxxxxxxxxxxxxxxxxxxxxxxxxxxxxxxxxxxxxxxxxxxxxxxxxxxxxxxxxxxxxxxxxxxxxxxxxxxxxxxxxxxxxxxxxxxxxxxxxxxxx
\section{Introduction}
\label{sec:intro_lunar_origin}

In the giant impact hypothesis for lunar origin \citep{Hartmann1975,Cameron1976}, the proto-Earth suffered a collision with another protoplanet near the end of accretion that ejected material into a circumterrestrial disk, out of which the Moon formed \citep[see reviews by][]{Stevenson1987, Asphaug2014,Barr2016}. Giant impacts are highly energetic events that vaporize a portion of the impacting bodies. Hence, the disk is a multiphase mixture of liquid and vapor \citep{Canup2001,Canup2004,Canup2008,Canup2012,Cuk2012}. Modeling the formation of the Moon from such a disk is challenging and the details of lunar accretion are still uncertain \citep[e.g.][]{Thompson1988,Machida2004,Salmon2012,Ward2012,Ward2014,Ward2017,Charnoz2015,Carballido2016,Gammie2016}.
To date, lunar origin studies have not demonstrated that a single giant impact can explain both the physical and chemical properties of our Moon \citep{Asphaug2014,Barr2016}.

Most studies of the origin of the Moon have focused on a narrow range of impact scenarios.
\citet{Cameron1976} proposed that the Moon-forming giant impact could have prescribed the present-day angular momentum (AM) of the Earth-Moon system. Numerical simulations have shown that a grazing collision with a Mars-mass impactor near the mutual escape velocity can impart the present-day AM and generate a silicate-rich disk composed of more than a lunar mass of material \citep{Canup2001,Canup2004,Canup2008}. This scenario, which we refer to as the canonical giant impact, has become the de facto working model for lunar origin. However, studies of the canonical impact and its aftermath have difficulty explaining some key observables of the Earth-Moon system, including: the isotopic similarity between Earth and the Moon; the lunar depletion in moderately volatile elements; the large mass of the Moon; and the present day lunar inclination.

Numerical simulations of giant impacts predict that the canonical lunar disk is derived primarily from the impactor \citep{Canup2001,Canup2004,Canup2008}. However, increasingly precise isotopic measurements of terrestrial and lunar samples have shown that Earth and the Moon share very similar initial isotopic ratios for a wide range of elements \citep{Lugmair1998,Wiechert2001,Zhang2012}. Because the isotope ratios of such elements are observed to vary significantly among planetary bodies \citep[][]{Clayton1996,Yin2002,Trinquier2007,Zhang2012}, the impactor is generally expected to have had a distinct isotopic composition, resulting in a measurable isotopic difference between Earth and the Moon \citep{Pahlevan2007,Melosh2014,Young2016}. 

Two classes of solutions to the problem of isotopic similarity have been proposed.
First, post-impact mixing between the planet and lunar disk could erase initial isotopic heterogeneities \citep{Pahlevan2007}, but the extent of mixing required to explain the observations is a problem in the canonical model \citep{Melosh2014}.
Second, the impactor and proto-Earth could have formed from the same source material and thus shared nearly identical isotopic signatures \citep{Jacobsen2013LPSC,Dauphas2014,Dauphas2017}. There is evidence for a reservoir of terrestrial precursor materials with fractionation-corrected isotopic ratios that are distinct from the meteorites and planetary samples in our collections \citep[e.g.,][]{Drake2002}. If the impactor and target accreted the majority of their mass from the same reservoir, the Earth and Moon would share similar stable isotopic ratios. Stable isotopic ratios are controlled by the source material only and not affected by processes within the body (e.g., O, Cr, Ti). Yet, even if bodies in the inner solar system were formed from material with similar isotopic signatures, this explanation for the isotopic similarity between Earth and the Moon relies upon a coincidence to explain tungsten, which is sensitive to the conditions and timing of core formation on each of the colliding bodies \citep{Touboul2015,Kruijer2015,Dauphas2014,Dauphas2017,Kruijer2017}.  As discussed in \citet{Melosh2014} and \citet{Kruijer2017}, neither of these proposals on their own provides a satisfactory explanation for the isotopic similarity between Earth and the Moon. Nevertheless, the possibility of a more homogeneous inner solar system relaxes the isotopic constraint on Moon-formation, as models need only produce similar enough tungsten isotopes in Earth and the Moon. The tungsten isotope constraint is weaker than that for stable isotopes due to the uncertainties in inferring the post-impact composition before the addition of late veneer \citep{Touboul2015,Kruijer2015}.

Study of lunar samples has revealed that the Moon is significantly depleted in moderately volatile elements (MVEs; e.g., K, Na, Cu, and Zn) relative to the bulk silicate Earth (BSE). For example, potassium and sodium are inferred to be depleted by factors of 5 to 10 compared to terrestrial abundances \citep[e.g.,][see \S\ref{sup:sec:BSM}]{Ringwood1977}. In a series of studies 
\citep[e.g.,][]{Ringwood1977,Ringwood1986}, Ringwood and colleagues argued that the lunar composition could be explained if the Moon was a partial condensate of vapor derived from Earth's mantle. The MVE depletion of the Moon is a key constraint on lunar origin models. In addition, the volatile depletion of the Moon has been used to argue for a process-based link between giant impacts and MVE loss. Indeed, the lunar depletion has been used to propose that Mercury would be depleted, if it formed by a giant impact \citep{Peplowski2011}. We must understand the physical processes that led to volatile depletion on the Moon in order to place its data in the context of other bodies in the solar system. Few studies have attempted to combine the dynamics, thermodynamics and chemistry of lunar origin, which is necessary to be able to test the proposed models. 
Recently, \citet{Canup2015} used the lunar disk models of \citet{Salmon2012} and physical chemistry calculations to link the dynamics and thermodynamics of accretion from a canonical disk.
They also suggested that the lunar volatile element depletion could be explained if the material that formed the observable Moon was a partial condensate of disk material.
\citet{Wang2016} recently reported that the potassium isotopes of the Moon are heavier than BSE, which supports the idea of partial condensation.
However, the model presented by \citet{Canup2015} does not quantitatively explain the magnitude, nor pattern, of moderately volatile element depletion observed for the Moon.
Small isotopic fractions could be produced if the canonical disk had a period of hydrodynamic loss \citep{Pritchard2000}, but, given the mean molecular weight of vapor in the disk, hydrodynamic escape is unlikely \citep{Nakajima2014LPSC}.
Further work is needed to fully integrate chemical and physical models of lunar origin.

Predicting the final mass of satellites formed by giant impacts is challenging.
The methods that are currently used for simulating giant impacts do not include the physics necessary for modeling lunar accretion; therefore, separate calculations of disk evolution are required to infer the mass of the satellite produced by a specific impact.
Typically, scaling laws fitted to $N$-body simulations of idealized circumterrestrial debris disks \citep{Ida1997,Kokubo2000} have been used to estimate the satellite mass from the total mass and AM of orbiting material \citep{Canup2001,Canup2004,Canup2008,Canup2012,Cuk2012}. 
Studies of canonical impacts have found that, over a narrow range of impact angles, sufficient mass is injected into orbit to produce a lunar mass Moon based on $N$-body scaling laws.
Because $N$-body simulations do not include the multiphase physics of the lunar disk, they overestimate the efficiency of satellite formation. Simulations that include a simplified one-dimensional model of Roche-interior multiphase disks have inferred much lower accretion efficiencies \citep{Salmon2012,Salmon2014}. The Roche limit is the closest distance a satellite can withstand the tidal forces from the planet (about 18500~km for silicate satellites orbiting Earth).
Using the scaling laws produced by these most recent models, very few of the disks produced in published canonical giant impact simulations inject the required mass and AM into orbit to produce a lunar mass satellite (supporting information $\S$\ref{sup:sec:canonical}). 
Furthermore, the simple Roche-interior disk model used by \cite{Salmon2012,Salmon2014} likely overestimated the efficiency of the spreading of material beyond the Roche limit. By incorporating more multiphase physics, \cite{Charnoz2015} showed that viscous spreading of material beyond the Roche limit is slower than calculated by \cite{Salmon2012,Salmon2014} and that more mass from the disk is lost to Earth.
Canonical impacts typically inject a large amount of mass directly beyond the Roche limit and \citet{Charnoz2015} suggested that the Moon could have largely formed from this material. The efficiency of accretion has not been quantified and such a model would still need to explain the isotopic similarity and moderately volatile element depletion.
Given the current results from giant impact calculations and available satellite accretion scaling laws, it is uncertain whether canonical giant impacts can form a sufficiently large moon.

The origin of the Moon's present-day orbital inclination, which is about $5^{\circ}$ from the ecliptic plane, has been a long standing problem in lunar tidal evolution. If a Moon-forming giant impact also determined Earth's present obliquity, then lunar origin in an equatorial disk and subsequent tidal evolution through the Laplace plane transition, from an orbit that precesses in Earth's equatorial plane to one that precesses in the ecliptic plane, should have led to a lunar orbit with near-zero inclination to the ecliptic. Therefore, dynamical processes subsequent to the impact are required to explain the present lunar inclination. Proposed solutions in the framework of the canonical model include a complex sequence of luni-solar resonances \citep{Touma1994}, resonant interactions between the Moon and the circumterrestrial disk \citep{Ward2000}, and encounters between large planetesimals and the newly formed Earth-Moon system \citep{Pahlevan2015}.  Recently, \citet{Chen2016} and \citet{Cuk2016} investigated inclination damping by lunar obliquity tides, and \citet{Cuk2016} found that lunar inclination must have been large ($\sim 30^{\circ}$) prior to the point in tidal recession where the lunar orbit transitions between Cassini states \citep[distinct dynamical solutions that govern the alignment of the lunar spin axis and orbital plane, see][]{Peale1969}. Such a large inclination prior to the Cassini state transition defies explanation by any of the previously proposed mechanisms to raise lunar inclination after a canonical giant impact. Connecting the canonical giant impact to the Moon's current orbit remains an unsolved problem.

Despite the fact that it has not yet explained major characteristics of the Earth-Moon system, the giant impact hypothesis has not been rejected, primarily due to the lack of another viable mechanism for the origin of the Moon. A range of alternative impact models have been proposed \citep{Reufer2012,Cuk2012,Canup2012,Rufu2017}, but each calls upon an additional process or a fortunate coincidence to better explain the Earth-Moon system. Hence, none of these recent variations on an impact origin have gained broad support. 

A substantial constraint on the canonical Moon-forming impact is that the AM of the Earth-Moon system has not changed significantly since the formation of the Moon. \citet{Cuk2012} showed that an evection resonance could drive significant AM loss from the Earth-Moon system after the Moon-forming impact. Since \citet{Cuk2012}, additional mechanisms have been found that could remove AM during lunar tidal evolution \citep{Wisdom2015,Cuk2016,Tian2017}. Allowing for a change of AM after the impact significantly expands the range of possible impact parameters for the Moon-forming collision. \citet{Cuk2012} and \citet{Canup2012} showed that high energy, high-AM impact events can inject much more material into orbit than canonical impacts. In addition, \citet{Cuk2016} showed that if the Earth after the impact had both high-AM and high obliquity, an instability during the Laplace plane transition could both remove AM from the Earth-Moon system and explain the Moon's present-day orbital inclination. With these promising dynamical results, high-AM giant impact scenarios for lunar origin warrant continued investigation.

Here, we present a new model for lunar origin within a terrestrial synestia, an impact-generated structure with Earth-mass and composition that exceeds the corotation limit (CoRoL). Synestias are formed by a range of high-energy, high-AM collisions during the giant impact stage of planet formation \citep[][hereafter LS17]{Lock2017}. 
A synestia is a distinct dynamical structure compared to a planet with a condensate-dominated circumplanetary disk, and, as a result, different processes dominate the early evolution of a synestia. Note that preliminary versions of this work \citep[e.g.,][]{Petaev2016LPSC,Lock2016LPSC} used different nomenclature than is used here. In particular, synestias were referred to as continuous mantle-atmosphere-disk (MAD) structures.

At present, no single calculation can fully capture the dynamics, thermodynamics and chemistry of lunar accretion.
Therefore, our approach is to link the physics and chemistry of satellite accretion from a terrestrial synestia by understanding the processes that control the pressure and temperature paths of the material that forms a moon. First, we determine the pressure--temperature conditions of a moon that grows by accretion of condensing silicate vapor. 
We then argue that the composition of the growing moon is set by equilibrium with BSE vapor over a specific range of pressures and temperatures determined by the structure and the phase relationships for material of BSE composition.
Finally, we demonstrate that a variety of high-energy, high-AM giant impacts can generate initial conditions that can potentially lead to the formation of a lunar mass moon with the observed geochemical characteristics of our Moon. 
Our model provides a promising pathway to explain all the key observables of the Moon discussed above: the isotopic similarity between Earth and the Moon; the magnitude and pattern of moderately volatile element depletion in the Moon; and the large mass of the Moon. If Earth had a large obliquity after the giant impact, then a single event may also explain the inclination of the lunar orbit and the present-day AM of the Earth-Moon system \citep{Cuk2016}.

This paper is organized as follows. First, we describe the structures that are generated by giant impacts (\S\ref{sec:dynamics_structure} and \S\ref{sec:dynamics_equilibration}). Next, we discuss the processes that dominate the evolution of a synestia as it cools (\S\ref{sec:dynamics_cooling}).
We present calculations of the pressure field of a synestia and estimate the mass and location of the moon that is formed (\S\ref{sec:cooling_methods} and \S\ref{sec:cooling_results}). 
Then, we present a calculation of the phase diagram for BSE composition material over the pressure and temperature range of the outer portions of the structure (\S\ref{sec:thermo}). 
In \S\ref{sec:comb}, we combine the results of the previous sections to propose a coupled dynamic and thermodynamic model for the formation of a moon from a terrestrial synestia and identify the pressure--temperature--spatial paths of condensates and growing moonlets. 
The chemical composition of the moon is estimated using the physical chemistry of the BSE system at the pressure and temperature predicted by our model. 
In \S\ref{sec:discussion}, we discuss formation of our Moon from a synestia as a new model for lunar origin. We discuss the consistency of such a model with observations and examine the possible range of giant impacts that may generate post-impact structures with the potential for forming our Moon.
We present a short unified synopsis of our model for the origin of the Moon in \S\ref{sec:summary}. Finally, we draw our major conclusions and describe future tests of our model (\S\ref{sec:conclusion}). 
This work includes online supporting information.

%xxxxxxxxxxxxxxxxxxxxxxxxxxxxxxxxxxxxxxxxxxxxxxxxxxxxxxxxxxxxxxxxxxxxxxxxxxxxxxxxxxxxxxxxxxxxxxxxxxxxxxxxxxxxxxxxxxxxxxxxxxxxxxxxxxxxxxxx
%xxxxxxxxxxxxxxxxxxxxxxxxxxxxxxxxxxxxxxxxxxxxxxxxxxxxxxxxxxxxxxxxxxxxxxxxxxxxxxxxxxxxxxxxxxxxxxxxxxxxxxxxxxxxxxxxxxxxxxxxxxxxxxxxxxxxxxxx
%xxxxxxxxxxxxxxxxxxxxxxxxxxxxxxxxxxxxxxxxxxxxxxxxxxxxxxxxxxxxxxxxxxxxxxxxxxxxxxxxxxxxxxxxxxxxxxxxxxxxxxxxxxxxxxxxxxxxxxxxxxxxxxxxxxxxxxxx
\section{Structure and dynamics of a synestia}
\label{sec:dynamics}
In this section, we examine the physical processes that occur during the formation and evolution of synestias. First, we describe the physics controlling the structure of post-impact states and demonstrate the magnitude of pressure support in synestias (\S\ref{sec:dynamics_structure}). Next, we look at the transition from the impact to a post-impact state (\S\ref{sec:dynamics_equilibration}). In \S\ref{sec:dynamics_cooling} we discuss the dominant processes that drive evolution of the synestia and argue that condensates formed at the photosphere can transport mass radially in the structure.
Based on these arguments, we construct a simple cooling model to calculate the temporal evolution of the vapor structure of a synestia (\S\ref{sec:cooling_methods}). In \S\ref{sec:cooling_results} we present an example calculation of the cooling of a potential Moon-forming synestia. Additional examples of cooling of potential Moon-forming synestias, formed by very different impact events, are provided in the appendix. 

%xxxxxxxxxxxxxxxxxxxxxxxxxxxxxxxxxxxxxxxxxxxxxxxxxxxxxxxxxxxxxxxx
%xxxxxxxxxxxxxxxxxxxxxxxxxxxxxxxxxxxxxxxxxxxxxxxxxxxxxxxxxxxxxxxx
\subsection{Structure of post-impact states}
\label{sec:dynamics_structure}

\begin{sidewaysfigure*}
\centering 
\includegraphics[scale=0.83333333]{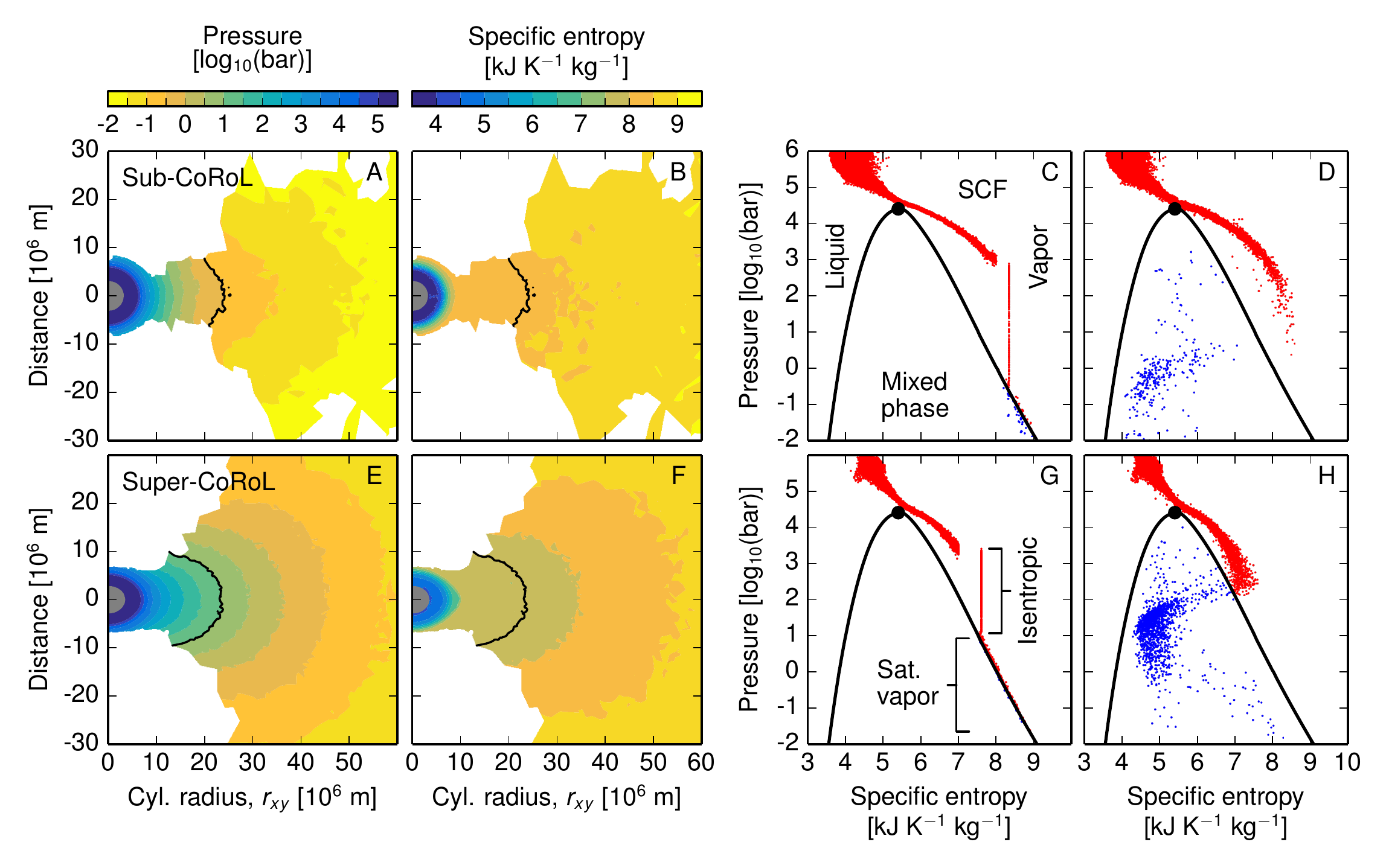}
\caption[]{Examples of sub-CoRoL and super-CoRoL post-impact structures generated by canonical (A-D) and high-energy, high-AM (E-H) giant impacts, showing axisymmetric pressure (A,E) and silicate specific entropy (B,F) contours perpendicular to the spin axis, and the thermal state of the silicate material in the midplane in pressure-specific entropy space (C,D,G,H). The liquid-vapor phase boundary is a dome-shaped curve in pressure-specific entropy space (black line). The black dot on the vapor dome (C,D,G,H) is the critical point for the equation of state used in these simulations ($S_{\rm crit}$~$=$~$5.40$~kJ~K$^{-1}$~kg$^{-1}$, $p_{\rm crit}$~$=$~$25.5$~kbar, $T_{\rm crit}$~$=$~$8810$~K, $\rho_{\rm crit}$~$=$~$1680$~kg~m$^{-3}$). Material to the left of the dome is liquid, material to the right of the dome and below the critical point is vapor, material above and to the right of the critical point is supercritical fluid (SCF), and material within the dome is a mixture of both liquid and vapor (blue points). The liquid-solid phase boundary is neglected in this EOS. The calculated pressure and specific entropy distributions (A-C, E-G) estimate the overall shape of gravitationally and thermally equilibrated post-impact structures, but the pressure structure at the photic surface was not resolved. During thermal equilibration, the Roche-exterior condensate fraction is removed and the post-impact distribution of specific entropy (D,H) is averaged such that the middle of the structure has a mass-weighted isentropic region (vertical set of points in C,G) which transitions to a saturated vapor region that follows the vapor side of the dome. In G, the isentropic and saturated vapor regions of the thermally equilibrated structures are labeled. The inner edge of the saturated vapor region is identified by the black line in A,B,E,F. The structure in A-D was generated by an impact of a 0.13~$M_{\rm Earth}$ projectile onto a 0.9~$M_{\rm Earth}$ target with an impact velocity of 9.2~km~s$^{-1}$ and an impact parameter of 0.75. For E-H, a 0.47~$M_{\rm Earth}$ projectile hit a 0.57~$M_{\rm Earth}$ target with an impact velocity of 9.7~km~s$^{-1}$ and an impact parameter of 0.55. Midplane profiles are presented in Figure~\ref{fig:linestructures}.}
\label{fig:contourstructures}
\end{sidewaysfigure*}

Figure~\ref{fig:contourstructures}A,E shows the approximate fluid pressure structure of two post-impact states, calculated using a smoothed particle hydrodynamics (SPH) code (see \S\ref{sec:cooling_methods} for details of methods), that evolves the position and thermal properties of particles of fixed mass using the governing forces and material equations of state (EOS). The two examples illustrate a post-impact structure that exceeds the corotation limit (CoRoL, LS17) and one that does not. Both impacts generate significant amounts of vapor. The fluid structures are controlled by a balance between the gravitational, pressure gradient and centrifugal forces. For a parcel of material in the midplane, the force balance is approximately
\begin{equation}
\frac{G M_{\rm bnd}}{r_{xy}^2} + \frac{1}{\rho}\frac{dp}{dr_{xy}} - \omega^2 r_{xy} =0 \, ,
\label{eqn:force_balance}
\end{equation}
where $G$ is the gravitational constant, $M_{\rm bnd}$ is the bound mass of the structure, $r_{xy}$ is the cylindrical radius, $\rho$ is the density of the parcel, $p$ is the gas pressure, and $\omega$ is the angular velocity. The vertical structure, in the direction parallel to the rotation axis, is also controlled by hydrostatic balance.

In debris disks, the effect of the pressure support term on the mass distribution is negligible as the gas fraction is small. The density of condensates is much larger than that of vapor, and consequently the effect of the vapor pressure term in Equation~\ref{eqn:force_balance} on condensed particles is small. In prior work on post-impact states \citep[e.g.,][]{Canup2001,Canup2004,Canup2008,Ida1997,Kokubo2000,Salmon2012,Salmon2014,Nakajima2014,Nakajima2015,Ward2012,Ward2014,Ward2017}, the disk was assumed to be dominated by condensates with negligible pressure support. 

The thermal structure of our two example post-impact states are shown in Figure~\ref{fig:contourstructures}B,F. There is a strong gradient in entropy from the inner to the outer regions of the structure. Figure~\ref{fig:contourstructures}C,G presents the thermal structure in the specific entropy-pressure phase space with each SPH particle shown as a colored dot. The black curve is the liquid-vapor phase boundary for the single component silicate equation of state used in the simulation. Material below the black line is a mixture of liquid and vapor, and material above the curve is pure vapor, supercritical fluid, or liquid. Post-impact structures are highly thermally stratified. Typically, the thermal profile is such that it does not intersect the liquid-vapor phase boundary until low pressures, and the silicate transitions smoothly from liquid to supercritical liquid to vapor [LS17]. Thus, the post-impact structures have no surface, and a liquid-vapor mixture is initially restricted to the outer regions and near the photosphere (see \S\ref{sec:dynamics_cooling}). In the examples shown in this work, the midplane is initially completely vapor to beyond the Roche limit (within the black lines in Figure~\ref{fig:contourstructures}A,B,E,F). At low pressures, the structure intersects the liquid-vapor phase boundary and follows a saturated adiabat. Beyond the black line in Figure~\ref{fig:contourstructures}A,B,E,F, the saturated adiabat extends to the midplane and a fraction of the silicate is condensed. In most of the structure, the condensed mass fraction is small, and we have neglected the gravitational effects of the condensed mass in calculating the vapor pressure structure. Due to the dominance of vapor in post-impact structures, there is substantial pressure support, and the pressure gradient term in Equation~\ref{eqn:force_balance} is not negligible.

\begin{figure*}
\centering
\includegraphics[scale=0.833333333]{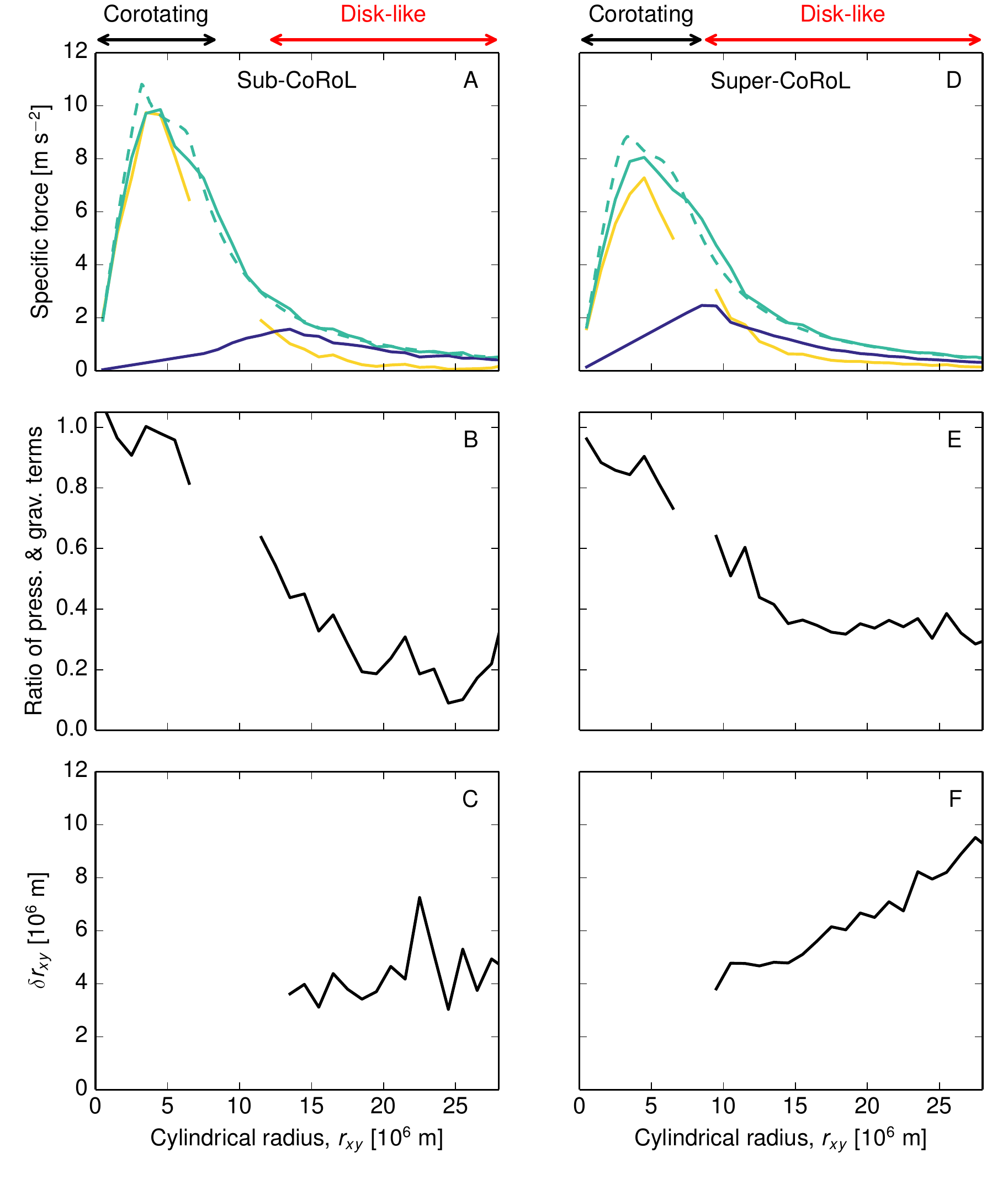}
\caption{The disk-like regions of post-impact states can be strongly pressure supported. We calculate the magnitude of the terms in Equation~\ref{eqn:force_balance} for the two example, post-impact structures shown in Figures~\ref{fig:contourstructures} and \ref{fig:linestructures}. A,D) Profiles of the specific gravitational force (solid green), specific pressure gradient force (yellow) and specific centrifugal force (navy) in the midplane. Quantities have been calculated by taking the average properties of particles in the midplane in 1~Mm bins and calculating the gradients of pressure and potential using these average properties. Dashed green lines shows the gravity assuming that the body is spherically symmetric, neglecting higher order terms. B,E) The ratio of the pressure gradient and gravitational forces. C,F) The difference between the orbit of the pressure supported mass and a circular Keplerian orbit of the same AM in the disk-like regions, $\delta_{xy}$. Sections of profiles are not shown due to model artifacts associated with the averaging of the isentropic region of the structures. The profiles for the sub-CoRoL example are more variable due to fewer particles and hence lower resolution in the disk-like region. The equivalent profiles for different example post-impact structures are shown in Figures~\ref{sup:fig:pressure_supportA}-\ref{sup:fig:pressure_supportp}.
}
\label{fig:pressure_support_main}
\end{figure*}

Figure~\ref{fig:pressure_support_main} shows the relative magnitudes of the gravitational, pressure gradient, and centrifugal terms in Equation~\ref{eqn:force_balance} for the example post-impact structures in Figure~\ref{fig:contourstructures}. In the corotating region, the pressure gradient term (Equation~\ref{eqn:force_balance}) is comparable to gravity as expected, but the pressure gradient force can also be the same order of magnitude as gravity in the outer regions. Additional examples of the magnitude of pressure support in synestias are given in Figures~\ref{sup:fig:pressure_supportA}, \ref{sup:fig:pressure_supportB}, and \ref{sup:fig:pressure_supportp}. Equation~\ref{eqn:force_balance} can be rearranged to give an expression of the angular velocity of the gas at a given cylindrical radius,
\begin{equation}
\omega = \sqrt{\frac{G M_{\rm bnd}}{r_{xy}^3} + \frac{1}{\rho_{\rm vap} r_{xy}}\frac{dp}{dr_{xy}}}  \, .
\end{equation}
A negative pressure gradient term reduces the angular velocity at a given radius. In other words, a parcel of material with a fixed specific AM may orbit at a larger radius with the addition of a pressure gradient force. If the pressure support was subsequently removed, this parcel of material would evolve to a Keplerian orbit closer to the rotation axis in order to conserve AM. Figure~\ref{fig:pressure_support_main}C,F shows the difference between the cylindrical radius of vapor in post-impact states and circular Keplerian orbits of the same AM, $\delta r_{xy}$. The presence of the pressure gradient in post-impact structures supports material many megameters farther away from the rotation axis than would be possible based solely on the specific AM of the structure and a balance between gravity and centrifugal forces.

\begin{figure*}
\centering 
\includegraphics[scale=0.83333333333]{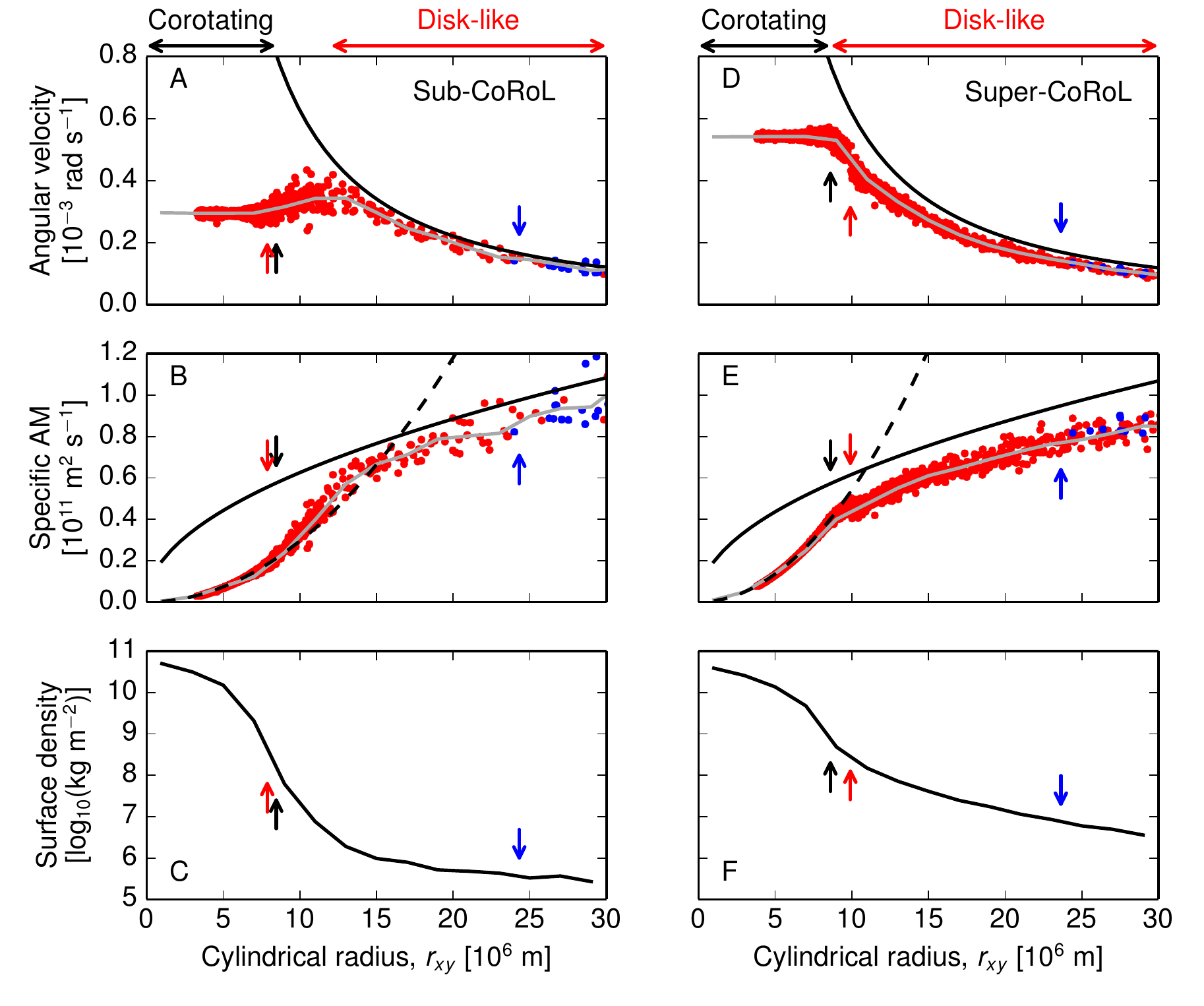}
\caption[]{The sub-CoRoL and super-CoRoL post-impact structures produced by canonical (A-C) and high-energy, high-AM (D-F) giant impacts have important differences. Midplane profiles of angular velocity (A,D), specific angular momentum (B,E) and surface density (C,F) correspond to the thermally and gravitationally equilibrated post-impact structures shown in Figure~\ref{fig:contourstructures}. Blue points indicate SPH particles that are within the vapor dome and are of mixed phase. Solid black lines denote the angular velocity or specific AM of a circular Keplerian orbit around a point mass, and dashed black lines indicate the specific AM for corotation with the inner region of the structure. Gray lines show the mass weighted average of the angular velocity and specific AM of particles in the midplane. Both surface density and average properties are calculated in $2$~Mm radial bins. Sub-CoRoL structures have a substantial shear boundary between the corotating region and disk-like region (points above the dashed line in B). Super-CoRoL structures, synestias, have a monotonically decaying angular velocity profile between the corotating and disk-like regions (D). Black arrows indicate the approximate outer edge of the corotating regions ($8.5$~Mm and $8.6$~Mm respectively). Red arrows indicate the inner edge of the isentropic regions ($7.9$~Mm and $9.9$~Mm). Blue arrows indicate the outer edge of the isentropic regions ($24.3$~Mm and $23.6$~Mm). The slight kinks in the profiles in A, B, D and E are caused by the entropy discontinuity at the inner edge of the isentropic region imposed during processing of post-impact structures (\S\ref{sec:cooling_methods}).}
\label{fig:linestructures}
\end{figure*}

All post-impact structures have a corotating inner region and a disk-like outer region (Figure~\ref{fig:linestructures}). The equatorial radius of the impact-heated corotating region is larger than the radius of a non-rotating, condensed planet of the same mass [LS17]. Different post-impact structures have corotating regions that rotate at different rates. In some cases, there is a difference between the corotating angular velocity and the angular velocity at the inner edge of the disk-like region, which requires the presence of a transition region between the corotating and disk-like regions (e.g., the portion of the gray line that lies above the dashed black line in Figure~\ref{fig:linestructures}B). The transition region in the fluid has a monotonically increasing angular velocity and, in the midplane, is similar to the profile for shear between two cylinders \citep[e.g.,][]{Chandrasekhar1969,Desch2013}. Structures with a significant transition region (such as the example in Figures~\ref{fig:contourstructures}A,B,C and \ref{fig:linestructures}A,B,C) tend to be below the corotation limit (CoRoL). The CoRoL is defined by where the angular velocity at the equator of a corotating planet intersects the Keplerian orbital velocity [LS17]. The CoRoL is a surface that depends on thermal state, AM, total mass, and compositional layering. In contrast, the structure shown in Figures~\ref{fig:contourstructures}E,F,G and \ref{fig:linestructures}D,E,F is above the CoRoL. The corotating region rotates much more rapidly than in the sub-CoRoL example. There is no transition region between the corotating and disk-like regions, and the angular velocity profile decreases monotonically with radius. Bodies above the CoRoL are called synestias [LS17].

The substantial pressure support in the disk-like region of post-impact structures leads to sub-Keplerian angular velocities (points below the black line in Figures~\ref{fig:linestructures}A,B,D,E). The effect of pressure support extends further away from the rotation axis in the example synestia. The surface density in the disk-like region of the synestia is an order of magnitude higher than in the sub-CoRoL case (Figure~\ref{fig:linestructures}), and consequently there are much higher pressures in the midplane (Figure~\ref{fig:contourstructures}). The surface density in the disk-like region of the sub-CoRoL structure is approximately constant (Figure~\ref{fig:contourstructures}C) as reported in previous studies \citep{Canup2001,Canup2004,Canup2008,Canup2008a}, but the constant surface density region begins at a larger radius than previously reported. The dynamic and thermodynamic structure of post-impact states depends strongly on the parameters of the impact that produced them. The synestia example shown here is typical of the impact-generated synestias found by LS17.

All studies of giant impacts in the regime of the Moon-forming event report substantial production of vapor. However, in previous work on post-impact structures \citep[e.g.,][]{Canup2001,Canup2004,Canup2008,Ida1997,Kokubo2000,Salmon2012,Salmon2014,Nakajima2014,Nakajima2015,Ward2012,Ward2014,Ward2017}, it was assumed that the vapor in the structure would cool rapidly and the pressure gradient term in Equation 1 could be neglected. Prior post-impact analyses generally simplified the structure to a point source planet and near-Keplerian disk. In general, the immediate post-impact structure cannot be analyzed in this manner [LS17]. 

Our conclusion that pressure has a significant effect in post-impact structure is not in conflict with the findings of prior work. For example, \citet{Nakajima2014} note the importance of the pressure gradient force. We analyzed the post-impact structures calculated by \citet{Nakajima2014} (pressure contours received from M. Nakajima by personal communication) to derive the relative magnitude of the force terms in Equation 1 and found that there are also strong radial pressure gradient forces. However, in calculating the surface density of their disk structures, \citet{Nakajima2014} generally neglected the pressure support.

Based on the results presented in this section, we find that post-giant impact structures must be analyzed as continuous rotating fluids and cannot be separated into a planet and near-Keplerian disk. The properties of the structure depend on the integrated effects of radial force balance (Equation~\ref{eqn:force_balance}), vertical hydrostatic equilibrium, and the thermal, mass, and angular momentum distributions imparted by the impact event, over the entire structure. 

%xxxxxxxxxxxxxxxxxxxxxxxxxxxxxxxxxxxxxxxxxxxxxxxxxxxxxxxxxxxxxxxx
%xxxxxxxxxxxxxxxxxxxxxxxxxxxxxxxxxxxxxxxxxxxxxxxxxxxxxxxxxxxxxxxx
\subsection{Transition from the impact to the post-impact structure}
\label{sec:dynamics_equilibration}

The methods that are currently used to simulate giant impacts do not include multiphase flow processes or thermal equilibration between parcels of material. In this section, we discuss the processes that govern the transition from the impact event to the post-impact state, the starting point for subsequent evolution and satellite formation. 

During and after the impact, condensates separate from the vapor and fall radially inwards and towards the midplane (\S\ref{sec:dynamics_cooling}). The separation of the condensate from the vapor in the outer regions of the post-impact structure changes the mass distribution and thermal structure. Condensates that fall into higher pressure, hotter regions of the structure vaporize and transfer their mass to vapor (\S\ref{sec:dynamics_cooling}). The fluid structure (Figure~\ref{fig:contourstructures}) adapts to the redistribution of mass and entropy. Condensates that have sufficient AM to remain beyond the Roche-limit can accrete to form satellites. These processes occur on dynamical timescales of hours to days.

Simultaneously, the vapor in the structure will convect, leading to rapid mixing vertically (parallel to the rotation axis). In all the example post-impact structures considered in this work, the specific entropy of the structure within the Roche limit is such that the equilibrium phase in the midplane is pure vapor (inside black lines in Figure~\ref{fig:contourstructures}A,B,E,F). The vertical thermal profile in these regions is adiabatic until it intersects the liquid-vapor phase boundary at lower pressures, at which point it follows a saturated adiabat. Farther out in the structure (beyond black lines in Figure~\ref{fig:contourstructures}A,B,E,F), the midplane pressure is lower, and the whole column is on a saturated adiabat. As noted by \citet{Nakajima2014}, a portion of each post-impact structure is approximately isentropic. Generally, the quasi-isentropic region has a specific entropy that exceeds the critical point [LS17]. This region corresponds to the approximately vertical subset of red SPH particles at pressures below the critical point in Figure~\ref{fig:contourstructures}D,H, which shows the thermal state of particles at the end of an SPH impact simulation.

The methods that are currently used to simulate giant impacts do not model the condensate separation or fluid convection that occurs in the hours after the impact. It is necessary to process the output from impact simulations to mimic the effect of these processes on the post-impact structure. We took the thermal profile from SPH impact simulations (Figure~\ref{fig:contourstructures}D,H), redistributed and removed condensate, and averaged the quasi-isentropic region to a single isentrope (Figure~\ref{fig:contourstructures}C,G). The details of this processing step are described in \S\ref{sec:cooling_methods} and \S\ref{sup:sec:SPH_cooling}. Following the rapid adjustment after the impact, there is a longer period of secular cooling of the post-impact structure that we discuss in the following sections.

%xxxxxxxxxxxxxxxxxxxxxxxxxxxxxxxxxxxxxxxxxxxxxxxxxxxxxxxxxxxxxxxx
%xxxxxxxxxxxxxxxxxxxxxxxxxxxxxxxxxxxxxxxxxxxxxxxxxxxxxxxxxxxxxxxx
\subsection{Cooling a synestia}
\label{sec:dynamics_cooling}

In \S\ref{sec:dynamics_structure} and \S\ref{sec:dynamics_equilibration}, we considered the structures of both sub-CoRoL structures and synestias. Hence forth, we focus solely on synestias and describe their evolution in the years following the impact. In particular, we consider the evolution of terrestrial synestias, those with Earth-like mass and composition. Some of the processes we consider are universal to all post-impact structures, but the evolution of sub-CoRoL structures is left to future work.  

A terrestrial synestia, as shown in Figures~\ref{fig:contourstructures}E-G, cools by radiation from the photosphere, where the structure is optically thin. We estimated the optical depth of the outer edge of the structure and found that the photic surface is at low pressures ($10^{-6}$ to $10^{-2}$~bar, supporting information \S\ref{sup:sec:photosphere}) and radiates at a temperature of about $T_{\rm rad}=2300$~K, determined by the liquid-vapor phase boundary of BSE composition material (\S\ref{sec:thermo}). 
At the photosphere, the majority of the energy lost by radiation is compensated for by condensation of vapor, and the material state is a mixture of vapor and condensates (liquid droplets and/or solid dust). Since condensates are not supported by the pressure gradient of the vapor structure, they are not dynamically stable at the photic surface and will fall. Because the photospheric temperature is far above equilibrium with the incoming solar radiation, radiative cooling is efficient and drives a torrential rain of condensates into the higher-pressure regions of the structure. Initially, radiative cooling leads to on the order of a lunar mass of silicate condensate per year, and the production rate of silicate rain near the photosphere of a synestia (a few centimeters an hour) is about an order of magnitude greater than heavy rain fall during hurricanes on Earth today.

The size of droplets forming at the photosphere and falling is controlled by a balance between shear from the vapor and surface tension. We approximate the drag force on condensates by assuming the droplets are in free fall such that
\begin{equation}
F_{\rm D} = m_{\rm cond} g \, ,
\end{equation}
where $F_{\rm D}$ is the drag force from the gas, $m_{\rm cond}$ is the mass of the condensate, and $g$ is the gravitational acceleration. The droplet size is then given by the balance between the drag force and surface tension over the droplet,
\begin{equation}
m_{\rm cond} g \sim \rho_{\rm cond} R_{\rm cond}^3 g \sim R_{\rm cond} \gamma \, ,
\end{equation}
where $\rho_{\rm cond}$ is the density of the condensate, $R_{\rm cond}$ is the radius of the droplet, and $\gamma$ is the surface tension. \citet{Boca2003} found that the surface tension for silica melts at 2000~K is 0.3~N~m$^{-1}$. From the balance of the forces, the radius of the droplet is 
\begin{equation}
R_{\rm cond} \sim \sqrt{\frac{\gamma}{\rho_{\rm cond} g}} \, .
\end{equation}
The gravitational acceleration at the photosphere ranges from 10 to 0.1~m~s$^{-2}$ due to the large spatial scale of the synestia. The corresponding range in droplet sizes is a few millimeters to a few centimeters.

As discussed in \S\ref{sec:dynamics_structure}, there is a substantial pressure gradient force perpendicular to the rotation axis acting on the vapor in synestias. 
When condensates form from the vapor at the photosphere, they do not have sufficient AM to remain in a circular orbit at the same location.
In the absence of vapor, the condensates would fall rapidly on significantly elliptical orbits in a plane through the center of mass. For the synestia shown in Figures~\ref{fig:contourstructures}-\ref{fig:linestructures}, the initial eccentricity varies from 0.6 at the inner edge of the disk-like region to 0.27 at a cylindrical radius of 25~Mm. As they fall into the synestia, condensates interact with the vapor and experience drag which perturbs their motion. 

To calculate the motions of condensates in the synestia, we used a simple orbital evolution model, including an acceleration due to gas drag of the form
\begin{equation}
\mathbf{a_{\rm D}} = - \frac{3 \rho_{\rm vap}}{8 R_{\rm cond} \rho_{\rm cond}}  C_{\rm D} \left | \mathbf{v} - \mathbf{v_{\rm vap}} \right | \left ( \mathbf{v} - \mathbf{v_{\rm vap}} \right ) \, ,
\label{eqn:gas_drag}
\end{equation}
where $\rho_{\rm vap}$ is the density of the vapor, $C_{\rm D}$ is the gas drag coefficient, $\mathbf{v}$ is the velocity vector of the condensate, and $\mathbf{v_{\rm vap}}$ is the velocity vector of the vapor. A set of simple classical dynamics equations were then integrated to find the position and velocity of the particle as a function of time. Full details of the calculation are given in \ref{sup:sec:gas_drag}.

\begin{sidewaysfigure*}
\centering
\includegraphics[scale=0.833333333]{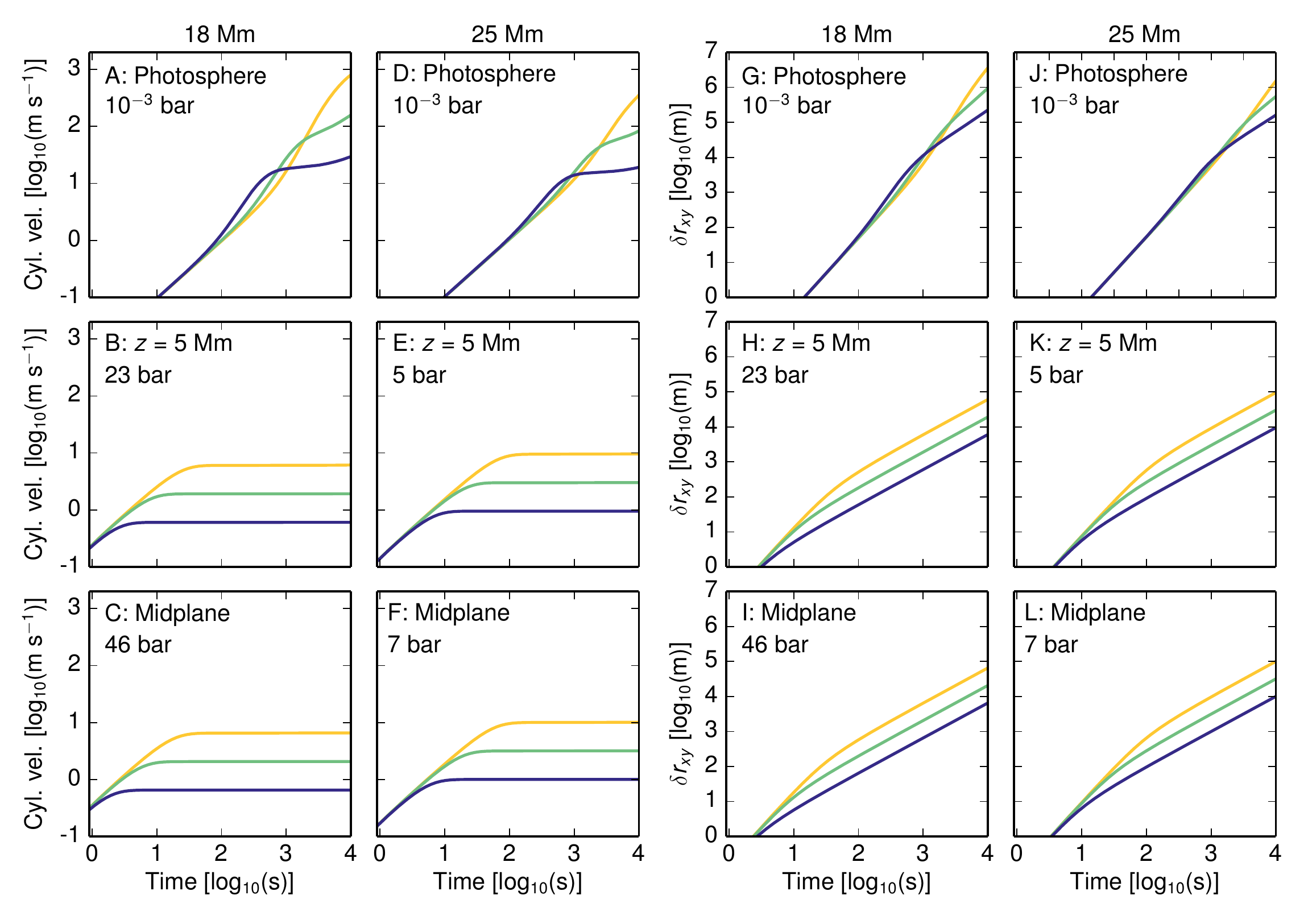}
\caption{Condensates formed at the photosphere of a synestia can rapidly spiral inwards in cylindrical radius, towards the rotation axis. Shown are the velocities of individual condensates towards the rotation axis (A-F) and the corresponding change in cylindrical radius ($\delta r_{xy}$, G-L) as a function of time after formation, for condensates in the post-impact synestia in Figures~\ref{fig:contourstructures} and \ref{fig:linestructures}. The condensate behavior is demonstrated at two different cylindrical radii (18 and 25~Mm) and at different heights in the structure, with correspondingly different gas densities. The top row shows the case of condensation at the approximate height of the photosphere of the structure (10 and 13~Mm for the two radii) with a gas density of $1.6$~$\times$~$10^{-4}$~kg~m$^{-3}$, corresponding to vapor on a saturated adiabat at $10^{-3}$~bar. The second row shows the dynamics at 5~Mm above the midplane at pressures interpolated from the post-impact synestia (23 and 5~bar). The corresponding gas densities were 2.1 and 0.6~kg~m$^{-3}$. The bottom row is for condensates formed in the midplane at pressures of 46 and 7~bar, and gas densities of 3.7 and 0.8~kg~m$^{-3}$. The condensates were initialized at the vapor velocity, and the gas density was assumed to be constant in each calculation. We considered condensates with radii of 1~mm (blue), 1~cm (green), and 10~cm (yellow). The density of the condensates was assumed to be 3000~kg~m$^{-3}$.
}
\label{fig:gas_drag}
\end{sidewaysfigure*}

Figure~\ref{fig:gas_drag} presents the velocity and distance traveled in the direction perpendicular to the rotation axis for small, isolated condensates starting at different points in the synestia shown in Figures~\ref{fig:contourstructures}E-G. We initialized the condensates at a velocity equal to the gas velocity, as if the particle had just condensed from the vapor. The gas velocity in the structure was calculated using the midplane angular velocity profile of the SPH structure, assuming that the angular velocity did not vary with height above the midplane in accordance with the Poincar\'{e}-Wavre theorem. We assumed the vapor motion was purely azimuthal. In these calculations, the gas density was constant. In the outer regions of the structure the scale height is large (on the order of megameters), and over the timescales shown in Figure~\ref{fig:gas_drag}, the particles do not typically move more than a scale height. The assumption of constant density does not have a significant effect on our results.

At the photosphere, the residual vapor after condensation is low density ($\sim 10^{-4}$~kg~m$^{-3}$) and offers little resistance to the falling particles. Condensates of the size calculated above (a few millimeters to a few centimeters) rapidly accelerate to velocities of several tens to hundreds of meters per second and both fall vertically towards the midplane and spiral in towards the rotation axis (Figure~\ref{fig:gas_drag}). The velocity initially increases rapidly and then plateaus at later times. Condensates originating closer to the midplane behave similarly to those formed at the photosphere, but, due to the higher gas density and hence amplified drag, reach lower terminal velocities (on the order of a few meters per second, Figure~\ref{fig:gas_drag}). The rate of radial infall is greater for larger condensates that experience less gas drag. The AM of the condensates formed from the vapor is such that they begin falling on highly elliptical orbits. The orbits of smaller condensates are rapidly circularized and the particles spiral inwards. Particles that experience less gas drag are less perturbed from their original elliptical orbits and fall more rapidly towards the rotation axis. 

We pause to confirm the typical droplet size using our condensate orbit calculations. Using the form in Equation~\ref{eqn:gas_drag}, we once again assume a balance between gas drag and surface tension
\begin{equation}
\frac{\rho_{\rm vap}}{R_{\rm cond} \rho_{\rm cond}}  C_{\rm D} \left | \mathbf{v} - \mathbf{v_{\rm vap}} \right |^2 \sim \frac{\gamma R_{\rm cond}}{m_{\rm cond}}\, .
\end{equation}
Rearranging, the droplet size is given by
\begin{equation}
R_{\rm cond} \sim \frac{\gamma }{ \rho_{\rm vap} C_{\rm D}  \left | \mathbf{v} - \mathbf{v_{\rm vap}} \right |^2 } \, .
\label{eqn:droplet_size_adv}
\end{equation}
In our calculations, the differential velocity varies depending on droplet size, with larger droplets having greater shear. At the later times shown in Figure~\ref{fig:gas_drag}, the 10~cm bodies reach velocities at which they would be sheared apart by the gas. For the velocities at $10^4$~s in Figure~\ref{fig:gas_drag}, the largest droplets that would travel slowly enough to not shear apart are on the order of a few millimeters to a few centimeters, in good agreement with our earlier size estimate. At earlier times, when the differential velocities are smaller, condensates could have been much larger. The equilibrium droplet size does not vary significantly with height in the structure as increased gas density is compensated for by lower differential velocities. 

Falling condensates are a radial mass transport mechanism in synestias.
Condensates originating at the photosphere rapidly accelerate to velocities comparable to the convective velocities of the vapor (hundreds of meters per second at the photosphere, see \S\ref{sup:sec:mixing}), and are likely to avoid entrainment by gas convection. As a result of rapid radial motion, condensates fall at relatively shallow angles to the photosphere. Due to the large scale height of the outer regions, condensates can move considerable distances at relatively low pressures and hence low vapor densities. In higher-density regions of the structure, the relative velocity between condensate and gas is lower, and it is possible that condensates could be entrained in the turbulent fluid. As a result, the bulk of the radial mass transport by condensates likely occurs near the photosphere of the synestia.

The efficiency of condensates as a radial mass transport mechanism is dependent on how long condensates can survive in the synestia before vaporizing. At the photosphere, condensates are in thermal equilibrium with the vapor and could persist indefinitely. However, as they fall into higher pressure regions of the structure, condensates are heated and begin to vaporize. Here, we approximate lower limits on the evaporation timescales for individual, isolated condensates. 
We assume that condensates are heated by blackbody radiation from the surrounding vapor.
The net power gained by a spherical particle is given by
\begin{linenomath*}
\begin{align}
P= 4 \pi R_{\rm cond}^2  \sigma \left (T_{\rm vap}^4 - T_{\rm cond}^4 \right ) ,
\label{eqn:cond_flux}
\end{align}
\end{linenomath*}
where $\sigma$ is the Boltzmann constant, $T_{\rm vap}$ and $T_{\rm cond}$ are the temperatures of the surrounding vapor and condensate respectively, and $R_{\rm cond}$ is the radius of the condensate.
Assuming a homogeneous condensate, the energy balance is given by
\begin{linenomath*}
\begin{align}
l\frac{d M}{d t} - c_{p} M \frac{d T_{\rm cond}}{d t} =- P ,
\label{eqn:cond_vap1}
\end{align}
\end{linenomath*}
where $M$ is the mass, $l$ is the latent heat, $c_{p}$ is the specific heat capacity, and $t$ is time.
Assuming that vaporization occurs linearly between the initial temperature of the condensate, $T_{\rm cond}^0$, and the temperature of complete vaporization, $\bar{T}_{\rm vap}$, 
\begin{linenomath*}
\begin{align}
M=\left ( \frac{\bar{T}_{\rm vap} - T_{\rm cond} }{\bar{T}_{\rm vap} - T_{\rm cond}^0} \right ) M_0 ,
\label{eqn:cond_M}
\end{align}
\end{linenomath*}
for $T_{\rm cond}^0 \leq T_{\rm cond} \leq \bar{T}_{\rm vap}$, where $M_0$ is the initial mass of the condensate. This is a reasonable approximation for moderate pressures and initial condensate temperatures close to the liquid-vapor phase boundary (see \S\ref{sec:thermo}). With this assumption, the temperature increases linearly with mass loss,
\begin{linenomath*}
\begin{align}
\frac{d T_{\rm cond}}{d t}=- \left ( \frac{\bar{T}_{\rm vap} - T_{\rm cond}^0 }{M_0} \right ) \frac{d M}{d t} .
\label{eqn:cond_dTdt}
\end{align}
\end{linenomath*}
Using Equations \ref{eqn:cond_vap1} and \ref{eqn:cond_dTdt},  the condensate mass changes at a rate
\begin{linenomath*}
\begin{align}
\frac{d M}{d t}= -P \left [ l + c_{p} M   \left ( \frac{\bar{T}_{\rm vap} - T_{\rm cond}^0 }{M_0} \right ) \right ]^{-1} .
\label{eqn:cond_dMdt}
\end{align}
\end{linenomath*}
We solve Equation \ref{eqn:cond_dMdt}, using Equations \ref{eqn:cond_flux} and \ref{eqn:cond_M}, for the mass of the condensate as a function of time.

We present two example calculations of the vaporization of condensates: at pressures close to the photosphere at $10^{-3}$~bar and at a midplane pressure of 20~bar. In \S\ref{sec:thermo}, we calculate the multicomponent phase diagram for silicates in a terrestrial synestia, and find that, at $10^{-3}$~bar, vaporization of initially fully condensed material occurs over a temperature range between $2300$ and $2700$~K (Figure~\ref{fig:cond}A). Based on this calculation, we used $T_{\rm cond}^0=2300$~K and $\bar{T}_{\rm vap}=2700$~K for our low pressure calculations. Similarly, we use $T_{\rm cond}^0=3700$~K and $\bar{T}_{\rm vap}=4200$~K for the midplane (Figure~\ref{fig:cond}B).
We do not consider the time for the condensate to heat up to $T_{\rm cond}^0$ since it is negligible compared to the vaporization time as the specific heat capacity is four orders of magnitude less than the latent heat of vaporization.
In the outer regions of a structure during cooling, the vapor is expected to be close to the liquid-vapor phase boundary (Figure~\ref{fig:contourstructures}). We hence assume reference vapor temperatures similar to $\bar{T}_{\rm vap}$, $2750$~K at $10^{-3}$~bar and $4250$~K in the midplane.
We used a latent heat of $l = 1.7 \times 10^{7}$~J~K$^{-1}$~kg$^{-1}$, which has been widely used in lunar disk studies and is consistent with the EOS used in the SPH simulations \citep[e.g.][]{Thompson1988,Ward2012}, a condensate density of 3000~kg~m$^{-3}$, and a specific heat capacity of 1000~J~K$^{-1}$~kg$^{-1}$, a typical value for silicate melts \citep{Stebbins1984}.

\begin{figure}[p]
\centering
\includegraphics[scale=0.8333333]{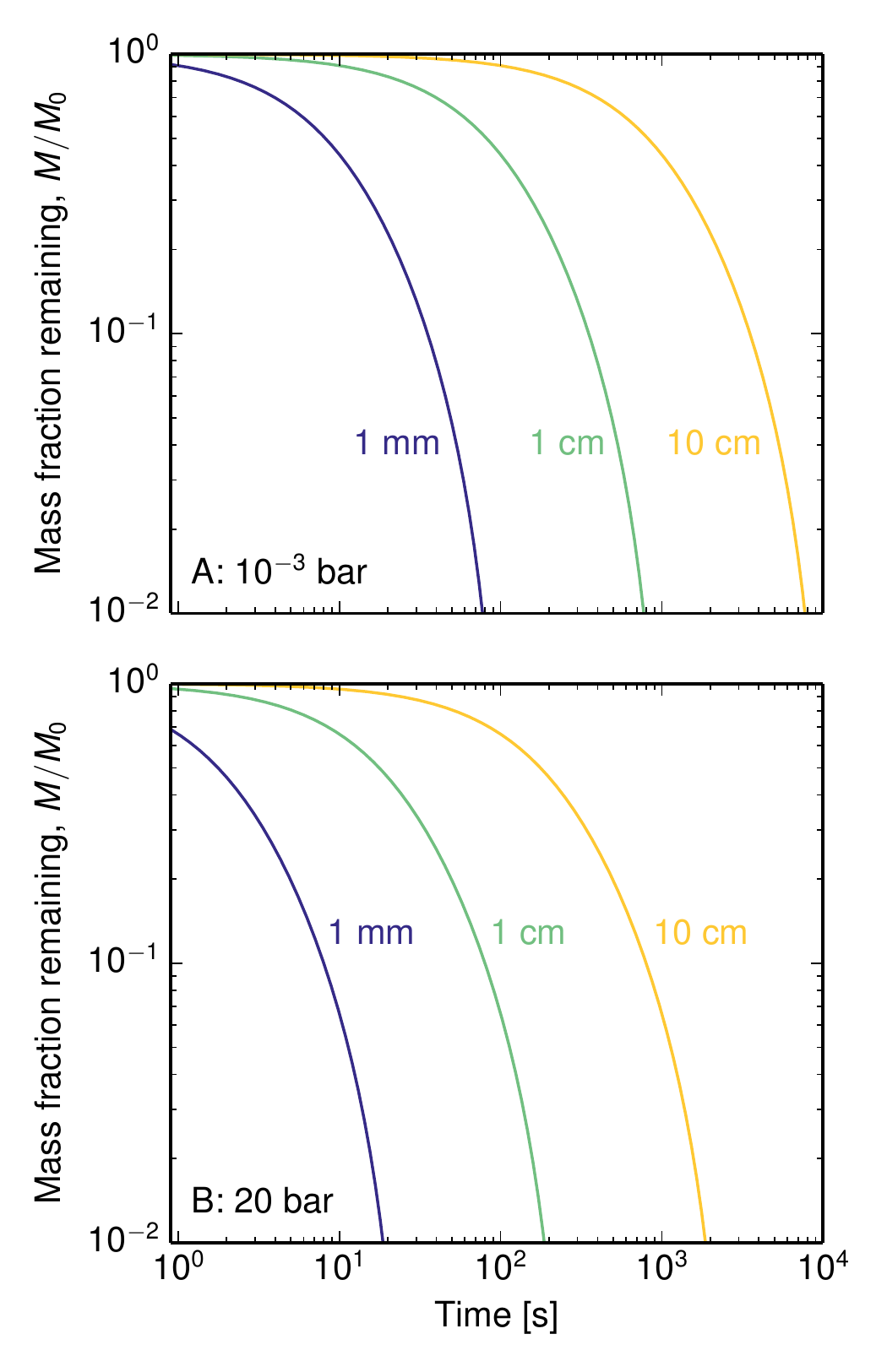}
\caption{Small, isolated droplets can survive for minutes to tens of minutes in the pure-vapor region of the synestia. Panels show (A) the loss of mass over time for small condensates of a given initial radius at low pressure and (B) midplane pressures. Mass loss is more rapid at the midplane due to the higher vapor temperature and larger temperature difference between the condensate and vapor.}
\label{fig:vap_droplets}
\end{figure}

Isolated condensates, of the typical sizes we calculated, can survive on the order of minutes to tens of minutes in the vapor of the synestia (Figure~\ref{fig:vap_droplets}). Our estimate for the vaporization timescales of condensates are lower limits, and a number of processes could increase the survival time of condensates. Energy exchange by black body radiation with pure vapor is the most efficient energy transfer process possible. The vapor produced from the condensate is colder than the gas of the surrounding structure. The colder gas will partly insulate the condensate from radiation and thermal exchange with the hotter vapor of the synestia. We did not account for the heating of condensates by thermal diffusion, due to the difficulty of accounting for this blanketing effect. Collisions and combination of small condensates would lower the effective surface area to mass ratio, again reducing the efficiency of vaporization. 

The short lifespan of small, isolated condensates in the synestia limits their ability to transport mass radially. An isolated condensate at the photosphere could only move on the order of 10~km towards the rotation axis before vaporizing. However, there is a substantial mass of condensate being continuously formed at the photosphere. When a fluid parcel rising in the synestia reaches the photosphere, a large mass fraction condenses almost instantaneously. Due to the high radiative temperature, a 1~km thick photosphere would condense in about a second. Condensates forming at the photosphere are not isolated but are falling as part of larger, condensate-rich downwellings.

Condensates dominate the optical depth of silicate vapor-melt mixtures (\S\ref{sup:sec:photosphere}), and the large mass-fraction of condensates in downwellings makes them initially optically thick. Thus, the radiation field within the downwelling is dominated by emission from the lower temperature condensates and not the hot vapor of the synestia. Heating from radiation within the mixture is much less efficient than for isolated condensates. Heating can also occur by thermal conduction between the condensates and the vapor that they are falling through. As condensates vaporize, they produce gas that is in thermal equilibrium with the condensate. The gas flow then advects the lost vapor away from the particle. In a downwelling, subsequent condensates are falling into vapor produced by partial vaporization of the leading condensates. The temperature difference between the condensates and the vapor is smaller and the condensates are not efficiently heated. The effects of self-shielding and self-buffering by condensates in the downwelling last until the downwelling is broken up by eddies in the fluid. The timescale for this process is uncertain but is likely to be on convective timescales. Thus, the lifetime of condensates in downwellings would be significantly longer than isolated condensates. 

Due to their longer lifetimes, condensates in condensate-rich downwellings from the photosphere could transport mass over substantial radial distances. As the radial velocity of condensates increases rapidly with time (Figure~\ref{fig:gas_drag}), a relatively small increase in survival time can significantly increase the radial distance traversed. For example, assuming the infall rates are the same for isolated condensates and those in downwellings, a survival time of $10^4$~s (a few hours) would allow for radial mass transport at the photosphere over $10^5$-$10^6$~m. The survival time for isolated condensates that we calculated above cannot be applied to groups of particles. The survival time of condensates in condensate-rich downwellings is much longer, and we expect that condensates can transport mass over megameter lengthscales before vaporizing.

Condensates could also transport mass in regions where condensates are stable in the midplane (region outside the black line in Figures~\ref{fig:contourstructures}E,F). These condensates would also experience drag and spiral inwards towards the rotation axis. As the midplane pressure is higher than at the photosphere ($0.1$-$10$~kg~m$^{-3}$), the infall velocities are somewhat smaller (several to tens of meters per second for centimeter sized bodies). However, where condensates are thermodynamic stable, they can fall for substantial distances. Outside the Roche limit, mutual collisions between condensates would lead to the accretion of larger bodies. The radial infall of moderate sized bodies can be more efficient than for smaller condensates depending on the gas density. Condensates must accrete onto moonlet sized bodies in order to avoid spiraling into higher density regions of the structure and revaporizing (\S\ref{sec:decoupling}).

A more sophisticated model of condensation and gas drag is needed to fully examine the dynamics of condensates in the synestia. However, our simple calculations suggest that falling condensates are a significant radial mass transport mechanism. During the evolution of synestias several lunar masses of material condenses at the photosphere (\S\ref{sec:cooling_results}) and could be advected substantial distances (e.g., hundreds of kilometers to megameters). Significantly, condensates can move mass perpendicular to the rotation axis, whereas mass transport radially in the vapor is difficult due to the substantial Coriolis force. As condensates are dragged by the gas, they also deposit AM into the vapor. Large scale mass and AM transport radially by condensates is a process that has not previously been considered in post-impact evolution models. 

The condensation of mass in the synestia is also a major component of the energy budget. The energy transported by condensates and lost by radiation from the photosphere is redistributed vertically by convection. We estimated the vertical convective mixing timescale for the structure using mixing length theory by making an analogy to purely thermal convection in rotating systems (\S\ref{sup:sec:mixing}).  
We found that convective velocities in the outer regions of the synestia are on the order of tens of meters per second in the midplane and hundreds of meters per second at the photosphere. The corresponding mixing timescales are on the order of days, much shorter than the cooling timescale for the disk-like region, which is on the order of years (\S\ref{sec:cooling_results}). The mass in each column of vapor is cooled simultaneously by radiation and condensate transport.

Radiation from the photosphere leads to rapid evolution of a synestia. As post-impact structures cool, the entropy of the vapor decreases, a fraction of the vapor condenses, and the pressure support is reduced. The production of condensates leads to a reduction in the vapor surface density. Material is no longer supported at such large radii and the structure radially contracts. Next, we describe a model for calculating the cooling of synestias and present example cooling simulations. 

%xxxxxxxxxxxxxxxxxxxxxxxxxxxxxxxxxxxxxxxxxxxxxxxxxxxxxxxxxxxxxxxx
%xxxxxxxxxxxxxxxxxxxxxxxxxxxxxxxxxxxxxxxxxxxxxxxxxxxxxxxxxxxxxxxx
\subsection{Cooling calculation: Methods}
\label{sec:cooling_methods}

Based on the results presented in previous sections, post-giant impact structures must be analyzed as continuous rotating fluids. For the early time evolution after the impact event, a fluid code calculation that approximates the redistribution of condensing material can capture the basic physics of the system. Here, we studied the early evolution of post-impact structures using SPH methods.

We constructed a simple model to calculate the physical structure (e.g., mass, pressure and entropy distribution) of a synestia during radiative cooling. This calculation is intended to assess the timescales for cooling, the potential mass and orbit of a primary satellite, and the magnitude of the vapor pressure around the growing satellite. In this work, we neglect or simplify a number of physical processes and, therefore, take a conservative approach and attempt to estimate the fastest possible timescale for cooling and condensing the Roche-exterior mass of a synestia. In this section, we describe the methods used in the calculation. Complete details of the code implementation of the model are provided in the supporting information (\S\ref{sup:sec:SPH_cooling}).

We adapted the GADGET-2 SPH code to calculate the cooling of post-impact synestias. In LS17, GADGET-2 was used to calculate the equilibrium structure of corotating planets and the results compared well with a potential field method. LS17 also used GADGET-2 to generate synthetic synestias, formed by heating isolated planets, as well as impact-generated synestias. Based on this previous work, GADGET-2 is able to solve for the pressure field of a synestia of a given distribution of mass, AM and thermal energy.

Because synestias evolve both by mass redistribution by condensates and viscous spreading, they are not static structures. The structure of a synestia with a given mass and AM is not unique and is dependent on the physical processes that generated the synestia and acted during its evolution. Therefore, our calculation includes a series of steps that model the major physical processes controlling the creation and evolution of a terrestrial synestia. First, we generate a synestia by a giant impact that is calculated until a near-axisymmetric structure is achieved (24 to 48 hours). Second, motivated by our discussion of the multiphase dynamics in the synestias during and immediately after the impact in \S\ref{sec:dynamics_equilibration}, the outer portion of the post-impact structure is thermally equilibrated. In this step, the fraction of condensates with sufficient AM to be rotationally supported in circular orbits beyond the Roche limit is removed from the SPH calculation under the assumption that this material quickly accretes onto a body that we refer to as the seed of the moon. Third, we mimic radiative cooling by decreasing the thermal energy of SPH particles while accounting for condensation and redistribution of mass and AM. We estimate the mass and orbit of the moon that is formed and determine the range of vapor pressures surrounding the moon from the calculated pressure structure of the cooling synestia.

In the first step, giant impacts were modeled in the same manner as in \cite{Cuk2012} and LS17. For this study, we drew examples of impact-generated synestias from the database presented in LS17.
The impacting bodies were differentiated (2/3 rocky mantle, 1/3 iron core by mass) with forsterite mantles and iron cores modeled using M-ANEOS equations of state \citep{Melosh2007,Canup2012}.
The forsterite EOS is a single component model, which provides a simple treatment of the liquid--vapor phase boundary.
In \S\ref{sec:comb}, we discuss the effect of a multicomponent phase boundary on the evolution of the structure and the formation of a moon.

In the second step, the outer portion of the post-impact structure was thermally equilibrated. The adjustment of the structure to the changes in entropy is calculated in a manner identical to the third step (cooling) but with no radiative heat loss. Based on the unique post-impact entropy distribution in the structure, a value of specific entropy was chosen to demarcate between an inner stratified region and a well-mixed outer region. SPH particles in the well-mixed region are divided into two groups: an isentropic pure-vapor group and a vapor-dome group. The three thermal groups are shown schematically in Figure~\ref{fig:methods_cartoon}A. Particles in the isentropic vapor group were all assigned the same entropy given by the mass-weighted mean value of the particles in that group (yellow particles). In the vapor dome group, the mass fraction of condensate was removed from each particle using the lever rule. The remaining mass was assigned the density and specific entropy of vapor on the phase boundary at the same pressure (blue particles on the saturated adiabat). Examples of the change in the entropy distribution in modeled synestias are shown in Figure~\ref{fig:contourstructures} (D-C, H-G) and \ref{sup:fig:thermalequil} (C-F). The position and velocity of the particles were not changed and hence they retained the same specific AM, $j$. 

The extracted condensate with specific AM greater than that of a body in a circular, Keplerian orbit at the Roche limit, $j_{\rm Roche}$, was assumed to accrete into a single body. The mass of condensate with specific AM less than $j_{\rm Roche}$ was distributed evenly in radial bins between the point of origin and the radius corresponding to a circular, Keplerian orbit for the specific AM of the condensate. The redistribution is shown schematically in Figure~\ref{fig:methods_cartoon}C. The mass of condensate that fell into a given bin was divided equally between each particle in that bin belonging to the isentropic group. The motivation for this simple mass distribution function is the expectation that condensates would be vaporized and reincorporated into the structure over a range of radii. Since the details of the dynamics of condensates in the synestia are not yet known, we chose a simple mass-distribution function in order to investigate the role of condensates in the evolution of the structure. For the purposes of this section, we refer to condensates with $j$~$<$~$j_{\rm Roche}$ as falling condensates. The thermal effect from falling condensates is discussed below. After the extraction of condensates and the redistribution of mass, the physical structure was evolved using the forces calculated by the SPH code under the constraint of the EOS. At each time step the identification of groups, and condensate extraction and redistribution were repeated. Typically, after a few time steps the production of condensate became negligible and the structure attained the desired thermal profile. The SPH calculation was continued for a few dynamical times (typically several hours) to form a quasi-steady structure. 

In the third step, the specific entropy of the well-mixed region was reduced in a process that approximates radiative cooling (Figure~\ref{fig:methods_cartoon}B). The time step in SPH codes is limited by the sound speed and the Courant criterion. In addition, the artificial viscosity in the SPH code causes unrealistically rapid viscous spreading of the structure. Thus, an SPH code cannot be used to directly simulate the 10 to 100 year evolution of a synestia. Here, our goal is to estimate the temporal distribution of condensates and the approximate pressure field of the cooling vapor structure. To overcome the timescale issue, radiative cooling was implemented using a large effective radiating temperature, $T_{\rm eff}$, to determine the energy loss per time step. Then the calculation time was scaled by a factor of $(T_{\rm eff}/T_{\rm rad})^4$, where $T_{\rm rad}$ is the true radiating temperature, to obtain the corresponding cooling time. The evolution of the synestia was broadly similar for a range of $T_{\rm eff}$, and we typically used 15,000 or 20,000~K. 

In each cooling time step, we identified the particle groups and recalculated the entropy of the isentropic group (Figure~\ref{fig:methods_cartoon}A). Next, we decreased the specific entropy of each well-mixed particle in each 1~Mm radial bin such that the total enthalpy removed equaled the radiative energy lost from the surface area of that bin. The process is illustrated by the cooling of a single bin, indexed by $k$, by $dQ_k$ in Figure~\ref{fig:methods_cartoon}B. For example, the $i$th particle initially on the saturated adiabat will cool, reducing its specific entropy by $dS_k$. As a result, a portion of the particle condenses, $dm_i$. This process is repeated for each particle in each bin.  

The condensate fraction was removed in the same manner as in the thermal equilibration step. The mass of condensates falling within the Roche limit, $a_{\rm R}$, was redistributed by adding mass to isentropic group particles and reducing the enthalpy of those particles by an amount determined by the latent heat of vaporization. In Figure~\ref{fig:methods_cartoon}C, an example particle $j$ in bin $l$ increases in mass by $dm_j$ and decreases in entropy by $dS_j$. This procedure is repeated for every particle in bin $l$. The addition of mass from falling and vaporizing condensates also mimics the transfer of the original condensate's AM to the gas over a range of radii. If the structure cooled to the point where a given Roche-interior bin no longer contained pure vapor particles, the condensates falling into that bin were removed. In this work, we did not attempt to model the evolution of any thermodynamically stable condensates within the Roche limit, and removal of this material follows our conservative approach of estimating the fastest possible cooling time for the vapor structure.

Based on the results of the thermal equilibration and cooling steps, we estimated the mass and angular momentum of a growing moon. We estimate the range of mass and AM for a primary satellite using two different assumptions. The first estimate (A) only includes condensates with specific AM larger than $j_{\rm Roche}$. The second estimate (B) adds falling condensates that fell beyond the Roche limit according to our simple condensate redistribution scheme. Both estimates of the satellite mass assume perfect accretion of Roche-exterior condensates into a single body. Perfect accretion is consistent with $N$-body simulations that show efficient accretion of condensates beyond the Roche limit \citep[e.g.,][]{Ida1997, Kokubo2000, Salmon2012,Salmon2014}. The growing moon has a large collisional cross section and would accrete some falling condensates, which motivates our inclusion of them for the moon B estimate. The addition of falling condensates increases the total mass and decreases the specific AM of moon B compared to moon A. We calculate the radii of the circular orbits of moons A and B and record the midplane pressure of the vapor structure at those radii. 

In our model, we make a number of simplifications. Condensates are assumed to perfectly separate from the vapor structure and the gravitational effects of condensates are neglected. Thus, neither the gravitational perturbations from the growing moon on the vapor structure nor the effect of gas drag from the structure on the satellite orbit are included. The gravitational field of the growing moon would increase the local vapor pressure and the vapor pressure around the growing moon in our calculation is a hard lower limit on the pressure around the moon (\S\ref{sec:boundary_layer}). Conversely, the pressure at the Roche limit provides a weak upper limit on the pressure around the moon, assuming the gravitational effect of moonlets are negligible. We neglect tidal forces and dynamical resonances, such as Lindblad resonances, as discussed in the supporting information. Internal heating by viscous dissipation is not included in the energy budget in order to determine the fastest cooling time. The contribution from viscous heating to the energy budget is uncertain as discussed in \citet{Charnoz2015} who showed that the viscous heating rate is much less than the radiative cooling rate in canonical disks.

%SPH method cartoon
\begin{figure*}
\centering
\includegraphics[scale=0.833333333]{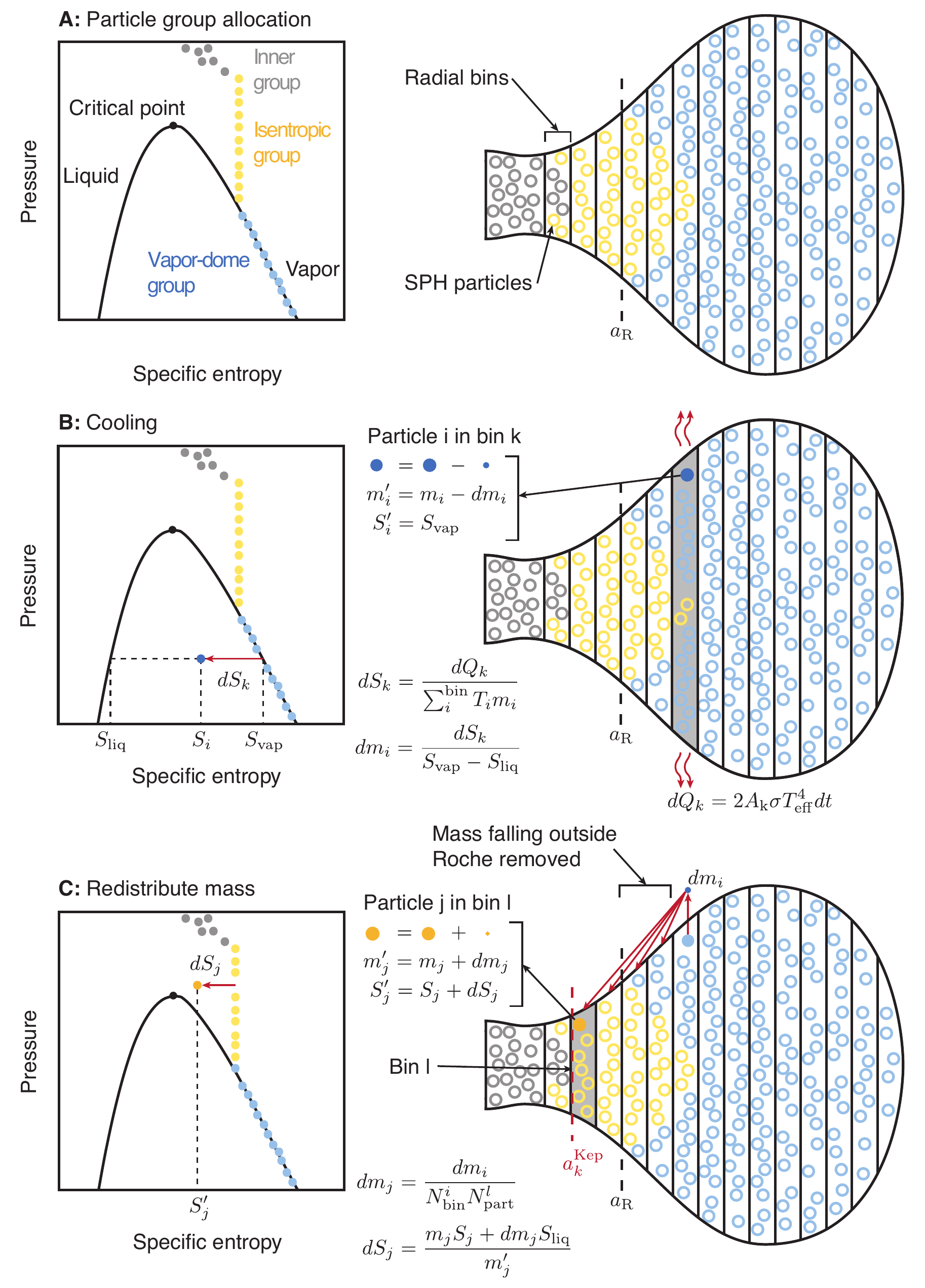}
\caption{Caption on next page.}
\end{figure*}
%additional figure for caption
\setcounter{figure}{5}
\begin{figure*}
\caption{A schematic of the radiative cooling model for a synestia. The left column shows the position of a select number of SPH particles in pressure-specific entropy space (colored points). The liquid-vapor phase boundary is shown in black. The right column presents a spatial schematic of the cooling synestia as a cross section parallel to the rotation axis. SPH particles in the inner group (gray), isentropic group (yellow), and vapor dome group (blue) are given as colored circles and the division of radial bins is shown by the black lines. The dashed line indicates the Roche limit, $a_{\rm R}$. A. At the beginning of each time step, particles are assigned to thermodynamic groups. B. Every particle $i$ in bin $k$ is cooled in proportion to radiative heat loss over the surface area $2A_k$. Cooling is calculated for each radial bin. C. The mass falling into bin $l$ is added to all isentropic particles and the enthalpy of the particles is reduced by the latent heat of vaporization. Mass addition and revaporization is repeated for each radial bin. The variables are defined as follows: $m_i$ is the original mass of particle $i$, $dm_i$ is the change in mass of particle $i$ upon condensation or upon addition of mass to a bin, $m_i'$ is the new mass of particle $i$ upon cooling or addition of mass, $S_{\rm liq}$ and $S_{\rm vap}$ are the specific entropies on the liquid and vapor side of the liquid-vapor phase boundary respectively at the pressure of particle $i$, $S_i'$ is the updated entropy of particle $i$, $dQ_k$ is the energy lost by bin $k$ due to cooling, $\sigma$ is the Stefan-Boltzmann constant, $dt$ is the time increment, $T_{\rm eff}$ is the effective radiative temperature, $T_i$ is the temperature of particle $i$, $a_k^{\rm Kep}$ is the radius of the circular Keplerian orbit corresponding to the specific AM of particle $k$, $N_{\rm bin}^i$ is the number of bins into which mass is being redistributed from particle $i$, and $N_{\rm part}^l$ is the number of isentropic particles in bin $l$.}
\label{fig:methods_cartoon}
\end{figure*} 

%xxxxxxxxxxxxxxxxxxxxxxxxxxxxxxxxxxxxxxxxxxxxxxxxxxxxxxxxxxxxxxxx
%xxxxxxxxxxxxxxxxxxxxxxxxxxxxxxxxxxxxxxxxxxxxxxxxxxxxxxxxxxxxxxxx
\subsection{Cooling calculation: Results}
\label{sec:cooling_results}

\begin{sidewaysfigure*}
\centering 
\includegraphics[scale=0.833333333]{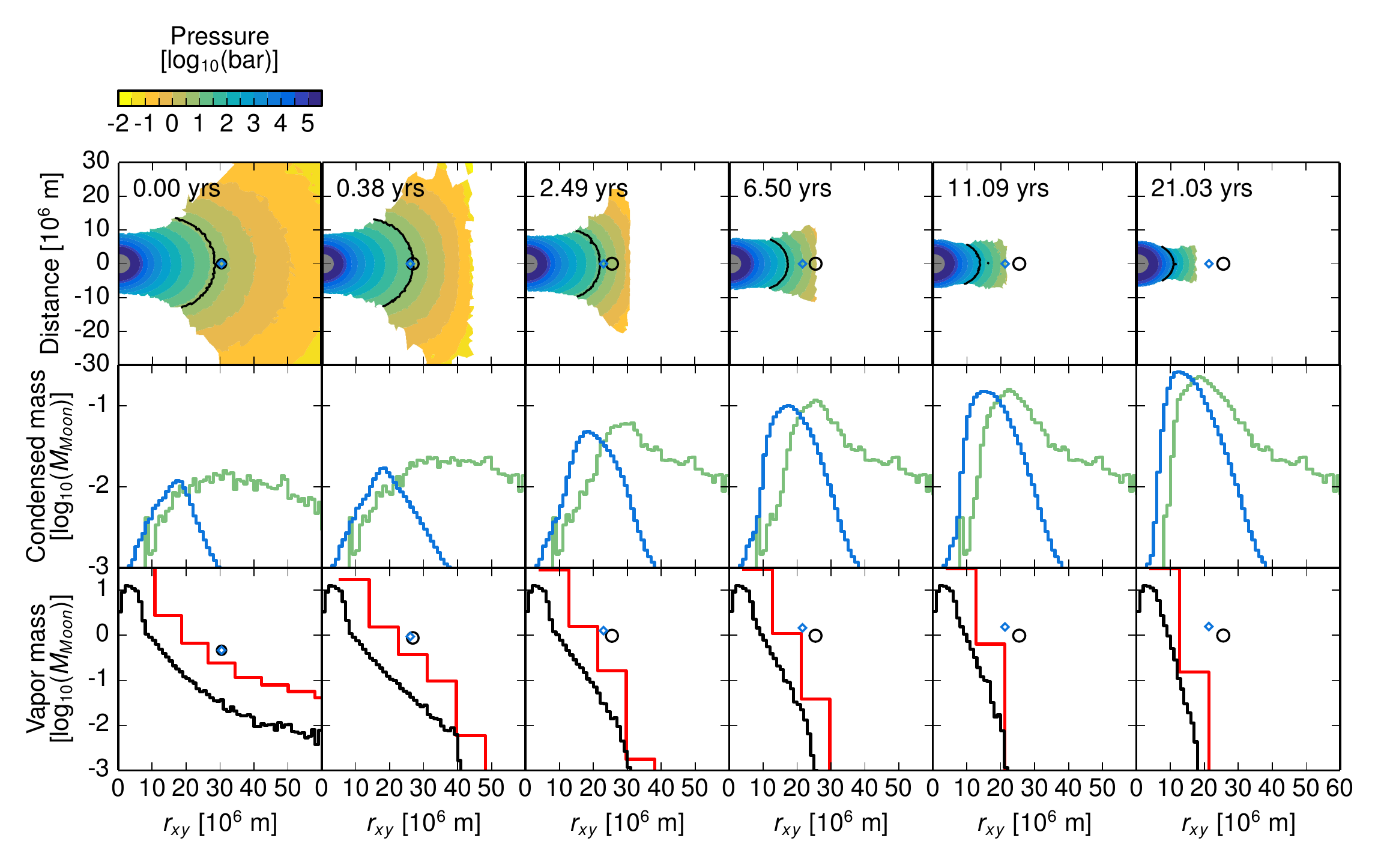}
\caption{Cooling of a synestia leads to contraction of the structure and formation of a moon. Columns present different time steps. Top row: pressure contours of the vapor structure, where the black line denotes the boundary between the isentropic and vapor-dome regions; middle row: cumulative histograms of the radii at which mass condensed (green) and the locations to which falling condensate was redistributed (blue); bottom row: histograms (red, black) of the instantaneous mass distribution in the synestia. Red histograms binned by Hill diameter of moon A at that time step; other histograms binned by $1$~Mm. The black circles represents moon A, an estimate based on the total mass of condensing material beyond the Roche limit. In the top row, the size of moon A is shown to scale assuming a bulk density of 3000~kg~m$^{-3}$. Blue diamonds represent moon B, an estimate which includes some falling condensate, not shown to scale. Moons A and B are plotted at the radius of a circular Keplerian orbit corresponding to the integrated angular momentum of their constituent mass. The initial synestia was produced by an impact where a $0.468M_{\rm Earth}$ body struck a $0.572M_{\rm Earth}$ body at 12.33 km~s$^{-1}$ and an impact parameter of 0.4. The seed of the moon in the first time step had a mass of 0.466~$M_{\rm Moon}$.}
\label{fig:SPH_cooling}
\end{sidewaysfigure*}

In this section, we demonstrate how a moon forms from a terrestrial synestia as it cools. Impact-generated synestias have a wide range of mass, AM and thermal energy distributions and can form a variety of different mass moons with varying properties. However, we did not conduct a comprehensive study of cooling synestias over the full parameter space of giant impacts for this work. Instead, we focus on synestias that initially have more than a lunar mass beyond Roche and that form a lunar-mass moon. We refer to such post-impact structures as potential Moon-forming synestias (see discussion in \S\ref{sec:impacts}). Here, we present one example calculation of the cooling of such a synestia and additional examples are provided in Figures~\ref{sup:fig:coolingA}-\ref{sup:fig:coolingC}. 

Figure~\ref{fig:SPH_cooling} presents an example calculation of a radiatively cooling synestia. In this example, the initial structure was generated by a collision between 0.572 and 0.468 $M_{\rm Earth}$ bodies at 12.33 km~s$^{-1}$ with an impact parameter of 0.4. Unlike a canonical post-impact structure with a quasi-constant surface density in the disk-like region (Figure~\ref{fig:linestructures}C), the surface density of the disk-like region of a typical synestia varies by orders of magnitude (Figures~\ref{fig:linestructures}F and \ref{fig:SPH_cooling} bottom row). Hence, the distal regions cool much more rapidly than the interior regions. The synestia is strongly pressure supported and cooling of the outer regions reduces the pressure, and hence the pressure gradient force, causing the structure to contract over time. However, the pure-vapor region of the structure provides substantial pressure support to beyond the Roche limit for tens of years. In Figure~\ref{fig:SPH_cooling}, the boundary between the pure-vapor and vapor-dome regions is shown by the black line superimposed over the pressure contours. The two estimates for the satellite mass and the corresponding orbital radii are shown by a black circle (moon A) and blue diamond (moon B). In the top row of Figure~\ref{fig:SPH_cooling}, the size of moon A is to scale assuming a density of 3000~kg~m$^{-3}$. Moon B is not shown to scale for clarity. 

The second row of Figure~\ref{fig:SPH_cooling} illustrates the source and redistribution of condensing mass. The green histogram presents the original location of all the condensing mass. The blue histogram shows where falling condensates, those with $j$~$<$~$j_{\rm Roche}$, were redistributed using our simple redistribution scheme. The portion of the blue histogram that falls beyond the Roche limit was removed from the calculation and incorporated into the mass of moon B. The histograms of condensed mass at time zero correspond to the material removed during thermal equilibration. In the thermal equilibration step, the mass of the initial Roche-exterior condensates was $0.466~M_{\rm Moon}$. We find that the mass transported by falling condensates is an important component of the mass budget throughout the disk-like region of the structure. The total mass of falling condensate over the time period shown in Figure~\ref{fig:SPH_cooling} is 8.0~$M_{\rm Moon}$. The peak of redistributed mass near Roche is a result of the combination of the strongly pressure supported rotational velocity profile (e.g., Figure~\ref{fig:linestructures}D,E) and the fact that most condensates are sourced from just beyond the Roche limit. As the synestia contracts upon cooling, the source and destination of condensates shifts inwards. The deposition of mass from condensates just inside the isentropic region is consistent with the idea of condensates spiraling inwards and vaporising rapidly in the pure vapor region.

\begin{figure}
\centering
\includegraphics[scale=0.833333333333]{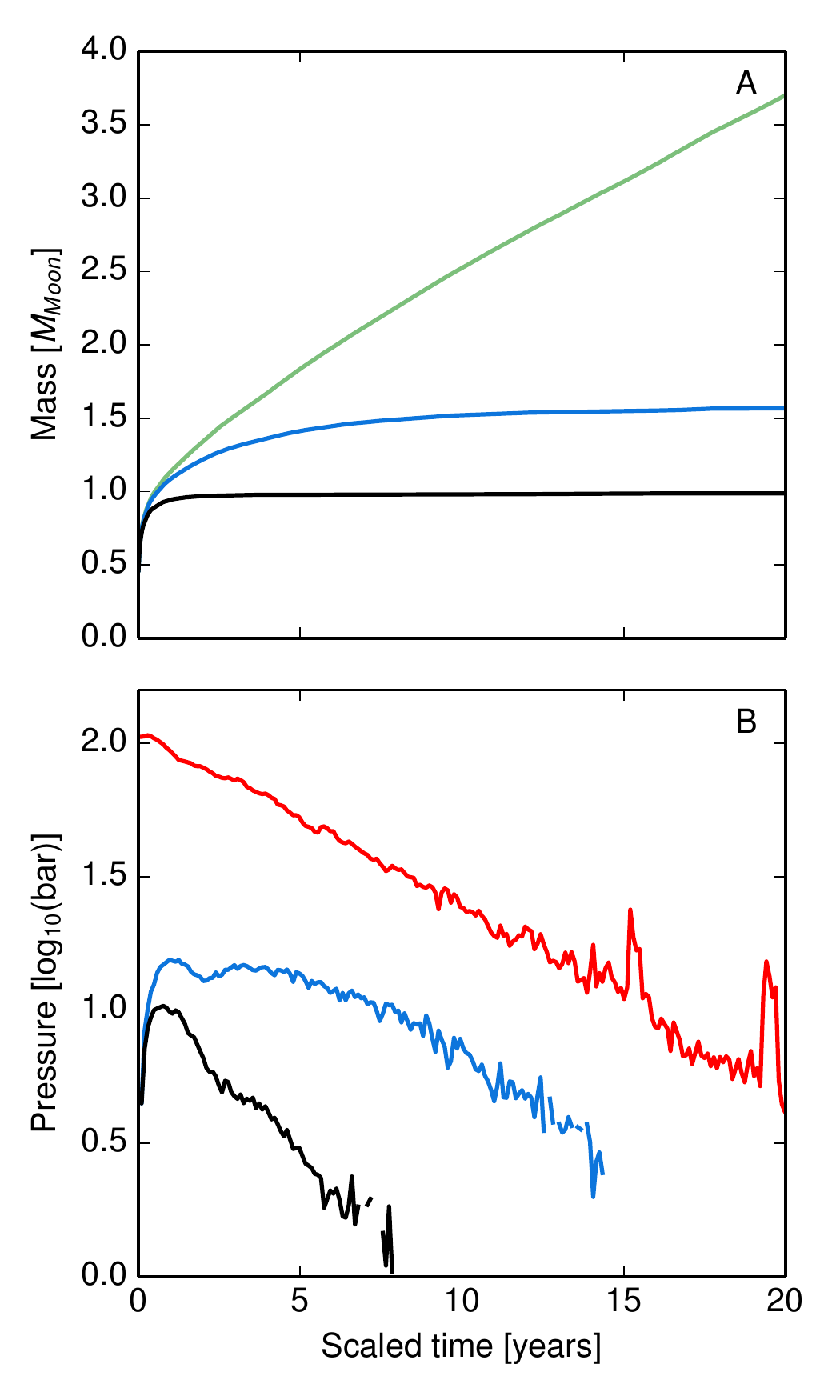}
\caption{Condensation in the outer regions of a synestia can lead to the rapid formation of a moon within the vapor structure, surrounded by approximately tens of bars of vapor. The mass (A) and vapor pressures (B) at moon A (black lines) and B (blue lines) are shown for the calculation in Figure~\ref{fig:SPH_cooling}. The green line is the total condensed mass (corresponding to the green histograms in Figure~\ref{fig:SPH_cooling}). The red line is the midplane vapor pressure at the Roche limit.}
\label{fig:SPH_cooling_time}
\end{figure}

The growth of the moon mostly occurs in the first year of cooling, as shown in Figure~\ref{fig:SPH_cooling_time}A. The black line corresponds to the moon A mass estimate which only includes material with sufficient AM to orbit beyond the Roche limit. The blue line corresponds to the moon B mass estimate, which also includes a portion of the condensates falling within the Roche limit. The green line is the total mass condensed in our calculation. The calculated condensed mass is a lower limit because the smaller scale height in the inner region is not resolved in the SPH simulations. Thus, the substantial mass of condensate that would be forming in these regions is not captured in our calculations. However, our simulations include cooling of the well-mixed inner regions as if condensates formed on the surface and revaporized at depth. The large mass of condensate formed in our calculations quantitatively supports the idea discussed above that falling condensates play a significant role in mass transport in the synestia. In our calculations, the mass transport by condensates is much greater than transported by viscous spreading (supporting information \S\ref{sup:sec:SPH_cooling}).

For all of our potential Moon-forming synestias, a moon forms within the vapor structure of the synestia. 
In the example in Figure~\ref{fig:SPH_cooling}, the satellite is enveloped by the vapor of the structure for several years to $>$$10$ years, depending on the radius of its orbit. The vapor pressures in the midplane at the location of moons A and B are shown in Figure~\ref{fig:SPH_cooling_time}B. The pressure at the Roche limit in the midplane is initially about 100~bars but steadily drops as the structure cools. 
The moon is initially surrounded by about 10~bars of vapor. 
As discussed in \S\ref{sec:boundary_layer}, the calculated pressure around the moon is a lower limit as the gravitational field of the moon will increase the local vapor pressure. The local pressure drops as the vapor structure cools and recedes within the orbit of the moon. When the edge of the structure is within the Hill sphere of the moon (about 6.5 years in Figure~\ref{fig:SPH_cooling}), some of the vapor will become bound to the moon and the moon separates from the synestia. The Hill sphere is the region of gravitational influence of a body. 
When the moon begins to dominate its local vapor environment, exchange with the vapor of the synestia is reduced. Thus, chemical exchange with the synestia will cease while the moon is still surrounded by a substantial vapor pressure. Future work will investigate this transition stage in more detail.

The timescale for cooling the structure within the Roche limit depends strongly on the initial distribution of mass, AM, and thermal energy within the structure. For the examples presented here, the time to cool to 10~bars of vapor pressure at the Roche limit is typically around 10 years, which is a strong lower limit. We did not model the evolution of structures once they had receded within the Roche limit. Continued cooling would lead to further contraction and eventually the structure would fall below the CoRoL.

In calculating the mass of the moon formed from a synestia, we did not consider addition of material once the vapor structure had receded inside the Roche limit. Typically, when the synestia has cooled to the point shown at 21 years in Figure~\ref{fig:SPH_cooling}, the Roche-interior region is largely pressure-supported. The formation of additional moonlets would require AM transport outwards, e.g., via viscous spreading. The structure is still a synestia at this time, and condensates will form at the cooling edge of the radially spreading structure. Any late moonlets will dynamically interact with the primary moon, leading to transfer of AM to the satellite. We expect any mass added in this late stage to be minimal. Due to the strong pressure support of the structure, the specific AM of mass that would be rotationally supported in the disk-like region is small. Scaling laws for lunar disks \citep{Salmon2012,Salmon2014} predict that the satellite formed from such a disk would be negligible. However, these scaling laws were not formulated for synestias and this stage of the evolution of the structure will require future investigation with an $N$-body code coupled to the evolution of the synestia \citep[e.g.,][]{Hollyday2017}. 

For the example synestias in this work (Figure~\ref{fig:SPH_cooling} and Figures~\ref{sup:fig:coolingA}-\ref{sup:fig:coolingC}), a greater than lunar-mass moon was formed. However, we assumed perfect accretion of Roche-exterior material. Although accretion of the Roche exterior material is unlikely to be wholly efficient, there is a substantial mass of falling condensate that could accrete to the moon. Based on our simple mass redistribution model, more than half a lunar mass of falling condensates would be available to be accreted to the moon, which is illustrated by the difference between the A and B mass estimates in Figure~\ref{fig:SPH_cooling_time}. Falling condensates may aid the production of large moons and help compensate for inefficiencies in accretion. Accretion of falling condensate to the moon will reduce the AM of the moon causing it to orbit closer to the central mass and in higher pressure regions of the synestia.

In summary, an impact-generated terrestrial synestia may form a lunar mass satellite that orbits within the vapor structure for several years. We find that vapor pressures of about 10~bars or more surround the moon for a range of giant impact scenarios. Here, we have described a different environment for satellite formation than that from standard circumplanetary disks that have been considered previously.

%xxxxxxxxxxxxxxxxxxxxxxxxxxxxxxxxxxxxxxxxxxxxxxxxxxxxxxxxxxxxxxxxxxxxxxxxxxxxxxxxxxxxxxxxxxxxxxxxxxxxxxxxxxxxxxxxxxxxxxxxxxxxxxxxxxxxxxxx
%xxxxxxxxxxxxxxxxxxxxxxxxxxxxxxxxxxxxxxxxxxxxxxxxxxxxxxxxxxxxxxxxxxxxxxxxxxxxxxxxxxxxxxxxxxxxxxxxxxxxxxxxxxxxxxxxxxxxxxxxxxxxxxxxxxxxxxxx
%xxxxxxxxxxxxxxxxxxxxxxxxxxxxxxxxxxxxxxxxxxxxxxxxxxxxxxxxxxxxxxxxxxxxxxxxxxxxxxxxxxxxxxxxxxxxxxxxxxxxxxxxxxxxxxxxxxxxxxxxxxxxxxxxxxxxxxxx
\section{Thermodynamics of Bulk silicate Earth material}
\label{sec:thermo}

The cooling model presented in \S\ref{sec:dynamics} used a single phase (forsterite) to represent 
the silicate portion of the synestia.
In reality, the synestia is a multicomponent system, and the thermodynamics of the bulk material will control key aspects of the formation of a moon. In this section, we first discuss the bulk chemical composition of the outer portions of a terrestrial synestia (\S\ref{sec:thermo_composition}). We then present calculations of the phase diagram for BSE material (\S\ref{sec:thermo_methods} and \S\ref{sec:BSE_phase_diagram}).

%xxxxxxxxxxxxxxxxxxxxxxxxxxxxxxxxxxxxxxxxxxxxxxxxxxxxxxxxxxxxxxxx
%xxxxxxxxxxxxxxxxxxxxxxxxxxxxxxxxxxxxxxxxxxxxxxxxxxxxxxxxxxxxxxxx
\subsection{Composition of a terrestrial synestia}
\label{sec:thermo_composition}

In giant impacts, the silicate portions of the colliding bodies are highly shocked. The material becomes a continuum of liquid, supercritical fluid, and vapor. Silicates sourced from both the impactor and target material are combined into a single continuous fluid. Shear in the impact results in Kelvin-Helmholtz instabilities, producing small scale eddies that can mix the fluid. Furthermore, the momentum of the impact leads to advection and hence mixing of material. Both of these processes drive the system towards local chemical and thermal equilibrium. The shear and flow velocity in the impact is spatially varying, and so the degree of mixing is highly heterogeneous \citep{Stewart2015LPSC}. Although most of the silicate is a continuous  fluid, differences in mixing length and time scales produce a synestia that is likely thermally and isotopically heterogeneous.

The degree to which the colliding bodies are mixed is sensitive to the precise impact configuration. The canonical Moon-forming collision is a graze-and-merge event \citep{Canup2001,Canup2004,Canup2008,Canup2008a}. A fraction of the projectile grazes the target and is subsequently disrupted and torqued into orbit. This fraction of the impactor does not have an opportunity to intimately mix with the target body, and the outer portions of the structure are enriched in impactor material. In contrast, the impact geometries of most high-AM, high-energy impacts \citep{Cuk2012,Canup2012,Lock2017} lead to much more contact between the silicates originating from the impactor and target, and there can be substantial shear and advective mixing during the impact event. In a large number of simulations of high-AM, high-energy impacts \citep{Cuk2012,Canup2012}, the outer regions of the structure out of which a moon would form have similar proportions of impactor and target material to the bulk (e.g., within about 10\%).

Simulations of giant impacts likely underestimate the degree of mixing in high-AM, high-energy impacts \citep{Deng2017}.
The methods that are predominantly used to model giant impacts do not calculate thermal equilibration, and the large spatial and temporal scale of planetary collisions make it unfeasible to numerically resolve small scale eddies. The lack of thermal exchange and local mixing in simulations results in parcels of material that should have locally mixed and thermally equilibrated, separating again by buoyancy to form a stable density profile. Material from the impactor is generally more shocked and hotter than the target and rises to lower pressures. As a result, the outer portions of post-impact structures are artificially enriched in impactor material.

In this work, we consider the formation of the Moon from a terrestrial synestia, a body with a BSE bulk composition. The mixture of silicate and metal from the impactor and target in the Moon-forming giant impact determines the composition of the BSE today, minus additions during late accretion. For simplicity, we assume that mixing during the impact was efficient enough that the high-entropy regions of the synestia have a roughly BSE composition immediately after the impact. This assumption will only be valid for a subset of high-AM, high-energy impacts. Further work will be required to determine the size of this subset. We only make this assumption for the bulk elemental composition and will return to examine the question of isotopic heterogeneity in \S\ref{sec:isotopes}. Mass transport by falling condensates and vertical fluid convection will ensure that the outer regions remain well mixed, and approximately BSE, during the early evolution of the synestia (\S\ref{sec:dynamics_cooling}). We continue under the assumption that the bulk composition of the outer regions is near BSE. 

%xxxxxxxxxxxxxxxxxxxxxxxxxxxxxxxxxxxxxxxxxxxxxxxxxxxxxxxxxxxxxxxx
%xxxxxxxxxxxxxxxxxxxxxxxxxxxxxxxxxxxxxxxxxxxxxxxxxxxxxxxxxxxxxxxx
\subsection{Calculation of BSE condensation}
\label{sec:thermo_methods}

In order to understand the thermochemistry of material in a terrestrial synestia, we calculated the phase diagram of BSE material.
The partitioning of matter between condensed (melt and/or crystals) and gaseous phases was calculated using the 20 element (H, He, C, N, O, Na, Mg, Al, Si, P, S, Cl, K, Ca, Ti, Cr, Mn, Fe, Co, Ni) and 34 element (20 + Cu, Ga, Ge, Mo, Ru, Pd, Hf, W, Re, Os, Ir, Pt, Au, Zn) versions of the GRAINS code \citep{Petaev2009}.
This code uses a Gibbs free energy minimization scheme to calculate equilibrium partitioning of these elements among gaseous, liquid, and solid phases.
The code contains thermodynamic data for 530 condensed and 245 gaseous species \citep[listed in][with additional species added for this study given in Table~\ref{sup:tab:added_species}]{Petaev2009} for the temperature range of 300-2500~K. 
In order to be able to consider the higher temperature regime required for post-impact states, the thermodynamic data for the condensed and gaseous species that could be stable above 2500 K were expanded to 5000 K using the tabulated values from JANAF \citep[][]{Chase1998} or standard thermodynamic data (enthalpy of formation, $\Delta H_{(f,298)}$; standard entropy, $S_{298}$; and heat capacity at constant pressure, $C_{p}(T)$, polynomials) from other sources (see Tables \ref{sup:tab:vap_species} and \ref{sup:tab:cond_species}). At the temperature and pressure near where condensed phases are stable, the fraction of ionized species is expected to be small and so they were not included in our model.

The GRAINS code has two built-in silicate melt models, ideal and CMAS \citep{Berman1983}, with the latter assuming activity coefficients of: Mg~$=$~Fe, Mn, Ni, Co; Si~$=$~Ti; Al~$=$~Cr; and Ca~$=$~Na, K. 
The two models yielded similar condensation curves but the condensation temperatures of the ideal melt are generally about 100~K lower. At the very high temperatures in post-impact states, the ideal model is preferable because the extrapolation of the CMAS, or any other, internally consistent model well beyond its calibration range could result in large and unpredictable errors. 
Compared to \cite{Petaev2009}, a non-ideal model of the Fe-Ni-Co-Cr-P-Si metal liquid was added and the Na$_2$O and K$_2$O end-members of the silicate melt were replaced with Na$_2$SiO$_3$ and K$_2$SiO$_3$, respectively, as the high-temperature thermodynamic data for the latter are more reliable than for the oxides. 

The code calculates the mole fraction of gaseous and condensed
species (larger than 10$^{-24}$). Using these data, we calculated the chemical compositions of the condensed phases, their molar, atomic, and weight concentrations, bulk compositions of the condensed and gaseous phases, and a number of other parameters.
Previous work on gas-condensate equilibria in a system of BSE composition \citep{Schaefer2012,Visscher2013} used different BSE compositions \citep{Kargel1993,Palme2007} than that used here \citep{McDonough1995}. Our modeling of condensation in systems with these alternative compositions agrees reasonably well with the results of previous studies. An important difference in calculations of systems with different BSE compositions is the stability of small amounts of Fe-Ni metal in a system of BSE composition given by \citet{McDonough1995}, and lack of it in other BSE systems \citep{Kargel1993,Palme2007}, consistent with previous work. For the calculations reported here, we considered equilibrium condensation and hold the bulk composition of the condensing system constant.

We used our condensation calculations to produce a phase diagram for material of BSE bulk composition at equilibrium in a range from $10^{-6}$ to $200$~bar
and 1000 to 5000~K.  
For BSE, we used the widely accepted
composition of \citet{McDonough1995}.  
We assumed a hydrogen concentration of 1000~ppm that is about 18 times higher than the BSE composition used by \citet{Schaefer2012}.  Similar to \citet{Schaefer2012}, we found that varying hydrogen content affects both the speciation of the gaseous phase and the position of the vapor curve. The vapor curves moves up by $\sim$200~K as the hydrogen content is decreased to $\le$100 ppm. Nevertheless, the elemental pattern of condensate -- the main focus of this study -- remains essentially unchanged.

\begin{figure}[p]
\centering
\includegraphics[scale=0.83333333]{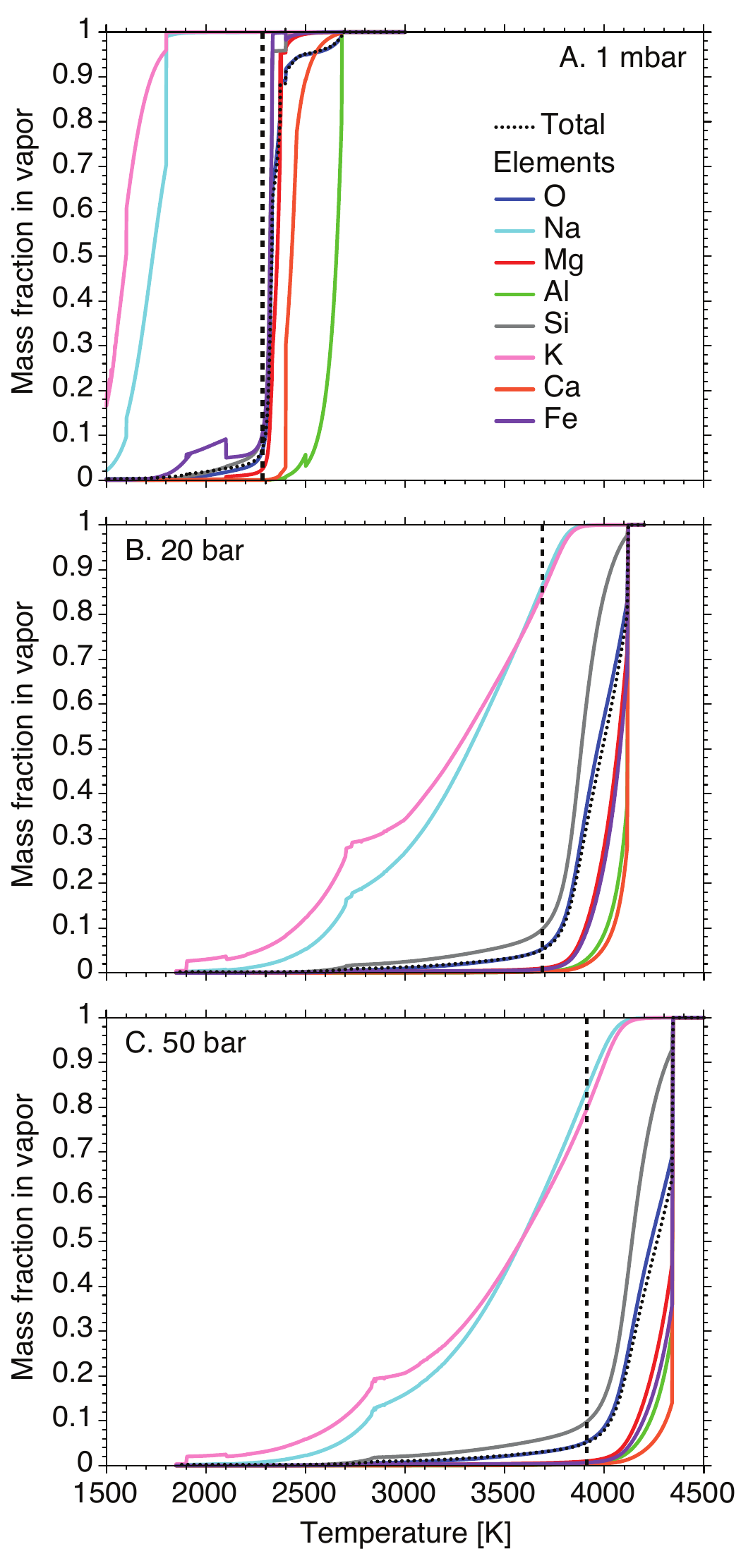}
\caption{Condensation curves for BSE composition vapor at different pressures. In each panel, the black dotted line is the mass fraction of the system that is vapor and the colored lines show the fraction of selected elements that are in the vapor. The vertical black dashed line in each panel shows the temperature at which 10\% silicon is in the vapor. The kinks in the condensation curves at $\sim$$2100$~K are due to inconsistencies between the ideal melt and non-ideal solid solution models used.}
\label{fig:cond}
\end{figure}

We calculated the phase equilibrium along isobaric cooling paths at 1~K intervals, starting from a purely vapor state.
As the temperature decreases, the fraction of condensate increases and the composition of both the condensate and the vapor evolve. Examples of such isobaric calculations are given in Figure~\ref{fig:cond}.
The phase boundaries were identified by the presence of significant amounts of liquid or solid phases.
We also identify the temperature lower than which a significant amount of free metal precipitates from the silicate melt.

%xxxxxxxxxxxxxxxxxxxxxxxxxxxxxxxxxxxxxxxxxxxxxxxxxxxxxxxxxxxxxxxx
%xxxxxxxxxxxxxxxxxxxxxxxxxxxxxxxxxxxxxxxxxxxxxxxxxxxxxxxxxxxxxxxx
\subsection{Phase diagram for bulk silicate Earth}
\label{sec:BSE_phase_diagram}

\begin{sidewaysfigure*}
\centering
\includegraphics[scale=0.833333333]{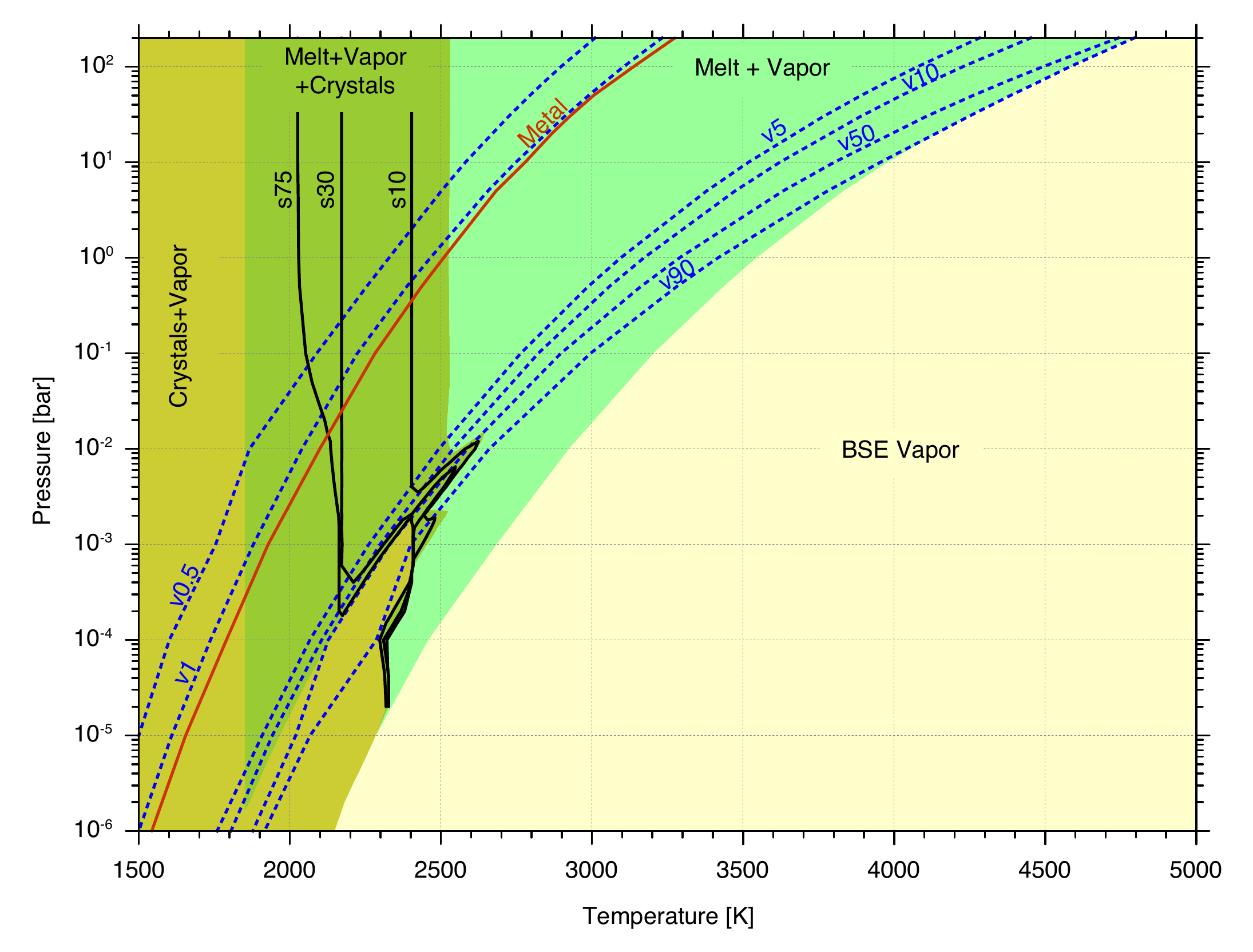}
\caption{The phase diagram for BSE material at the pressures and temperatures relevant for moon formation from a terrestrial synestia. Colored regions indicates where different combinations of phases are stable. Blue dashed lines indicate lines of constant vapor fraction, labeled as percentage of vapor by mass. Black lines indicate the fraction of the condensate that is solid, labeled as percent solid condensate by mass. The metal precipitation line is shown in red. To the left of this line, a substantial amount of free metal is stable in the condensate.}
\label{fig:BSEphase_diagram}
\end{sidewaysfigure*}

Figure~\ref{fig:BSEphase_diagram} shows the calculated phase diagram for BSE material.
The majority of the mass of vapor ($\sim$~$ 90$\%) condenses over a narrow temperature range.  At higher pressures, vapor condenses to liquid, but at lower pressures, vapor can condense directly to solid phases. 
In the melt stability field, decreasing temperature results in progressive crystallization of solid phases until the silicate melt completely solidifies.
At high temperatures, the condensate is a single silicate liquid, but at lower temperatures a small amount of free metal, mainly Fe-Ni alloy, precipitates (red line in Figure~\ref{fig:BSEphase_diagram}).
Note that this diagram holds for a closed system of BSE composition in equilibrium and does not include any effects of phase separation.

At high temperatures the vapor is dominated by refractory species (e.g., SiO, SiO$_2$, Mg, MgO, Fe, FeO, etc.).
At lower temperatures, the refractory species condense, leaving a vapor increasingly dominated by volatile species (e.g., H$_2$, CO, H$_2$S, N$_2$).
A vapor fraction is present for the whole range of pressures and temperatures considered here.

Most previous work on lunar origin has relied upon equilibrium condensation calculations
appropriate for the solar nebula \citep[solar composition at 10$^{-4}$ bar; ][]{Lodders2003} to aid interpretation of lunar data. Such comparisons are the source of the widely quoted estimates of
$\sim$~1000~K condensation temperatures for the depletion of moderately
volatile elements in lunar material. Other studies \citep[e.g.,][]{Schaefer2012,Canup2015} have calculated the vapor species that would be in equilibrium with a silicate melt of BSE composition. Such calculations can give insights into the composition of the gas in condensate dominated systems, such as in models of the canonical Moon-forming disk, but do not fully describe the physical chemistry of the the bulk BSE system. Figure \ref{fig:BSEphase_diagram} presents the first calculation of a bulk BSE phase diagram over the pressure and temperature range relevant to lunar origin.

The liquid-vapor phase boundary of multicomponent systems is described in terms of dew and bubble points.
Although described as points, more completely dew and bubble points refer to curves in pressure-temperature space.
The dew point is the temperature at a given pressure below which the first condensates appear in the system.
For BSE, the dew point is given by the low temperature edge of the BSE vapor field (cream region in Figure~\ref{fig:BSEphase_diagram}).
The bubble point is the temperature at a given pressure above which the first vapor phases are stable.
In the BSE system, this definition of a bubble point is not particularly useful as some species remain in the vapor phase over all temperatures of interest.
The first major element (the elements that constitute more than 5~wt\% of the BSE, i.e., O, Mg, Si, and Fe) to vaporize is silicon, and there is a narrow range of temperatures where the fraction of silicon in the vapor rises from a few wt\% to nearly 100~wt\% (Figure~\ref{fig:cond}). For this work, we define a major element bubble point as the temperature at which 10~wt\% of silicon is in the vapor (vertical dashed line in Figure~\ref{fig:cond}). This corresponds to about 5~wt\% total vapor (see dotted line in Figure~\ref{fig:cond} and the v5 blue dashed line in Figure~\ref{fig:BSEphase_diagram}).

%xxxxxxxxxxxxxxxxxxxxxxxxxxxxxxxxxxxxxxxxxxxxxxxxxxxxxxxxxxxxxxxxxxxxxxxxxxxxxxxxxxxxxxxxxxxxxxxxxxxxxxxxxxxxxxxxxxxxxxxxxxxxxxxxxxxxxxxx
%xxxxxxxxxxxxxxxxxxxxxxxxxxxxxxxxxxxxxxxxxxxxxxxxxxxxxxxxxxxxxxxxxxxxxxxxxxxxxxxxxxxxxxxxxxxxxxxxxxxxxxxxxxxxxxxxxxxxxxxxxxxxxxxxxxxxxxxx
%xxxxxxxxxxxxxxxxxxxxxxxxxxxxxxxxxxxxxxxxxxxxxxxxxxxxxxxxxxxxxxxxxxxxxxxxxxxxxxxxxxxxxxxxxxxxxxxxxxxxxxxxxxxxxxxxxxxxxxxxxxxxxxxxxxxxxxxx
\section{Combined dynamic and thermodynamic model for Moon formation}
\label{sec:comb}

Next, we combine our calculations of the structure of a cooling synestia and the phase diagram for BSE to produce a coupled dynamic and thermodynamic model for satellite accretion. We examine the thermodynamic paths of condensates in a synestia (\S\ref{sec:paths}) and assess the ability of moonlets to survive within the vapor structure (\S\ref{sec:moonlet_vap} and \S\ref{sec:decoupling}). We then consider the accretion of large moonlets and the processes that govern the exchange between the vapor of the synestia and the moonlets through a boundary layer (\S\ref{sec:boundary_layer}). Based on the thermodynamics of BSE material, we then determine the conditions for equilibration between the vapor and the moonlets (\S\ref{sec:buffer}) and predict the composition of the moon that is formed from a terrestrial synestia (\S\ref{sec:Moon_comp}).

%xxxxxxxxxxxxxxxxxxxxxxxxxxxxxxxxxxxxxxxxxxxxxxxxxxxxxxxxxxxxxxxx
%xxxxxxxxxxxxxxxxxxxxxxxxxxxxxxxxxxxxxxxxxxxxxxxxxxxxxxxxxxxxxxxx
\subsection{Thermodynamic paths of droplets and moonlets}
\label{sec:paths}

First, we investigate the thermodynamic paths and phase relationships for parcels of BSE material in a terrestrial synestia as it cools and evolves. As a parcel of pure vapor convectively rises to lower pressures, it will follow an adiabat and begin to partially condense once it intersects the dew point.
We used the BSE phase diagram to approximate adiabats in the structure assuming no physical phase separation by treating the vapor and condensate as two phases with a single latent heat of vaporization, $l$ (see supporting information \S\ref{sup:sec:adiabats}). Parcels rising from the pure-vapor regions of the synestia reach the dew point and then follow the phase boundary to low pressures.

\begin{sidewaysfigure*}
\centering
\includegraphics[scale=0.83333333333]{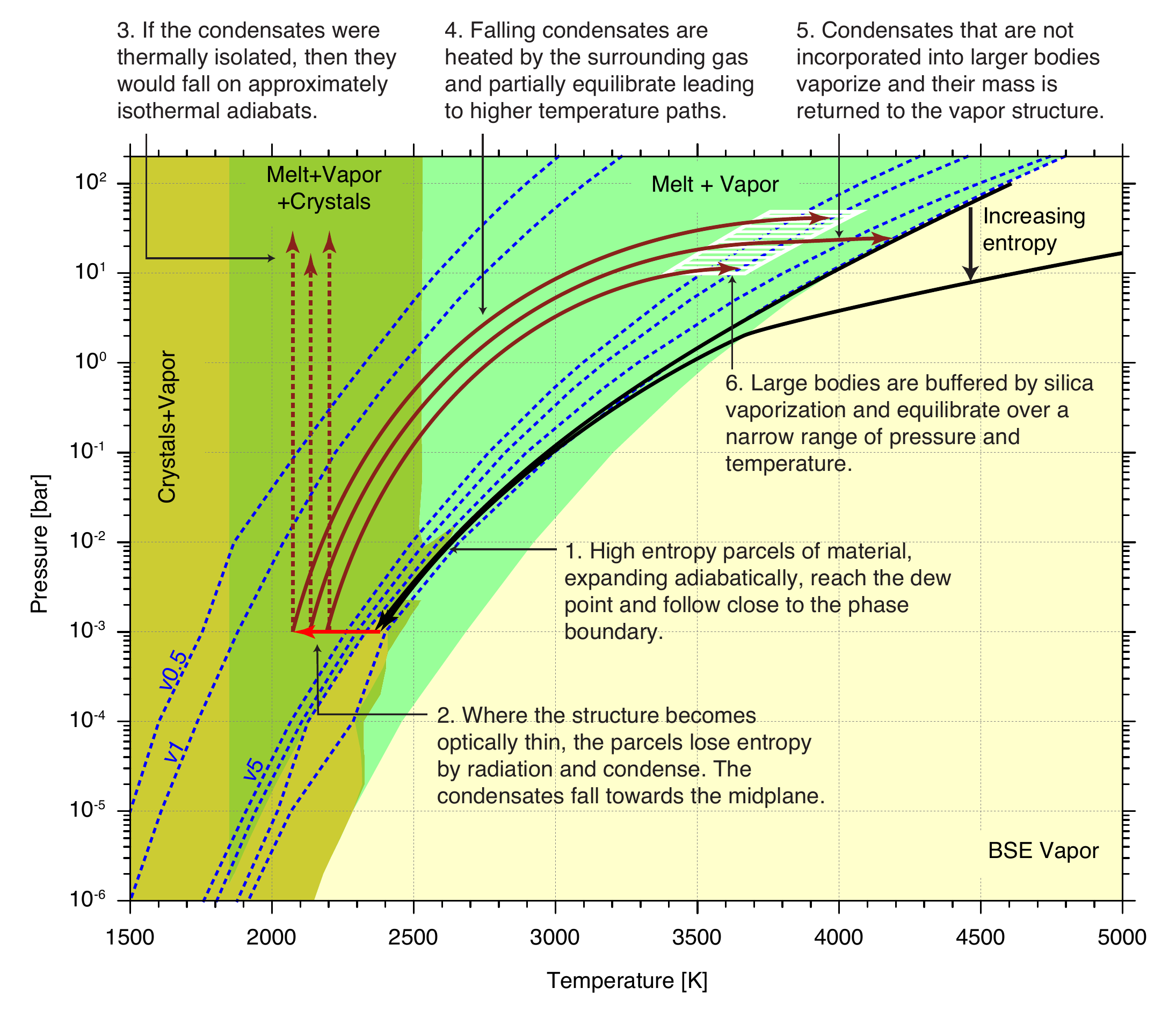}
\caption{Thermodynamic paths of condensates and vapor in a terrestrial synestia (see \S\ref{sec:paths}). The phase regions and constant vapor fraction lines are the same as in Figure~\ref{fig:BSEphase_diagram}. Black arrows schematically show adiabats of rising BSE material with different entropies, as calculated in \S\ref{sup:sec:adiabats}. Cooling at the photosphere in this figure is approximated as isobaric and shown by the red arrow. Example paths for condensates that fall adiabatically back into the structure without equilibration are shown as dashed maroon arrows. Solid maroon arrows show schematic paths of falling condensates that experience some equilibration with the surrounding vapor. The white dashed box approximates the region of equilibration for moonlets.}
\label{fig:paths}
\end{sidewaysfigure*}

At low pressures (10$^{-6}$ to 10$^{-2}$~bar, see supporting information \S\ref{sup:sec:photosphere}), the structure becomes optically thin.
Here, at the photosphere of the structure, material cools by radiating energy, reducing the specific entropy of the parcel and moving it off the original upwelling adiabat (red arrow and label 2 in Figure~\ref{fig:paths}).
In Figure~\ref{fig:paths}, we schematically show condensation  along an isobaric path, but the exact behavior of the material at the photosphere, both dynamically and thermodynamically, will require a more detailed study as there is a strong feedback between condensation and the ability of the structure to radiate (supporting information \S\ref{sup:sec:photosphere}).
However, given the high photospheric temperature ($\sim$2300~K), we expect cooling to be catastrophic, leading to almost complete condensation of the radiating parcel.
For example, a 1-km thick photosphere at 10$^{-3}$~bar would fully condense in approximately a second.
The condensates formed at the photosphere will thus inherit an unfractionated major element composition from the BSE vapor.
However, the condensate will likely not include substantial amounts of moderately volatile elements because, at the pressures of the photosphere, these elements condense at much lower temperatures than the radiative temperature (Figure~\ref{fig:cond}A).
Depending on the pressure of the photosphere, the initial condensates may be a mixture of solid and liquid (Figure~\ref{fig:paths}).

The condensate formed at the photosphere is about seven orders of magnitude denser than the surrounding vapor. As a result, a torrential rain of condensates will fall into the structure.
As discussed in \S\ref{sec:dynamics_cooling}, condensates will fall rapidly from the photosphere as part of condensate-rich downwellings, enabling rapid transport of condensates to the higher pressure regions of the structure.
The falling condensates, as a consequence of condensing at low pressure, are substantially colder than the surrounding vapor.
Condensates that fall adiabatically, without thermal or chemical equilibration with the surrounding vapor, would follow approximately isothermal paths (dashed maroon arrows and label 3 in Figure~\ref{fig:paths}).
However, condensates will likely partially thermally equilibrate with their surroundings and follow higher temperature paths (solid maroon arrows and label 4 in Figure~\ref{fig:paths}).

The ultimate fate of condensates depends on both the thermodynamics and dynamics of condensed bodies within the structure.
As discuss in \S\ref{sec:dynamics_cooling}, thermal exchange with the vapor will lead to the progressive vaporization of the condensates.
The rate at which this occurs depends on the size of the condensates as well as the process governing thermal exchange. For condensates falling into regions of the structure that have midplane pressures and temperatures below the major element dew point (approximately those regions outside the black line in Figure~\ref{fig:SPH_cooling}) a fraction of condensate will be stable in the midplane. As discussed in \S\ref{sec:dynamics_cooling}, small condensates will rapidly spiral inwards and revaporize in higher temperature regions of the structure. In order to grow a moon, larger condensates that are formed outside the Roche limit (moonlets), must dynamically decouple from the synestia to avoid being dragged inwards and tidally disrupted inside the Roche limit.
In the following sections, we apply simple physical models to assess the ability of moonlets to survive within the vapor structure.

%xxxxxxxxxxxxxxxxxxxxxxxxxxxxxxxxxxxxxxxxxxxxxxxxxxxxxxxxxxxxxxxx
%xxxxxxxxxxxxxxxxxxxxxxxxxxxxxxxxxxxxxxxxxxxxxxxxxxxxxxxxxxxxxxxx
\subsection{Calculation of the evaporation timescale of moonlets}
\label{sec:moonlet_vap}

In \S\ref{sec:dynamics_cooling}, we present a calculation of the vaporization of small condensates in the pure-vapor regions of the synestia. We use the same approach here to place a lower limit on how long moonlets can survive in the outer regions of the structure. We consider moonlets in the midplane at about 20~bar surrounded by vapor that is close to the major element dew point (given approximately by the v5 blue dashed line in Figure~\ref{fig:BSEphase_diagram}), using the same parameters as in \S\ref{sec:dynamics_cooling}.

\begin{figure}[p]
\centering
\includegraphics[scale=0.8333333]{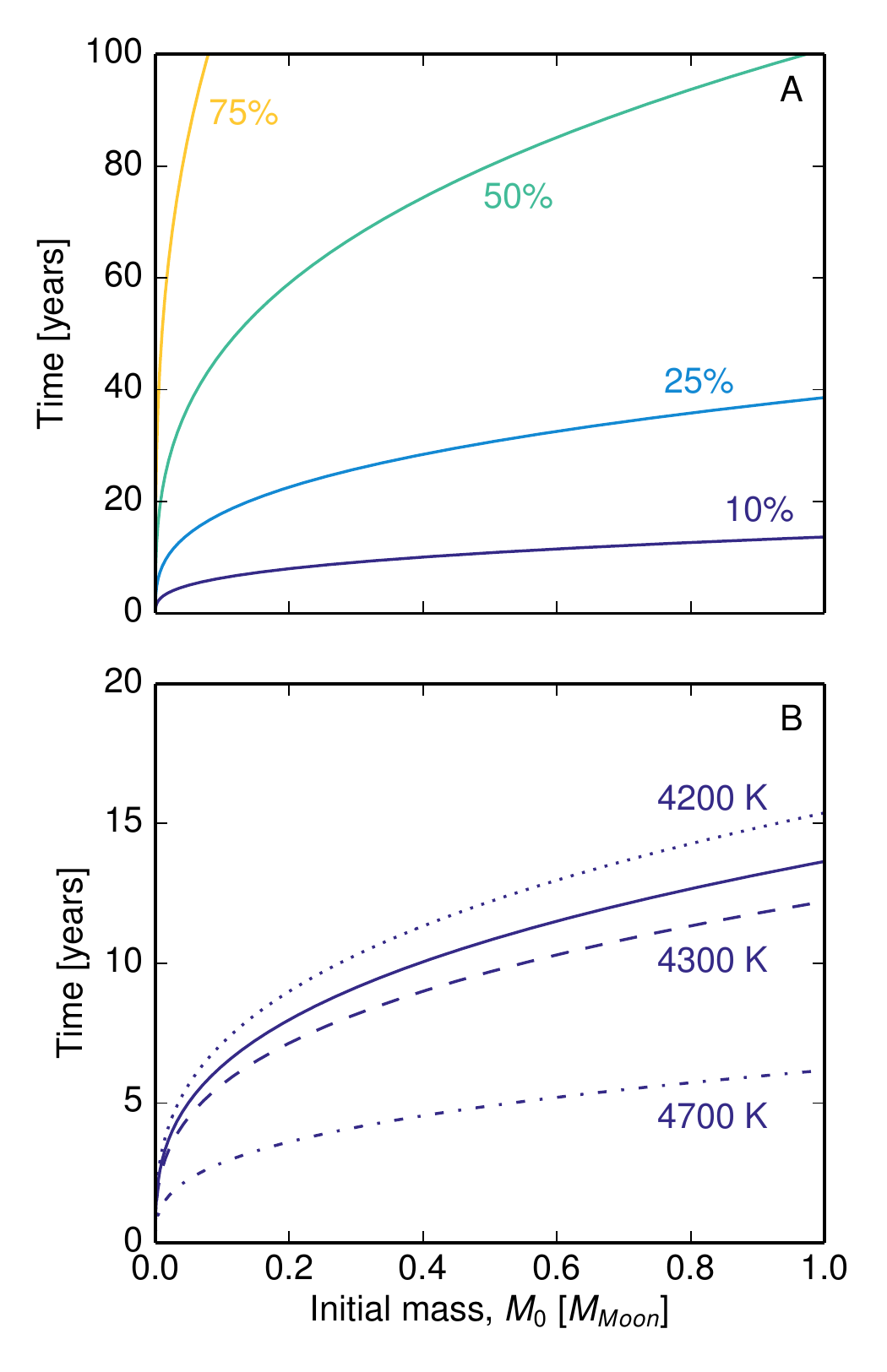}
\caption{Large moonlets can survive for a long time within the vapor structure. Panels show (A) the time taken to vaporize a given fraction of moonlets of varying initial mass; and (B) the effect of varying the temperature of the surrounding vapor on the time taken to vaporize 10~wt\% of moonlets of varying mass. The solid line in B is for a vapor temperature of 4250~K as used in A.}
\label{fig:vap_moonlets}
\end{figure}

Figure~\ref{fig:vap_moonlets}A shows how long it takes to vaporize different mass fractions of moonlets as a function of initial mass. Moonlets that are a significant fraction of a lunar mass can survive for long periods in the vapor structure. Bodies that are tens of percent of a lunar mass can survive with $<25$\% mass loss for tens of years in the midplane. This is longer than the time that the moon spent in the vapor structure in our synestia evolution calculations (\S\ref{sec:cooling_results}).

Our calculations of vaporization time for condensates are likely lower limits. We discuss the reasons for this in \S\ref{sec:dynamics_cooling} and describe mechanisms that could prolong the lifetime of smaller condensates. Moonlets in the structure will continually accrete smaller condensates that were produced at the photosphere. The addition of cooler condensates to the surface of moonlets will prolong their survival time. Also, the thermal exchange between moonlets and the vapor is mediated through a boundary layer (see \S\ref{sec:boundary_layer}) and is less efficient than we have assumed in our calculation.

The calculated survival timescales for moonlets are dependent on the temperature assumed for the surrounding vapor. In our evolution calculations, the moon grows in regions of the structure where the midplane temperature is close to the dew point (e.g., black arrows in Figure~\ref{fig:paths}). We therefore expect $T_{\rm vap} \sim \bar{T}_{\rm vap}$ for most of the lifetime of the structure. Initially, there are regions in the synestia that are pure vapor above the dew point (Figure~\ref{fig:SPH_cooling}). A larger $T_{\rm vap}$ only changes the survival times for moonlets by approximately a factor of two (Figure~\ref{fig:vap_moonlets}B).

%xxxxxxxxxxxxxxxxxxxxxxxxxxxxxxxxxxxxxxxxxxxxxxxxxxxxxxxxxxxxxxxx
%xxxxxxxxxxxxxxxxxxxxxxxxxxxxxxxxxxxxxxxxxxxxxxxxxxxxxxxxxxxxxxxx
\subsection{Calculation of moonlet decoupling from the vapor structure}
\label{sec:decoupling}

We estimate the mass at which moonlets can decouple from the gas and survive for extended periods of time in the Roche-exterior structure by calculating when the mass of vapor encountered by a freely orbiting body,  $M_{\rm encounter}$, is less than the mass of the body itself, $M$.
The moonlet mass that meets the condition, $M$~$\sim$~$M_{\rm encounter}$, can be determined from thermally equilibrated SPH post-impact structures.

The mass of vapor that a decoupled body encounters over one orbit is given by,
\begin{linenomath*}
\begin{equation}
M_{\rm encounter}=\pi \left (\frac{D}{2} \right )^2 (v_{\rm cond} - v_{\rm vap}) \, \frac{2 \pi a}{v_{\rm cond}} \, \rho_{\rm vap},
\end{equation}
\end{linenomath*}
where $D$ is the diameter of the body, $v_{\rm cond}$ and $v_{\rm vap}$ are the rotational velocity of the body and vapor respectively, $a$ is the orbital semi-major axis, and $\rho_{\rm vap}$ is the density of the vapor.
For a body on a circular Keplerian orbit around a central point mass,
\begin{linenomath*}
\begin{equation}
v_{\rm cond} = \sqrt{\frac{G M_{\rm bnd}}{a}},
\end{equation}
\end{linenomath*}
where $G$ is the gravitational constant and $M_{\rm bnd}$ is the bound mass of the synestia.
Here, we assumed that the density of vapor is constant throughout the orbit.
The body decouples when
\begin{linenomath*}
\begin{equation}
M_{\rm encounter}\sim M = \frac{4}{3} \pi \left (\frac{D}{2} \right )^3 \rho_{\rm cond},
\end{equation}
\end{linenomath*}
where $\rho_{\rm cond}$ is the density of the condensed body. 
The approximate size at which a body decouples is
\begin{linenomath*}
\begin{equation}
D_{crit}\sim \frac{ 3 \pi a \rho_{\rm vap}}{v_{\rm cond} \rho_{\rm cond}} (v_{\rm cond} - v_{\rm vap}).
\end{equation}
\end{linenomath*}
We calculated the decoupling size of moonlets using the parameters from several thermally equilibrated post-impact structures from the suite of impacts from LS17. The rotational velocity of the vapor was taken directly from the SPH structures, and the densities of the gas and condensate, $\rho_{\rm vap}$ and $\rho_{\rm cond}$ respectively, correspond to the values on the forsterite vapor dome at the pressure of the structure. 

The critical size for decoupled bodies ranges from meters to hundreds of kilometers, depending on the location of the orbit. In the midplane just beyond the Roche limit, the decoupling size is on the order of $10^5$~m, or equivalently $10^{-5}$ to $10^{-2}$~$M_{\rm Moon}$, for the example synestia shown in Figures \ref{fig:contourstructures} and \ref{fig:linestructures}. 
The decoupling size at the Roche limit decreases somewhat as the structure cools and the density of vapor in the outer structure decreases.

Decoupled moonlets are large enough to be on Keplerian orbits, but their orbits may still evolve due to gas drag or by gravitational interactions with the vapor structure and other condensates. Studying the migration of the moon within the structure is left to future work.

%xxxxxxxxxxxxxxxxxxxxxxxxxxxxxxxxxxxxxxxxxxxxxxxxxxxxxxxxxxxxxxxx
%xxxxxxxxxxxxxxxxxxxxxxxxxxxxxxxxxxxxxxxxxxxxxxxxxxxxxxxxxxxxxxxx
\subsection{The boundary layer of the growing moon}
\label{sec:boundary_layer}

As discussed in \S\ref{sec:cooling_results}, a satellite of a few tenths of a lunar mass forms very rapidly after a giant impact \citep[e.g., on a timescale of weeks,][]{Kokubo2000,Salmon2012}. Such a large body is decoupled from the vapor and dynamically stable outside the Roche limit.
This body could survive with only partial vaporization (less than 10~wt\%) for over ten years within the vapor structure (Figure~\ref{fig:vap_moonlets}A) and act as a nucleus onto which other condensates accrete. Generally, a single moonlet grows quickly and becomes the seed body for later stages of accretion. 

\begin{figure}[t]
\centering
\includegraphics[scale=0.833333]{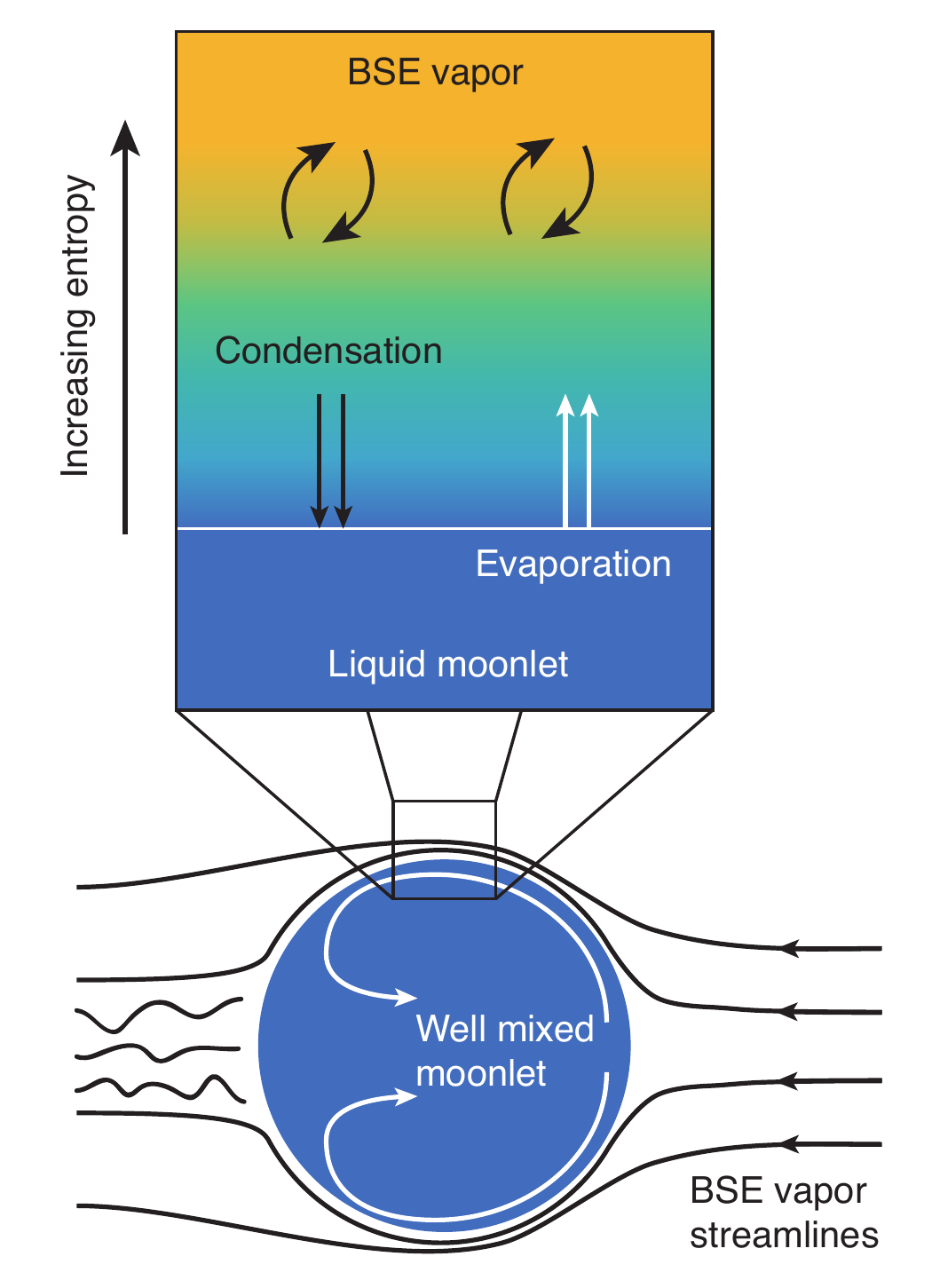}
\caption{Moonlets chemically equilibrate with the structure through a boundary layer. This cartoon illustrates the main features of this process. The sub-Keplerian vapor of the structure continuously flows around the moonlet, which could drive a flow field within the liquid moonlet and aid mixing. The boundary layer of the moonlet exchanges mass with the surrounding structure through turbulent mixing and by the condensation and evaporation of more volatile species. Colors on the diagram indicate specific entropy; the liquid moonlet acts as a cold trap for moderately volatile elements in the vapor, increasing the efficiency of mass exchange. }
\label{fig:Moon_cartoon}
\end{figure}

The seed body (and other moonlets) accretes smaller moonlets and falling condensates.
We expect accretion of small condensates to be relatively efficient because of the large gravitational cross section of moonlets. Gravitational capture could also be enhanced by aerodynamic drag from the vapor, in a manner similar to pebble accretion \citep{Lambrechts2012}. The falling condensates accreted onto the seed will be colder than the vapor in the midplane of the structure. These condensates formed at low pressures and are devoid of moderately volatile elements (Figure~\ref{fig:cond}A). The growing moon is out of thermal and chemical equilibrium with the vapor in the midplane; however, we argue that the surface of a liquid moonlet equilibrates with the vapor within a boundary layer (shown schematically in Figure~\ref{fig:Moon_cartoon}).

Since moonlets are decoupled from the gas, sub-Keplerian vapor continuously flows past the moonlets at velocities of 100s~m~s$^{-1}$.
The boundary layer mediates the thermal and chemical exchange between the hot vapor and the colder surface of the liquid moonlet.
The physical processes acting in the boundary layer are key to determining the chemical composition of the final moon. 
The details of the fluid dynamics of the boundary layer are complex and a numerical treatment is beyond the scope of this paper.
Here we discuss some of the basic properties of the boundary layer and focus on the processes governing equilibration of moderately volatile elements (MVEs).

Condensates forming at the pressures and temperature of the photosphere have negligible concentrations of volatile elements (e.g., Figure~\ref{fig:cond}A). 
However, in the higher pressures of the midplane, a portion of MVEs are stable in the condensate (Figure~\ref{fig:cond}B,C). The mass of the synestia is significantly larger than the mass of the moonlets; therefore, the vapor in the synestia acts as a compositionally homogeneous reservoir interacting with liquid moonlets that have a lower abundance of volatile elements.
As BSE vapor flows past the moonlets, a portion of the MVEs in the gas is cold-trapped onto the liquid surface. 
At the same time, the surface of the moonlet is heated and partially vaporized by radiation and thermal exchange with the flowing vapor.
Therefore, there is a continuous exchange of material between the moonlet and the vapor of the synestia.

Despite being heated from the top, the moonlets are likely to be well mixed.
During the period of rapid accretion, the addition of colder, denser condensates onto the surface of the moonlet will create a gravitationally unstable layer that will continually overturn.
Furthermore, the relative velocity between the moonlet and the vapor is 100s~m~s$^{-1}$. The shearing flow of vapor could drive a surface current on the moonlet (white arrows in Figure~\ref{fig:Moon_cartoon}).
Since the shear is unidirectional across the moonlet, aligned with the flow of gas, there would be a deep return flow within the body. 
Such a flow pattern would force mixing.

The vapor pressure at the surface of large moonlets is larger than the background pressure in the synestia due to the gravitational attraction of the moonlets themselves.
We approximate the pressure at the surface of moonlets by assuming that the boundary layer is bounded by the pressure of the structure at some spherical radius and that the boundary layer is polytropic.
The gas is treated as ideal and diatomic with a polytrope defined by the ratio of heat capacities, $\sim$1.4.
The increase in pressure is small (up to a factor of about 2) for moonlets $<\sim$$0.25M_{\rm Moon}$ but can be significant (factor of about 10) for larger moonlets. 
Although this calculation demonstrates the gravitational effect of moonlets on the pressure in the boundary layer, it does not capture the complexities of the boundary layer.
For example, the radii of larger moonlets can be a significant fraction of the scale height of the structure, so the bounding pressure varies around the moonlet, and the dynamic nature of the boundary layer can cause variations in pressure across the surface. 
For the purposes of this work, it is sufficient to demonstrate that we expect the surface pressure of moonlets could be a few times higher than the pressures given in Figure~\ref{fig:SPH_cooling_time}B.

The boundary layer will change as the structure cools. As the primary moon grows and the pressure in the synestia at the location of the moon drops, the gravity of the moon will begin to dominate the flow.
When the Hill radius of the moon dominates the edge of the synestia (e.g., around 6.5 years in Figure~\ref{fig:SPH_cooling}), the boundary layer will quickly transition to become a captured atmosphere.
At this point, the moon's atmosphere will be a closed system, ending the period of chemical exchange with the terrestrial synestia. 

%xxxxxxxxxxxxxxxxxxxxxxxxxxxxxxxxxxxxxxxxxxxxxxxxxxxxxxxxxxxxxxxx
%xxxxxxxxxxxxxxxxxxxxxxxxxxxxxxxxxxxxxxxxxxxxxxxxxxxxxxxxxxxxxxxx
\subsection{The silicate vaporization temperature buffer}
\label{sec:buffer}
 
As discussed above, there is thermal and chemical exchange between the condensate of the moonlet and the BSE vapor through the boundary layer. Since the temperature is high and the thermal and chemical exchange is rapid, we expect the condensate and vapor at the surface of the moonlet to reach equilibrium before the vapor is swept past the moonlet and replaced by more BSE vapor. The thermodynamics and chemistry of the boundary layer can thus be estimated using an equilibrium calculation. 

In this work, we make the assumption that the flow of gas past the surface of the moonlet chemically equilibrates with a thin skin of the moonlet. In this case, the interacting chemical system that is interacting at the surface of the moonlet is approximately BSE.

The moonlets are heated by thermal exchange with the vapor in the synestia. The temperature of the moonlets rises to the major element bubble point, where the first major element begins to vaporize. For BSE composition, the first major element to substantially vaporize is silicon (Figure~\ref{fig:cond}) as SiO. SiO$_2$ comprises about half the mass of moonlets, and the work required to raise the temperature of the moonlet beyond the major element bubble point is substantial.
Thus, the surface temperature of moonlets will be buffered by the major element bubble point, and we refer to this important physiochemical process as the silicate vaporization temperature buffer.

The fraction of silicon in the vapor is not a linear function with temperature. As presented in Figure~\ref{fig:cond}, the vapor contains a few weight percent of the silicon budget over a wide range of temperatures. In contrast, there is a much narrower range of temperatures where the fraction of silicon in the vapor rises from a few wt\% to nearly 100~wt\%. At the temperature corresponding to about 10~wt\% silicon vaporization, which is approximately the major element bubble point, the energy required to vaporize additional mass is a significant barrier to further increases in temperature.
Based on our estimates that less than 10~wt\% of large moonlets can be vaporized while residing in the vapor structure, we expect the temperature of the moonlets to be buffered close to the major element bubble point. 
In Figure~\ref{fig:cond}, the vertical dashed line denotes the temperature of 10~wt\% silicon vaporization. Based on our example synestias that could form a lunar-mass moon, the pressures expected around the moonlets are tens of bar or more. For 10~wt\% Si in the vapor and 10 to 50 bars, the temperature of the silicate vaporization buffer ranges from about 3400 to 4000~K.

We expect the precise pressure-temperature history for each moonlet to be variable. The value of 10~wt\% Si vaporization, and the corresponding range of pressures and temperatures, are an approximation of the mean conditions of equilibration for the bulk moon. 
An approximate range of possible equilibration pressures and temperatures for moonlets are shown by the white hashed region in Figure~\ref{fig:paths}. This schematic region encompasses temperatures below and above the major element bubble point. The final composition of the moon is likely a nonlinear average of equilibration conditions during accretion.

%xxxxxxxxxxxxxxxxxxxxxxxxxxxxxxxxxxxxxxxxxxxxxxxxxxxxxxxxxxxxxxxx
%xxxxxxxxxxxxxxxxxxxxxxxxxxxxxxxxxxxxxxxxxxxxxxxxxxxxxxxxxxxxxxxx
\subsection{Predicted composition of the moon}
\label{sec:Moon_comp}

\begin{figure}[t]
\centering
\includegraphics[scale=0.833333333]{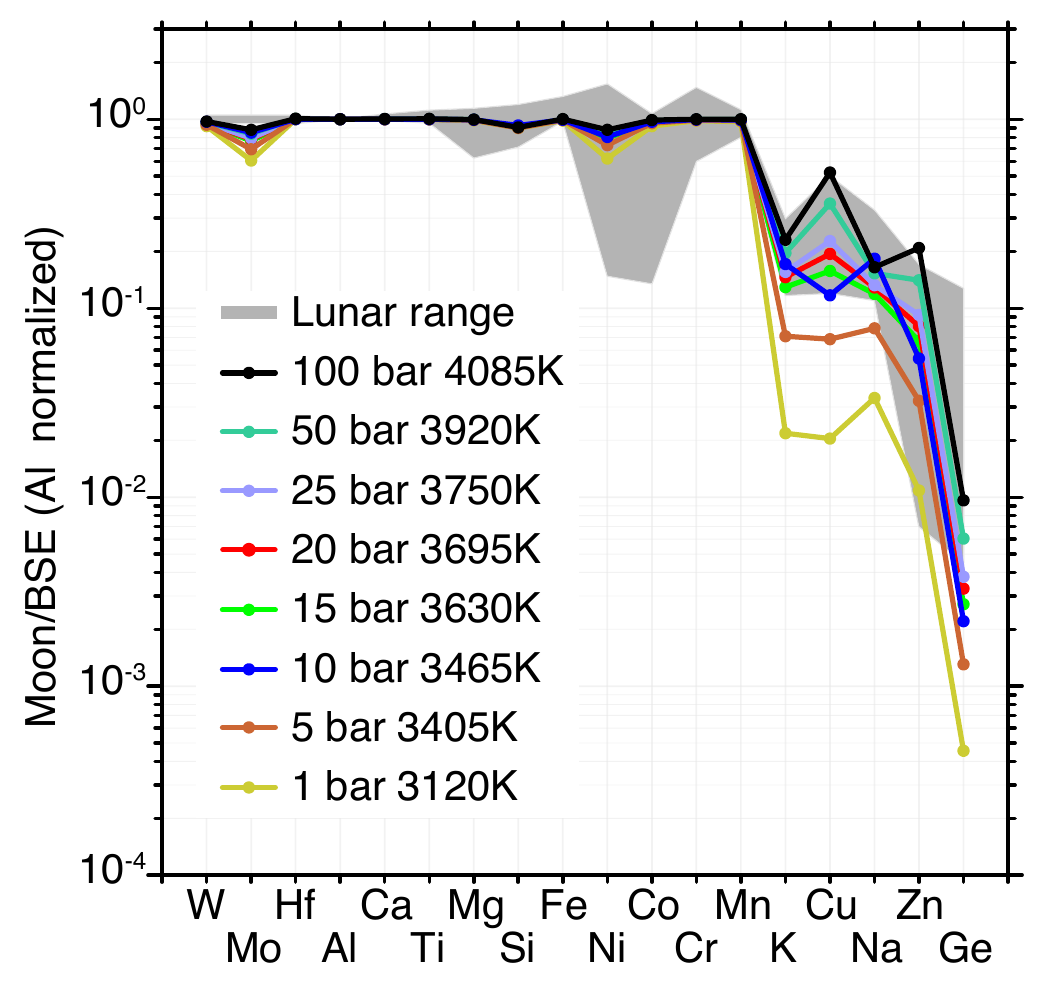}
\caption{Equilibration of moonlets within the vapor structure can reproduce the bulk composition of our Moon. 
The calculated composition of condensate at different pressures, with the temperature set by the silicate vaporization buffer (10~wt\% silicon in the vapor). The range of published estimates for the bulk lunar composition is shown by the gray band. We find a good match to lunar composition for pressures in the range of tens of bar. Equilibration pressures of tens of bar are predicted by our calculations of moon formation from a terrestrial synestia. Lower pressures of equilibration lead to too large a depletion in the moderately volatile elements to match our Moon and conversely higher pressures give an excess of moderately volatile elements.}
\label{fig:lunar_comp}
\end{figure}

To approximate the composition of the moon formed from a synestias, we used our physical chemistry model (\S\ref{sec:thermo}) to calculate the composition of the liquid condensate in a BSE chemical system over a broad range of pressures, with the temperature set by the silicate vaporization buffer (colored lines in Figure~\ref{fig:lunar_comp}). For all cases, the refractory major elements in the condensate are relatively unfractionated from BSE. Our condensation calculations show that the partitioning of MVEs between the condensate and vapor varies with pressure and temperature. As a result, the magnitude and pattern of MVE in the condensate varies substantially with different equilibration conditions. For example, at 10~wt\% silicon vaporization, a negligible amount of MVE is contained in the condensate at 1~mbar (Figure~\ref{fig:cond}A, vertical dashed line). However, at tens of bar, a larger fraction of MVEs are incorporated into the condensate (Figure~\ref{fig:cond}B,C, vertical dashed lines). 

In Figure~\ref{fig:lunar_comp}, we assumed that the silicate vaporization buffer fixes the temperature of equilibration to the point where 10~wt\% of the silicon is in the vapor. 
Varying the amount of silicon in the vapor changes the MVE composition of the equilibrium condensate; however, we find that the condensate compositions at slightly lower and higher temperatures (and lower and higher pressures) are somewhat complementary (Figure~\ref{sup:fig:lunar_comp_buffer}).
Combining material that equilibrated at different temperatures and pressures could still produce an average composition similar to our single pressure-temperature calculations.

For our calculated range of pressures (about 10 bars or more) during satellite accretion from a terrestrial synestia (Figure~\ref{fig:SPH_cooling_time} and supporting information \S\ref{sup:sec:BSM}), the predicted composition of a moon falls within the observed uncertainties for lunar composition (gray band, supporting information). 
At higher and lower pressures, the relative pattern of MVEs does not agree with the observations of our Moon.
Thus, the composition of our Moon could be explained by formation from a terrestrial synestia.

%xxxxxxxxxxxxxxxxxxxxxxxxxxxxxxxxxxxxxxxxxxxxxxxxxxxxxxxxxxxxxxxxxxxxxxxxxxxxxxxxxxxxxxxxxxxxxxxxxxxxxxxxxxxxxxxxxxxxxxxxxxxxxxxxxxxxxxxx
%xxxxxxxxxxxxxxxxxxxxxxxxxxxxxxxxxxxxxxxxxxxxxxxxxxxxxxxxxxxxxxxxxxxxxxxxxxxxxxxxxxxxxxxxxxxxxxxxxxxxxxxxxxxxxxxxxxxxxxxxxxxxxxxxxxxxxxxx
%xxxxxxxxxxxxxxxxxxxxxxxxxxxxxxxxxxxxxxxxxxxxxxxxxxxxxxxxxxxxxxxxxxxxxxxxxxxxxxxxxxxxxxxxxxxxxxxxxxxxxxxxxxxxxxxxxxxxxxxxxxxxxxxxxxxxxxxx
\section{Discussion}
\label{sec:discussion}

In this work, we show that an impact-generated synestia leads to a new environment for satellite formation.
We identified example potential Moon-forming synestias, and the predicted chemical composition of the moon is in excellent agreement with lunar data. 
Therefore, we propose a new model for lunar origin: our Moon accreted within a terrestrial synestia. In this section, we discuss the finer points of this new model, focusing on consistency with other properties of the Earth-Moon system.

%xxxxxxxxxxxxxxxxxxxxxxxxxxxxxxxxxxxxxxxxxxxxxxxxxxxxxxxxxxxxxxxx
%xxxxxxxxxxxxxxxxxxxxxxxxxxxxxxxxxxxxxxxxxxxxxxxxxxxxxxxxxxxxxxxx
\subsection{The fate of volatile elements}

Forming the Moon from a terrestrial synestia provides a physical mechanism to quantitatively explain the well-documented MVE depletion of the Moon compared to Earth. The magnitude and pattern of MVE depletion is determined by equilibration chemistry over a relatively narrow range of pressures and temperatures. Through the silicate vaporization buffer, the composition of the moon is not controlled by an absolute temperature but by the relative partitioning of elements between phases. The MVEs that are not incorporated into the Moon remain in the synestia. As the synestia cools and contracts within the lunar orbit, the remaining MVEs are destined to be incorporated into the bulk silicate Earth. The complement of the Moon's MVEs is a negligible fraction of the terrestrial budget (approximately 2\%), less than the uncertainty in the composition of BSE \citep{McDonough1995}.

When the Moon separates from the synestia, it has a thin silicate atmosphere that cools by radiation. As the pressure drops, the temperature of the lunar surface remains buffered by silicate vaporization and also drops. As a result, the surface conditions quickly reach pressures and temperatures where solid phases are stable. To estimate the speciation of the vapor in equilibrium with the lunar magma ocean as it cools, we used the same methods as in \S\ref{sec:thermo} and calculated the vapor species as bulk lunar composition material cooled isobarically at 10~bar. At the time of separation from the synestia, the lunar atmosphere is dominated by heavy species (e.g., SiO, MgO, H$_2$S, CO, etc., with a mean molecular mass of $>25$~g~mol$^{-1}$). A collisional atmosphere with a mean molecular mass of $>25$~g~mol$^{-1}$ is gravitationally bound to the Moon even at the temperatures of the lunar magma ocean. 
As the atmosphere cools and condenses, the vapor becomes increasingly enriched in volatile species. Simultaneously, the total mass of the atmosphere decreases. This small residual atmosphere may condense or be lost from the Moon. 
Therefore, in our model, the observed isotopic variability of some volatile elements (e.g., Zn, Cl) in lunar samples \citep{Sharp2010,Paniello2012,Day2014} is unlikely to reflect bulk isotopic fractionation due to atmospheric loss \citep{Paniello2012,Day2014} and must arise from later, potentially localized, processes. 

Recent measurements of melt inclusions in lunar samples \citep[e.g.][]{Saal2008,Hauri2011} have been interpreted as evidence for significant amounts of hydrogen (1 to 1000~ppm H$_2$O equivalent by weight) in the Moon's interior. In any giant impact model, the lunar inventory of hydrogen is likely controlled by solubility in the silicate liquid, which may explain the observations \citep[see e.g.,][]{Pahlevan2016}. As our melt model does not include water, our condensation calculations cannot be used to infer the hydrogen budget of the Moon; however, inclusion of hydrogen in the Moon is not inconsistent with our model.

%xxxxxxxxxxxxxxxxxxxxxxxxxxxxxxxxxxxxxxxxxxxxxxxxxxxxxxxxxxxxxxxx
%xxxxxxxxxxxxxxxxxxxxxxxxxxxxxxxxxxxxxxxxxxxxxxxxxxxxxxxxxxxxxxxx
\subsection{Core formation and metal-silicate equilibration}
\label{sec:discussion_core}

Our model has implications for understanding core formation and metal-silicate equilibration after the giant impact. At the pressures and temperatures of the Roche-exterior region of the synestia, the condensate is a silicate liquid. There is no metal phase, and iron is incorporated in the liquid primarily as FeO. The boundary between a mixture of metal and silicate and a pure silicate is shown by the red line in Figure~\ref{fig:BSEphase_diagram}.

During mixing within the synestia and equilibration with moonlets, both lithophile and siderophile elements are homogenized and equilibrated. The lunar core is produced by exsolution of a metal phase from the silicate liquid as it cools. In our calculations a few wt\% of metal is precipitated at moderate pressures, which is consistent with the small size of the lunar core \citep{Garcia2011, Weber2011, Williams2014, Matsuyama2016}. Because exsolved metal droplets would be small, the surface area between the metal and silicate is large, and we expect efficient metal-silicate equilibration at this stage of cooling in both the Moon and the terrestrial synestia.

%xxxxxxxxxxxxxxxxxxxxxxxxxxxxxxxxxxxxxxxxxxxxxxxxxxxxxxxxxxxxxxxx
%xxxxxxxxxxxxxxxxxxxxxxxxxxxxxxxxxxxxxxxxxxxxxxxxxxxxxxxxxxxxxxxx
\subsection{Isotopic composition of Earth and the Moon}
\label{sec:isotopes}

As discussed in the introduction, any complete lunar formation model must explain the similarity of Earth and the Moon in W isotopes \citep{Touboul2015,Kruijer2015,Kruijer2017}. If material in the inner solar system had widely varying stable isotopic compositions, then a model need also reproduce the observed similarity in stable isotopes, such as O \citep[e.g.,][]{Young2016}.

In our lunar origin model, the Moon forms from, and is equilibrated with, the high-entropy regions of a terrestrial synestia. Due to the continuous cycling of mass by falling condensates and vertical fluid convection, the high-entropy layers are likely to be well mixed (\S\ref{sec:dynamics_cooling}). Other mechanisms (such as large scale fluid instabilities, turbulent eddies driven by shear in the structure, and scattering of condensates and gravitational perturbations by the Moon) could also enhance mixing. At the initially high temperatures in the synestia, both lithophile and siderophile elements are incorporated into a single silicate fluid and there is no metal phase (\S\ref{sec:discussion_core}). Therefore, the outer portions of the synestia and the forming Moon share very similar isotopic compositions for both lithophile and siderophile elements.

\begin{figure}
\centering
\includegraphics[scale=0.833333333]{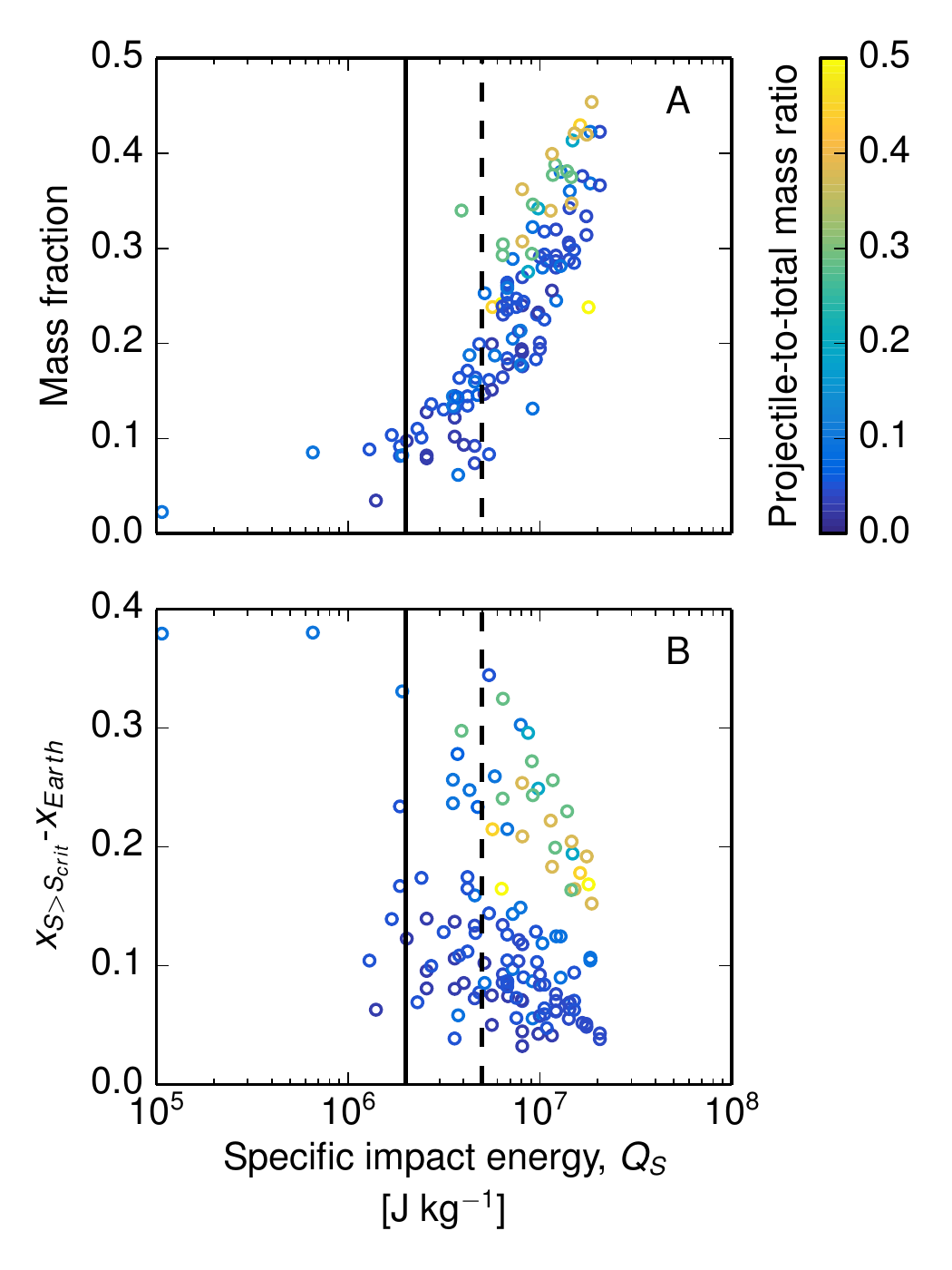}
\caption{Forming a moon from a synestia can potentially produce an isotopically similar satellite. Satellites that form within a synestia could equilibrate with a high-entropy region that is a substantial mass fraction of the silicate portion of the structure (A). For a range of different impact scenarios, the compositional difference between the final moon that forms from the high-entropy region and the bulk Earth (B) are small. $x_{S > S_{\rm crit}}$ and $x_{\rm Earth}$ are the mass fraction of the high entropy regions and the bulk Earth derived from impactor material. In comparison, the Earth-Moon difference in the canonical giant impact is large and $x_{\rm Moon}$~$-$~$x_{\rm Earth}$~$>$~$\sim$~$0.6$, where $x_{\rm Moon}$ is the mass fraction of the Moon derived from impactor material (the equivalent of $x_{S > S_{\rm crit}}$ in B). Colors indicate the projectile-to-total mass ratio, $M_{\rm p}/(M_{\rm t} + M_{\rm p})$, where $M_{\rm p}$ and $M_{\rm t}$ are the mass of the projectile and target respectively. Vertical lines indicate specific impact energies, $Q_{\rm S}$, of 2 and $5$~$\times$~$10^6$~J~kg$^{-1}$.
}
\label{fig:mixing_isotopes}
\end{figure}

Figure~\ref{fig:mixing_isotopes}A shows the mass fraction of the silicate portion of synestias that constitute the high-entropy regions from the suite of SPH simulations from LS17. We have approximated the mass fraction of the synestia that would form the high-entropy region of the synestia, or be included into the lunar seed, as the mass that is either initially mixed phase or has an entropy above the critical point entropy, $S_{\rm crit}$, in the unprocessed SPH simulation output (e.g., the blue particles, and red particles to the right of the critical point in Figure~\ref{fig:contourstructures}D). 
The thermal structure varies substantially with different impact geometries, but we find that up to 50\% of the silicate mass can have specific entropies above the critical point (Figure~\ref{fig:mixing_isotopes}A). We find that the mass fraction of the high-entropy regions scales well with a modified specific impact energy, $Q_{\rm S}$, which is defined in LS17. As expected, higher energy impacts lead to more massive high-entropy regions. 

Formation of the Moon from such massive high entropy regions, combined with enhanced mixing during high-AM, high-energy impacts, may explain the isotopic similarity of Earth and the Moon. Figure~\ref{fig:mixing_isotopes}B shows the approximate difference in the mass fractions of the high-entropy regions and the bulk synestia that are derived from the impactor, denoted $x_{S>S_{\rm crit}}$ and $x_{\rm Earth}$ respectively, for the synestias found by LS17. The values shown assumes that the Moon inherits the composition of the high-entropy region of the structure and that the Earth inherits the bulk composition of the synestia. Lower values of $x_{S>S_{\rm crit}}-x_{\rm Earth}$ indicate smaller isotopic differences. The isotopic difference for moons formed from synestias is small compared to the Earth-Moon difference in the canonical giant impact, which typically result in values of $x_{\rm Moon} - x_{\rm Earth} >\sim0.6$, where $x_{\rm Moon}$ is the mass fraction of the Moon inferred to come from the impactor. For reference, assuming an initial O isotope distribution in the solar system and using the results from $N$-body simulations of terrestrial planet formation, \citet{Young2016} found that $\sim$40\% of terminal giant impacts would reproduce the similarity in O isotopes if $x_{\rm Moon} - x_{\rm Earth} =0.1$ and $\sim$20\% if $x_{\rm Moon} - x_{\rm Earth} =0.2$. There is a subset of potential Moon-forming synestias that could satisfy the O isotope constraint without the need for a homogeneous inner solar system. The subset of impacts that could match the W constraint is uncertain, but it will be larger. 

W isotopes are sensitive to the timing of core formation, and core formation on Earth and the Moon ends with the Moon-forming impact. If the Moon formed $<\sim50$~Myr after the start of the solar system, then the Hf-W system would have still been alive at the time of the impact.  However, the Hf/W ratio of the mantles of Earth and the Moon are probably very similar as the La/W ratios are comparable in a variety of lunar and terrestrial rocks \citep{Wieczorek2006}. Thus, any subsequent $^{182}$W evolution from $^{182}$Hf decay would still result in similar tungsten isotope signatures for both bodies.

Several studies of the deep terrestrial mantle sample a reservoir that is isotopically distinct from the upper mantle \citep{Yokochi2004,Holland2006,Mukhopadhyay2012,Tucker2012,Parai2012,Peto2013,Tucker2014,Rizo2016,Mundl2017}.
In particular, measurements of Xe and W isotopes suggest that there is a mantle reservoir that was formed early in Earth's accretion and has persisted to the present day, presumably surviving the Moon-forming giant impact \citep{Mukhopadhyay2012,Tucker2012,Parai2012,Peto2013,Rizo2016,Mundl2017}. The majority of post-impact structures are strongly thermally stratified [LS17] and the whole Earth is unlikely to be perfectly mixed during or after the impact. The layer at the base of the mantle has much lower entropy than the rest of the structure and generally contains a lower mass fraction of impactor. This portion of the body may have a distinct isotopic composition to the rest of the synestia. The observed isotopic anomalies may be the remnants of the low-entropy layer.

Some isotopic fractionation between Earth and the Moon would be expected in our model. Fractionation can be produced during the condensation of lunar material and while moonlets equilibrate with the synestia. Equilibration between moonlets and the vapor of the synestia would produce an isotopically heavy Moon compared to Earth, but since the equilibration occurs at relatively high pressures and temperatures, the degree of fractionation is expected to be small. Recently, a potassium isotope fractionation has been measured between lunar and terrestrial samples that is consistent with condensation at high pressures and temperatures \citep{Wang2016}. Further development of our model could make predictions as to the degree of fractionation of other elements.

Formation of the Moon from a terrestrial synestia is a promising mechanism to explain the isotopic similarity between Earth and the Moon.

%xxxxxxxxxxxxxxxxxxxxxxxxxxxxxxxxxxxxxxxxxxxxxxxxxxxxxxxxxxxxxxxx
%xxxxxxxxxxxxxxxxxxxxxxxxxxxxxxxxxxxxxxxxxxxxxxxxxxxxxxxxxxxxxxxx
\subsection{Potential Moon-forming impacts}
\label{sec:impacts}

In the canonical giant impact model, reproducing the present-day AM of the Earth-Moon system placed stringent constraints on the impact parameters. Only a narrow range of mass ratios, impact velocities, and impact angles produced the desired post-impact structure \citep{Canup2001,Canup2004,Canup2008}. Thus, the canonical giant impact has been a logical focus for studies related to lunar origin. 

In this work, we present examples of potential Moon-forming synestias that were generated by very different giant impact configurations, from small impactors to approximately equal-mass bodies (\ref{sup:sec:extra_examples}). Note that pre-impact rotation is not required to produce a Moon-forming synestia. However, these example impact-generated synestias have some similar characteristics and provide an initial guide to understanding the range of giant impacts that have the potential to form our Moon.

In our new model, the key features for a potential Moon-forming giant impact are: (i) a terrestrial synestia is formed; (ii) the synestia has a distribution of mass and thermal energy such that the Roche-exterior region has sufficient sustained vapor pressure for chemical equilibration of moonlets to produce the observed depletion in MVEs (tens of bars in our calculations); (iii) the synestia has the requisite mass and AM to accrete a lunar mass satellite. 

LS17 defined the conditions necessary to form a terrestrial synestia. 
An Earth-mass body must exceed the corotation limit, which is a function of AM and thermal energy (specific AM of the outer silicate material). 
LS17 found that the specific entropy of the outer silicate portions of bodies after giant impacts scales well with a modified specific impact energy, $Q_{\rm S}$. $Q_{\rm S}$ is a variation of the parameter developed in \citet{Leinhardt2012} that adjusts for the geometry of collisions between similarly sized bodies to estimate the relative deposition of impact energy into the post-impact body for different collision scenarios. LS17 showed that Earth-mass planets typically experience several giant impacts with $Q_{\rm S}$~$>$~$2\times10^6$~J~kg$^{-1}$ during accretion and over half experience at least one impact with $Q_{\rm S}$~$>$~$5\times10^6$~J~kg$^{-1}$.
Impacts with $Q_{\rm S}$~$>$~$2\times10^6$~J~kg$^{-1}$ generally deposit sufficient energy such that the upper 25~wt\% of the silicate portion of a body has an average specific entropy that exceeds the critical point value, $S_{\rm crit}$.
For impacts with $Q_{\rm S}$~$>$~$5\times10^6$~J~kg$^{-1}$, the upper 50~wt\% of silicates typically attain mean specific entropies above the critical point.

Post-impact bodies with high specific entropy ($\ge S_{\rm crit}$) outer layers exceed the corotation limit when they also have a total AM greater than about $1.5 L_{\rm EM}$. 
The AM of terrestrial bodies during accretion is difficult to track. \citet{Kokubo2010} have conducted the best assessment to date of the spin state of rocky planets during accretion. 
They used an $N$-body simulation of planet formation with bimodal impact outcomes, either perfect merging or hit-and-run, and tracked the AM of each of the bodies in the simulation. 
They found that the mean AM of rocky planets is large, e.g., 2.7~$L_{\rm EM}$ for Earth-mass planets, and that the distribution is broad. About 84\% of their final Earth-mass bodies had an AM greater than 1.6~$L_{\rm EM}$.
Based on the results of \citet{Kokubo2010}, most post-impact planets at the end of accretion would have sufficient AM to be above the corotation limit after a high energy impact. Given the prevalence of high energy impacts, LS17 concluded that synestias were common during accretion.

\begin{figure}
\centering
\includegraphics[scale=0.83333333333]{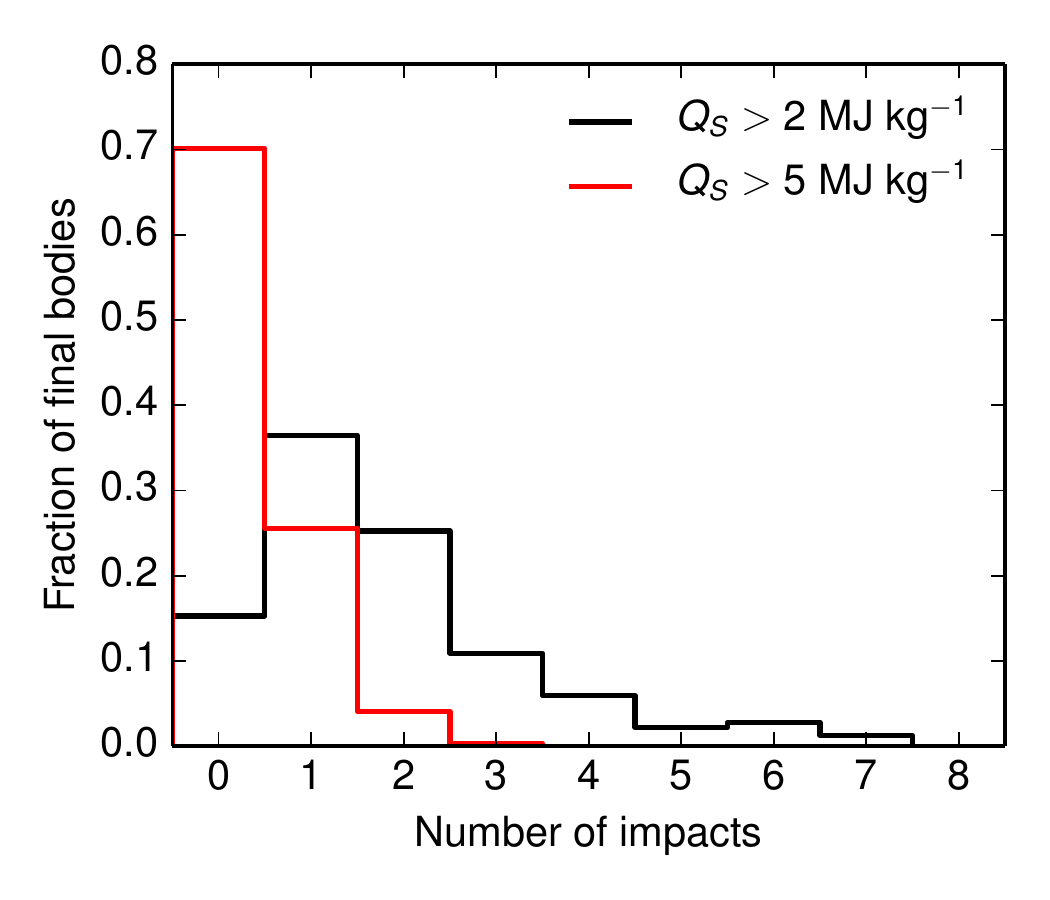}
\caption{In $N$-body simulations of planet formation, a majority of Earth-like bodies experience a number of late, high-energy giant impacts. Histograms show the fraction of final bodies in the simulations by \citet{Quintana2016} with masses $>0.5M_{\rm Earth}$ that experience a given number of impacts with a specific energy $Q_{\rm S}$ $>2 \times 10^6$~J~kg$^{-1}$ (black) and $>5 \times 10^6$~J~kg$^{-1}$ (red) late in their accretion. Late impacts are defined as those where the largest remnant after the impact is at least 50\% of the mass of the final body.}
\label{fig:impact_likelihood}
\end{figure}

In order to satisfy the observed chemical relationships between Earth and the Moon, the Moon-forming giant impact was likely the last major accretionary event on the growing Earth.
Hence, we consider the probability of a high-energy terminal giant impact. 
We analyzed the $N$-body simulations of \citet{Quintana2016} in a manner similar to LS17, but we only considered impacts onto almost fully formed bodies.
Figure~\ref{fig:impact_likelihood} presents the number of late high-energy giant impacts onto bodies with a final mass of $>$~$0.5 M_{\rm Earth}$. To include only late impacts, we only considered cases where the post-impact body had a mass $>$~$50$\% of the final mass of the planet.
High-energy giant impacts are prevalent at the end of accretion.
About 85\% of bodies experienced at least one impact with $Q_{\rm S}$~$>$~$2\times10^6$~J~kg$^{-1}$ and 30\% experienced at least one impact with $Q_{\rm S}$~$>$~$5\times10^6$~J~kg$^{-1}$ late in their formation. Other $N$-body studies without fragmentation and with different initial conditions and alternative configurations of the giant planets \citep[e.g.,][]{Obrien2006,Hansen2009,Raymond2009,Walsh2011,Levison2015a} all produce high-energy impacts in the final stages of accretion although the frequency of different energy impacts varies between these various scenarios. 
Given that the AM of terrestrial bodies are expected to be high at the end of planet formation \citep{Kokubo2010}, we conclude that synestias are a common outcome of terminal giant impacts. 

\begin{figure}[p]
\centering
\includegraphics[scale=0.833333333]{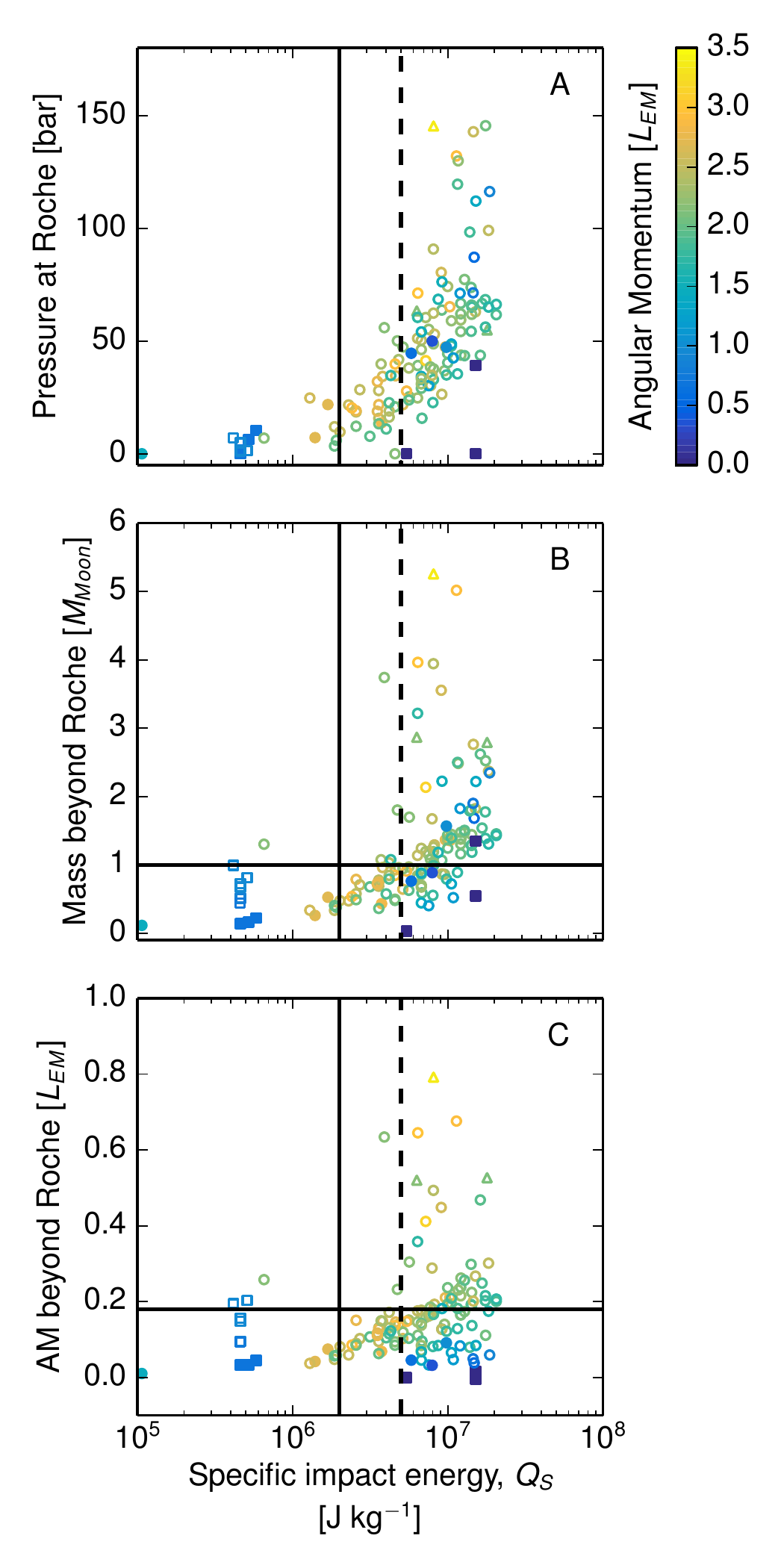}
\caption{A range of synestias formed at the end of planet formation may be suitable Moon-forming structures. Using a suite of SPH impact simulations from LS17, we calculated the pressure in the midplane at the Roche limit in the unprocessed post-impact structure (A), the mass beyond the Roche limit (B), and the AM beyond the Roche limit (C) as a function of the specific energy, $Q_{\rm S}$. Symbols denote the dynamic class of each post-impact structure as defined by LS17; super-CoRoL (circles), sub-CoRoL (squares), and co-CoRoL (triangles). Filled symbols denote structures with limited mass in their disk-like regions. The solid and dashed vertical lines indicate impact energies of 2 and $5\times10^6$~J~kg$^{-1}$ respectively. Horizontal lines indicate one lunar mass (B) and the AM of a moon orbiting at the Roche limit around an Earth-mass body (C).}
\label{fig:QS_scaling}
\end{figure}

We find that the pressure in the midplane at the Roche limit is roughly correlated with the specific impact energy $Q_{\rm S}$, shown in Figure~\ref{fig:QS_scaling}A. This figure presents characteristics of the unprocessed post-impact structures from the database presented in LS17. 
Impact-generated synestias are plotted as circles.
Most impacts with $Q_{\rm S}>2\times10^6$~J~kg$^{-1}$ produce structures that have initial pressures of at least 10~bars at the Roche limit. Note that the midplane pressure in the raw SPH calculation is a lower limit for disk-like regions with large condensate fractions, and thermal equilibration and removal of condensate would provide a more accurate vapor pressure estimate (supporting information \S\ref{sup:sec:SPH_cooling}). Nevertheless, with the direct SPH values, most impact-generated synestias have Roche-exterior vapor pressures consistent with our inferred conditions for lunar origin for satellites that form near the Roche limit.

The mass and AM that is injected beyond the Roche limit during each impact varies dramatically (Figure~\ref{fig:QS_scaling}B,C). The distribution of mass and AM is very sensitive to the exact impact parameters and not just the specific impact energy.
The current data suggests a trend to higher pressures at the Roche limit, and more mass and AM beyond Roche with increasing total AM of the post-impact structure.
Given this wide variety of initial mass and AM distributions and the uncertainty in the evolution of post-impact structures, it is not possible to predict which of these synestias could form a lunar mass satellite from $Q_{\rm S}$ alone. Because both mass and AM maybe transported outward during viscous evolution of a synestia, it is not required that the giant impact emplace all the mass and AM for the Moon beyond the Roche limit. However, the suite of impacts from LS17 contains many promising candidates that satisfy these criteria, in addition to many that nearly reach these levels.
Further work is needed to robustly determine the final mass of a moon formed from a synestia in order to better define the requirements for a Moon-forming post-impact structure. 

Here we examined the outcomes from single giant impacts; however, about 1/3 of giant impacts are hit-and-run events \citep{Asphaug2006,Stewart2012}. It is possible that a potential Moon-forming synestia could be formed by a hit-and-run event followed by an accretionary event with the same impactor. Because the two bodies would have a subsequent encounter on orbital timescales, the proto-Earth would not have recovered completely from the first event. Thus, not all of the thermal energy and AM of a Moon-forming synestia needs to be delivered in one event and a two part sequence may be considered in future work.

Based on this work, we anticipate that very different impact scenarios can generate Moon-forming synestias. Thus, our model for lunar origin shifts the focus from a specific impact event to the properties of the post-impact structure.

%xxxxxxxxxxxxxxxxxxxxxxxxxxxxxxxxxxxxxxxxxxxxxxxxxxxxxxxxxxxxxxxxxxxxxxxxxxxxxxxxxxxxxxxxxxxxxxxxxxxxxxxxxxxxxxxxxxxxxxxxxxxxxxxxxxxxxxxx
%xxxxxxxxxxxxxxxxxxxxxxxxxxxxxxxxxxxxxxxxxxxxxxxxxxxxxxxxxxxxxxxxxxxxxxxxxxxxxxxxxxxxxxxxxxxxxxxxxxxxxxxxxxxxxxxxxxxxxxxxxxxxxxxxxxxxxxxx
%xxxxxxxxxxxxxxxxxxxxxxxxxxxxxxxxxxxxxxxxxxxxxxxxxxxxxxxxxxxxxxxxxxxxxxxxxxxxxxxxxxxxxxxxxxxxxxxxxxxxxxxxxxxxxxxxxxxxxxxxxxxxxxxxxxxxxxxx
\section{Synopsis of the origin of the Moon}
\label{sec:summary}

Here, we summarize our new model for lunar origin. The trigger is a high-energy, high-angular momentum (high-AM) giant impact that forces the post-impact structure to exceed the corotation limit (CoRoL), forming a previously unrecognized planetary object called a synestia. The outer regions of synestias are largely vapor and there is strong pressure support, causing the disk-like regions to be substantially sub-Keplerian. A variety of high-energy, high-AM giant impact events can exceed the CoRoL with sufficient mass and AM in the outer structure to accrete the Moon. High-energy, high-AM impacts typically lead to more mixing of impactor and target material than occurs during the canonical impact.

\begin{figure*}
\centering
\includegraphics[scale=0.833333333]{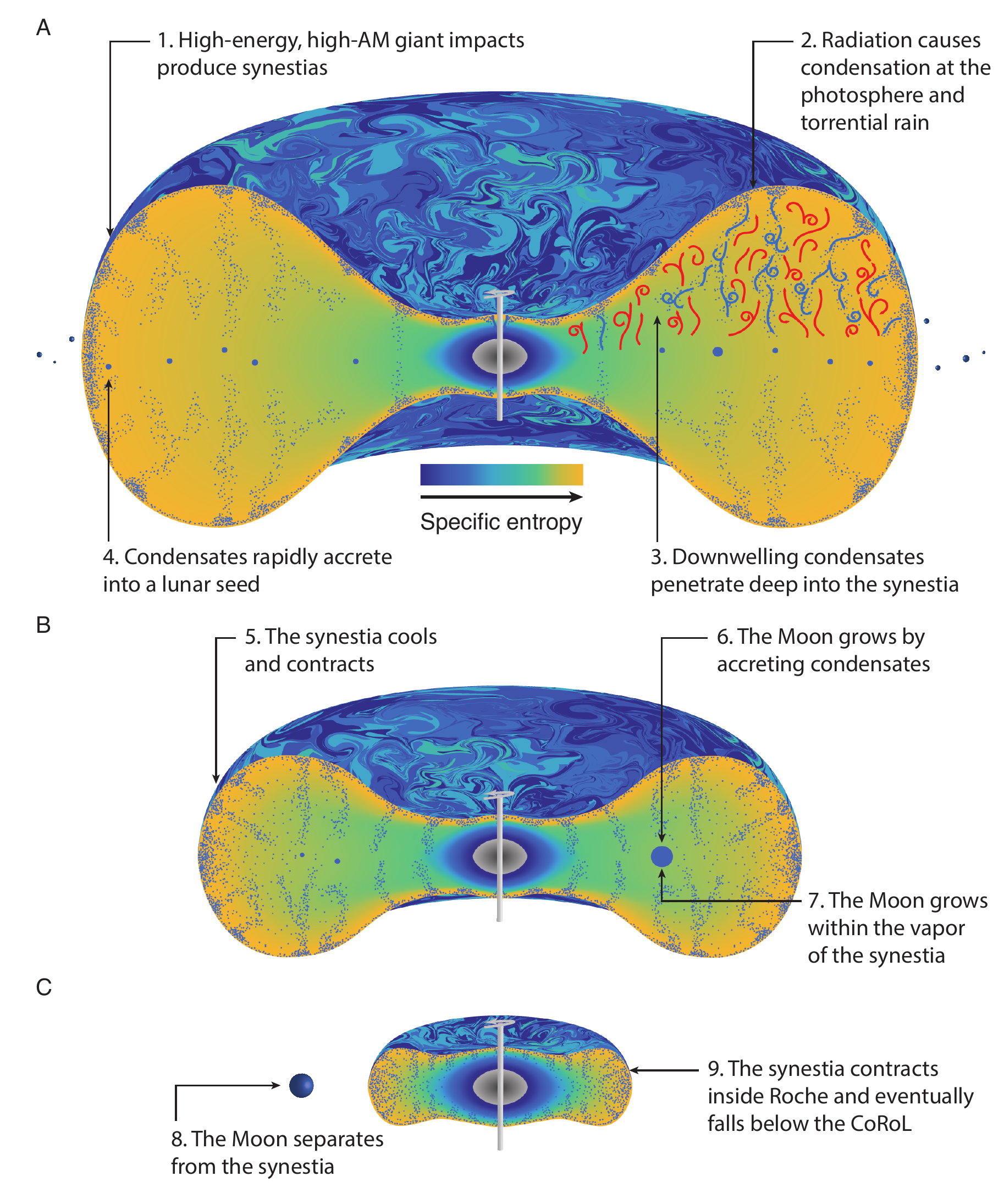}
\caption{Schematic illustration of the formation of the Moon within a terrestrial synestia. The key steps are: (A) Formation of a terrestrial synestia by a high-energy, high-AM giant impact and accretion of a lunar seed that orbits within the vapor structure. (B) The moon grows on the time scale of about a year, accreting condensates and moonlets as the synestia radiatively cools. Moonlets chemically equilibrate with bulk silicate Earth vapor at the silicate vaporization temperature buffer. (C) As the system cools and contracts, after several to tens of years, the Moon separates from the synestia with a small captured atmosphere and begins to cool and solidify. The synestia contracts within the Roche limit and eventually falls below the CoRoL. For each step, half of the 3-dimensional structure is shown with colors denoting the specific entropy of silicate material. The phase varies with specific entropy and pressure (see Figure~\ref{fig:contourstructures}C-D, G-H). At lower pressures, further out in the structure, condensates are shown in blue. The gray region is the core and the rotation axis is shown as a vertical column in the center of the structure. The photosphere of the synestia is dominated by condensates. The top right quadrant of A schematically shows turbulent convection in the terrestrial synestia with condensate-rich downwellings (blue arrows) cutting across the convective return flow (red arrows). All the stages are to scale.}
\label{fig:MAD_cartoons}
\end{figure*}

As the system radiatively cools, silicate droplets condense from the vapor in the low-pressure, optically thin outer layers of the structure and fall inwards (Figure~\ref{fig:MAD_cartoons}A,B). The torrential rain of condensates drives radial mass and AM transport. Initially, condensates that fall within the Roche limit are vaporized in the hotter, higher-pressure interior of the structure, and their mass is mixed with the silicate vapor in the structure. However, condensates with sufficient AM to remain outside the Roche limit accrete to form moonlets. The initial moonlets, formed by rapid cooling of the low-surface-density outer disk, accrete subsequent condensates to form the Moon (Figure~\ref{fig:MAD_cartoons}B).

The Moon inherits its composition by forming from, and equilibrating with, BSE vapor. Moonlets chemically equilibrate at a pressure set by the mass of the vapor disk and a temperature set by the onset of silicate vaporization. In our calculations, the moonlets equilibrate at tens of bar and between about 3400 and 4000~K. Equilibration between the lunar liquid and BSE vapor within a boundary layer has the potential to quantitatively produce the observed bulk lunar composition, including the magnitude and pattern of depletion in moderately volatile elements. At the high temperatures in the synestia, both lithophile and siderophile elements are incorporated into a single silicate fluid and there is no metal phase. Metal precipitates from the silicate fluid of the Moon to form the small lunar core. The outer portions of the synestia and the Moon share very similar isotopic compositions for both lithophile and siderophile elements. Formation from a sufficiently well-mixed synestia has the potential to explain the similarity between the Earth and the Moon in their tungsten, and potentially other, isotopic compositions.
	
The Moon is nearly fully grown before the synestia cools and contracts within the lunar orbit (Figure~\ref{fig:MAD_cartoons}C). Eventually, the structure cools below the CoRoL and forms a corotating planet. Thus, Earth inherits the Moon's complement of moderately volatile elements.

To complete our model, the AM of the Earth-Moon system must be reduced to the present-day value after lunar accretion. Multiple physical processes during lunar tidal evolution have been proposed to remove the required amount of AM from the Earth-Moon system \citep{Cuk2012,Wisdom2015,Cuk2016,Tian2017}.

%xxxxxxxxxxxxxxxxxxxxxxxxxxxxxxxxxxxxxxxxxxxxxxxxxxxxxxxxxxxxxxxxxxxxxxxxxxxxxxxxxxxxxxxxxxxxxxxxxxxxxxxxxxxxxxxxxxxxxxxxxxxxxxxxxxxxxxxx
%xxxxxxxxxxxxxxxxxxxxxxxxxxxxxxxxxxxxxxxxxxxxxxxxxxxxxxxxxxxxxxxxxxxxxxxxxxxxxxxxxxxxxxxxxxxxxxxxxxxxxxxxxxxxxxxxxxxxxxxxxxxxxxxxxxxxxxxx
%xxxxxxxxxxxxxxxxxxxxxxxxxxxxxxxxxxxxxxxxxxxxxxxxxxxxxxxxxxxxxxxxxxxxxxxxxxxxxxxxxxxxxxxxxxxxxxxxxxxxxxxxxxxxxxxxxxxxxxxxxxxxxxxxxxxxxxxx
\section{Conclusions and future tests of the model}
\label{sec:conclusion}

Most high-energy, high-AM giant impacts can produce synestias. The formation of the Moon within a terrestrial synestia can potentially reproduce the lunar bulk composition, the isotopic similarity between Earth and the Moon, and the large mass of the Moon. 
If the post-impact body also had high obliquity, the same giant impact may trigger a tidal evolution sequence that explains the present day lunar inclination and the AM of the Earth-Moon system \citep{Cuk2016}.

In our model, the properties of the post-impact structure constrain the timescale and pressure-temperature conditions for lunar accretion. In particular, the physical environment within the synestia controls the composition of the Moon. Based on our calculations, a range of giant impacts could produce post-impact synestias with appropriate structures to reproduce the physical and chemical properties of our Moon. Thus, our model shifts the focus for lunar origin from a specific impact scenario to the properties of the terrestrial synestia.

In this work, we have demonstrated the feasibility of our model and its potential to explain key observational constraints on the Earth-Moon system, but multiple processes require further investigation. In particular, dedicated study of the multiphase fluid dynamics of the terrestrial synestia is necessary to understand the environment for lunar accretion and the degree of isotopic mixing. Furthermore, development of impact codes that allow thermal equilibration will be required to more accurately determine the degree of mixing in giant impacts. Satellite accretion processes within a synestia and feedbacks between the Moon and the synestia have not been modeled directly. Here we have simply coupled our physical and chemical models using the pressure-temperature-time paths of material in the synestia. A fuller understanding of the dynamics and thermodynamics in the boundary layers of moonlets, and the feedback of moonlets on the synestia, will be required to more accurately determine the coupling of physical and chemical processes.

Further development of our model will allow us to make stronger predictions about the compositions of Earth and the Moon. Improved constraints on lunar composition, particularly improving the estimates for moderately volatile elements, will provide a test of our model by confirming the pattern of depletion and extending possible comparison to elements not considered here. 
Improved estimates of volatile depletion will also further refine the constraints on the pressure and temperature of equilibration.
The pressure-temperature conditions in our model also enable predictions of isotopic fractionation between Earth and the Moon.

%xxxxxxxxxxxxxxxxxxxxxxxxxxxxxxxxxxxxxxxxxxxxxxxxxxxxxxxxxxxxxxxxxxxxxxxxxxxxxxxxxxxxxxxxxxxxxxxxxxxxxxxxxxxxxxxxxxxxxxxxxxxxxxxxxxxxxxxx
%xxxxxxxxxxxxxxxxxxxxxxxxxxxxxxxxxxxxxxxxxxxxxxxxxxxxxxxxxxxxxxxxxxxxxxxxxxxxxxxxxxxxxxxxxxxxxxxxxxxxxxxxxxxxxxxxxxxxxxxxxxxxxxxxxxxxxxxx
%xxxxxxxxxxxxxxxxxxxxxxxxxxxxxxxxxxxxxxxxxxxxxxxxxxxxxxxxxxxxxxxxxxxxxxxxxxxxxxxxxxxxxxxxxxxxxxxxxxxxxxxxxxxxxxxxxxxxxxxxxxxxxxxxxxxxxxxx
\section*{Acknowledgments}
This work was supported by NESSF grant NNX13AO67H (SJL), NASA grant NNX11AK93G (STS, SJL), NASA grant NNX15AH54G (STS), DOE-NNSA grants DE-NA0001804 and DE-NA0002937 (STS, MIP, SBJ), NASA Emerging Worlds grant NNX15AH66G (SBJ), NASA grant NNX15AH65G (M\'C), NERC grant NE\/K004778\/1 (ZML, MTM). The authors would like to thank Dave Stevenson and Denton Ebel for thoughtful reviews that greatly improved the clarity of this manuscript, and Steven Hauck for excellent editorial guidance and comments. We also acknowledge Razvan Caracas, Yohai Kaspi, Maylis Landeau, Don Korycansky, Kevin Zahnle, S\'{e}bastian Charnoz, Lindy Elkins-Tanton, Jay Melosh, Jon Aurnou, Alex Grannan, Richard Kerswell, Rick Kraus, Rita Parai, Sujoy Mukhopadhyay, Nicolas Dauphas, Phil Carter, Erik Davies, Gigja Hollyday, Kaitlyn Amodeo, Miki Nakajima, and Francis Nimmo for useful comments and discussions that helped improve this paper and shaped our thinking on lunar origin. We thank Sean Raymond and Elisa Quintana for providing N-body simulation data. A \textsc{python} script for calculating the orbital evolution of condensates including gas drag is included in the supporting information. The modified version of GADGET-2 and the EOS tables used in this paper are contained in the supplement of \cite{Cuk2012}. The additional cooling routine for GADGET-2 is provided in the supporting information. The GRAINS code and the associated thermodynamic tables are also provided in the supporting information.

%xxxxxxxxxxxxxxxxxxxxxxxxxxxxxxxxxxxxxxxxxxxxxxxxxxxxxxxxxxxxxxxxxxxxxxxxxxxxxxxxxxxxxxxxxxxxxxxxxxxxxxxxxxxxxxxxxxxxxxxxxxxxxxxxxxxxxxxx
%xxxxxxxxxxxxxxxxxxxxxxxxxxxxxxxxxxxxxxxxxxxxxxxxxxxxxxxxxxxxxxxxxxxxxxxxxxxxxxxxxxxxxxxxxxxxxxxxxxxxxxxxxxxxxxxxxxxxxxxxxxxxxxxxxxxxxxxx
%xxxxxxxxxxxxxxxxxxxxxxxxxxxxxxxxxxxxxxxxxxxxxxxxxxxxxxxxxxxxxxxxxxxxxxxxxxxxxxxxxxxxxxxxxxxxxxxxxxxxxxxxxxxxxxxxxxxxxxxxxxxxxxxxxxxxxxxx
%%%%%%%%%%%%%%%%%%%%%APPENDIX%%%%%%%%%%%%%%%%%%%%%%%%%%%
%xxxxxxxxxxxxxxxxxxxxxxxxxxxxxxxxxxxxxxxxxxxxxxxxxxxxxxxxxxxxxxxxxxxxxxxxxxxxxxxxxxxxxxxxxxxxxxxxxxxxxxxxxxxxxxxxxxxxxxxxxxxxxxxxxxxxxxxx
%xxxxxxxxxxxxxxxxxxxxxxxxxxxxxxxxxxxxxxxxxxxxxxxxxxxxxxxxxxxxxxxxxxxxxxxxxxxxxxxxxxxxxxxxxxxxxxxxxxxxxxxxxxxxxxxxxxxxxxxxxxxxxxxxxxxxxxxx
%xxxxxxxxxxxxxxxxxxxxxxxxxxxxxxxxxxxxxxxxxxxxxxxxxxxxxxxxxxxxxxxxxxxxxxxxxxxxxxxxxxxxxxxxxxxxxxxxxxxxxxxxxxxxxxxxxxxxxxxxxxxxxxxxxxxxxxxx

%make all the sections labelled with an S for supp
%\renewcommand{\thepage}{A\arabic{page}}  
\renewcommand{\thesection}{Appendix \arabic{section}}   
\renewcommand{\thetable}{A\arabic{table}}   
\renewcommand{\thefigure}{A\arabic{figure}}
\renewcommand{\theequation}{A\arabic{equation}}

%reset page count and section count
\setcounter{section}{0}
\setcounter{figure}{0}
\setcounter{table}{0}
\setcounter{equation}{0}

%xxxxxxxxxxxxxxxxxxxxxxxxxxxxxxxxxxxxxxxxxxxxxxxxxxxxxxxxxxxxxxxxxxxxxxxxxxxxxxxxxxxxxxxxxxxxxxxxxxxxxxxxxxxxxxxxxxxxxxxxxxxxxxxxxxxxxxxx
%xxxxxxxxxxxxxxxxxxxxxxxxxxxxxxxxxxxxxxxxxxxxxxxxxxxxxxxxxxxxxxxxxxxxxxxxxxxxxxxxxxxxxxxxxxxxxxxxxxxxxxxxxxxxxxxxxxxxxxxxxxxxxxxxxxxxxxxx
%xxxxxxxxxxxxxxxxxxxxxxxxxxxxxxxxxxxxxxxxxxxxxxxxxxxxxxxxxxxxxxxxxxxxxxxxxxxxxxxxxxxxxxxxxxxxxxxxxxxxxxxxxxxxxxxxxxxxxxxxxxxxxxxxxxxxxxxx
%%%%%%%%%%%%%%%%%%%%%%%%%%%%%%%%%%%%%%%%%%%%%%%%%%%%%%%%%%%%%%%%%%%%%%
\section{Orbital integration with gas drag}
\label{sup:sec:gas_drag}

In \S\ref{sec:dynamics_cooling} we presented results for the orbital evolution of condensates at different heights in a vapor synestia including gas drag. Here we give more details of the implementation of that simple model.

Condensates in the structure experience acceleration by both gravitational forces, $\mathbf{g}$, and drag due to differential velocity, $\mathbf{a_{\rm D}}$. The acceleration of the body is given by
\begin{equation}
\ddot{\mathbf{r}} = \mathbf{g} + \mathbf{a_{\rm D}} \, ,
\label{sup:eqn:gas_drag_acc}
\end{equation}
where $\mathbf{r}$ is the position vector of the condensate in the non-rotating center of mass frame. In a synestia, due to the flattening of the structure the gravitational field is not spherical symmetric. However, in the outer regions of the synestia the majority of the mass is significantly further inwards in the structure and a simple gravitational term of the form
\begin{equation}
\mathbf{g} = \frac{G M_{\rm int}}{\left | \mathbf{r} \right |^3} \mathbf{r} 
\end{equation}
is sufficient to model the gravitational acceleration to within a few percent. The suitability of this approximation can be seen by comparing the green dashed and solid lines in Figure~\ref{fig:pressure_support_main}. $G$ is the gravitational constant, and $M_{\rm int}$ is the mass of the structure with a radius less than $ \left | \mathbf{r} \right |$. The acceleration due to gas drag was calculated using Equation~\ref{eqn:gas_drag}.

The equation of motion (Equation~\ref{sup:eqn:gas_drag_acc}) can be solved by dividing it into a series of first order, ordinary differential equations. The components of equation~\ref{sup:eqn:gas_drag_acc} in the $r$, $\theta$, and $\phi$ directions in spherical coordinates are
\begin{equation}
\ddot{r} = g + a_{\rm D}^r + r \dot{\phi}^2 + r \sin^2{(\phi)} \dot{\theta}^2 \, ,
\end{equation}
\begin{equation}
\ddot{\theta}=\frac{1}{r\sin{(\phi)}} 
\left ( a_{\rm D}^{\theta} - 2 \sin{(\phi)}  \dot{\theta}\dot{r} - 2r \cos{(\phi)}\dot{\theta}\dot{\phi}
\right ) \, , \, {\rm and}
\end{equation}
\begin{equation}
\ddot{\phi} = \frac{1}{r} \left ( a_{\rm D}^{\phi} - 2 \dot{r}\dot{\phi} + r \sin{(\phi)} \cos{(\phi)}  \dot{\theta}^2 \right ) \, ,
\end{equation}
where $a_{\rm D}^{r}$, $a_{\rm D}^{\theta}$, and $a_{\rm D}^{\phi}$ are the components of the gas drag acceleration. $\theta$ is the angle about the rotation axis and $\phi$ is the angle away from the positive rotation axis. By defining $\xi = \dot{r}$, $\Theta = \dot{\theta}$, and $\Phi = \dot{\phi}$ the above set of equations can be written as
\begin{equation}
\dot{\xi} = g + a_{\rm D}^r + r \Phi^2 + r \sin^2{(\phi)} \Theta^2 \, ,
\end{equation}
\begin{equation}
\dot{\Theta}=\frac{1}{r\sin{(\phi)}} 
\left ( a_{\rm D}^{\theta} - 2 \sin{(\phi)}  \Theta\xi - 2r \cos{(\phi)}\Theta\Phi
\right ) \, , \, {\rm and}
\end{equation}
\begin{equation}
\dot{\Phi} = \frac{1}{r} \left ( a_{\rm D}^{\phi} - 2 \xi\Phi + r \sin{(\phi)} \cos{(\phi)}  \Theta^2 \right ) \, .
\end{equation}
We solved this set of six first order, coupled, ordinary differential equation using the {\it odeint} function from the {\it SciPy} package in \textsc{python}. 

The condensates were initialized at a given position in the structure with a velocity equal to the azimuthal gas velocity taken from the equilibrated SPH synestia shown in Figures~\ref{fig:contourstructures} and \ref{fig:linestructures}. The mean angular velocity of particles in the midplane was calculated in 1~Mm bins and it was assumed, based on the Poincare-Wavre theorem, that the angular velocity did not vary with height above the midplane. The gas velocity was assumed to be only in the azimuthal, $\theta$, direction. The gas density, condensate density, condensate size, and drag coefficient were all held constant in each simulation.

The gas drag of spheres is dependent on the Reynolds number, ${\rm Re}$. The Reynolds number in this situation is defined as
\begin{equation}
{\rm Re} = \frac{ 2 R_{\rm cond} \rho_{\rm vap} \left | \mathbf{v} - \mathbf{v_{\rm vap}} \right |}{\mu_{\rm vap}}   \, ,
\end{equation}
where $\mu_{\rm vap}$ is the dynamic viscosity of the gas. The relative velocities ($ | \mathbf{v} - \mathbf{v_{\rm vap}} |$) for centimeter sized bodies for the simulations shown in Figure~\ref{fig:gas_drag} range from 0.1-100~m~s$^{-1}$ depending on gas density. Assuming a viscosity $\mu_{\rm vap} \sim 10^{-5}$, the corresponding range of Reynolds number is $10-100$. Lab experiments of gas flow around spheres \citep[e.g.,][]{Roos1971} have shown that the drag coefficient in this range of Reynolds numbers is of order 1. Based on these results, we took a simple approach and used a constant gas drag coefficient, $C_{\rm D}=2$.

%xxxxxxxxxxxxxxxxxxxxxxxxxxxxxxxxxxxxxxxxxxxxxxxxxxxxxxxxxxxxxxxxxxxxxxxxxxxxxxxxxxxxxxxxxxxxxxxxxxxxxxxxxxxxxxxxxxxxxxxxxxxxxxxxxxxxxxxx
%xxxxxxxxxxxxxxxxxxxxxxxxxxxxxxxxxxxxxxxxxxxxxxxxxxxxxxxxxxxxxxxxxxxxxxxxxxxxxxxxxxxxxxxxxxxxxxxxxxxxxxxxxxxxxxxxxxxxxxxxxxxxxxxxxxxxxxxx
%Extra examples
\section{Examples of potential Moon-forming synestias}
\label{sup:sec:extra_examples}

\begin{figure}
\centering
\includegraphics[scale=0.833333333]{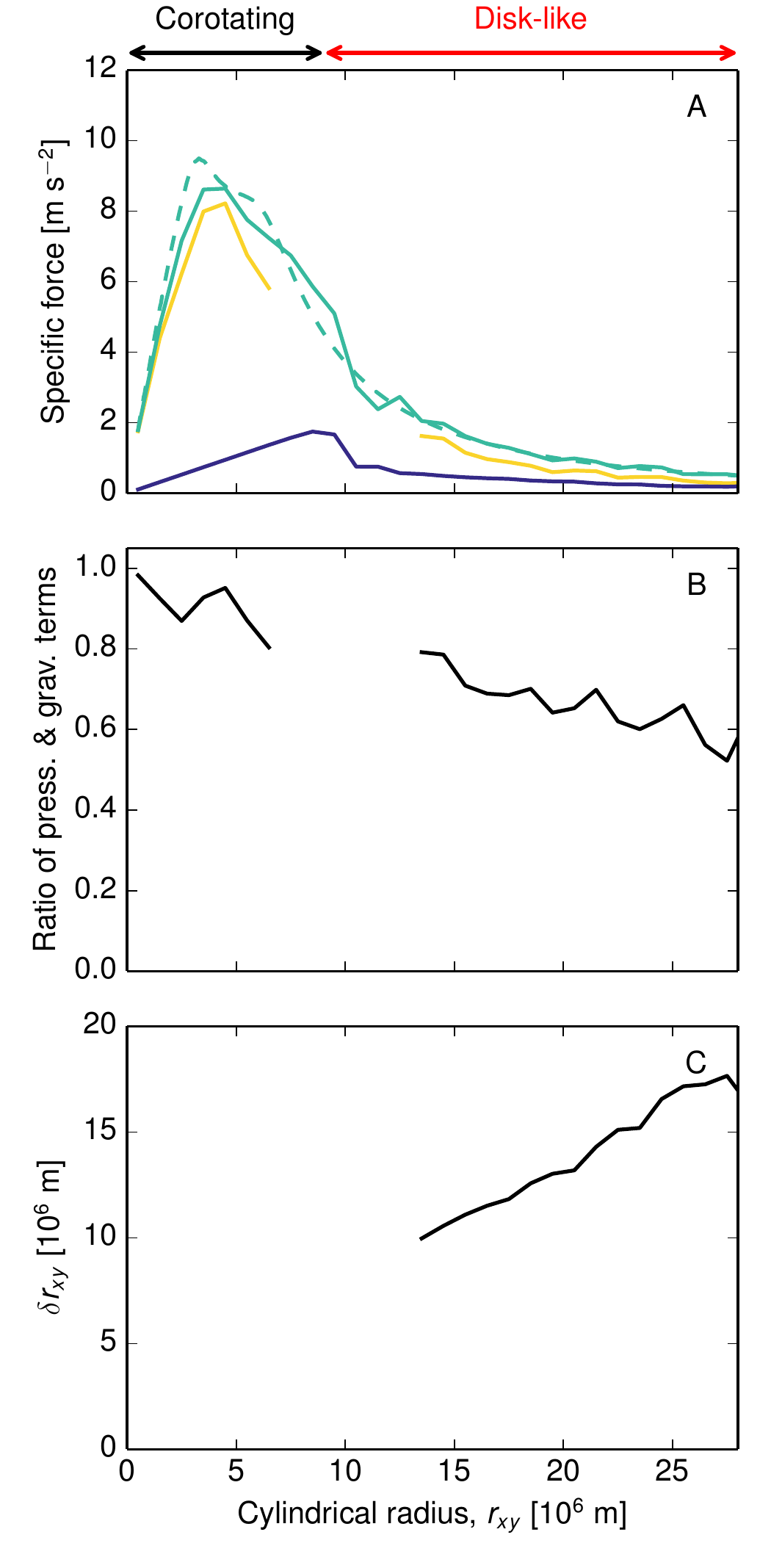}
\caption{Same information as presented in Figure~\ref{fig:pressure_support_main}, but for the initial processed, post-impact synestia shown in Figure~\ref{sup:fig:coolingA}. Note the change of the y-axis scale in panel C.
}
\label{sup:fig:pressure_supportA}
\end{figure}

\begin{figure}
\centering
\includegraphics[scale=0.833333333]{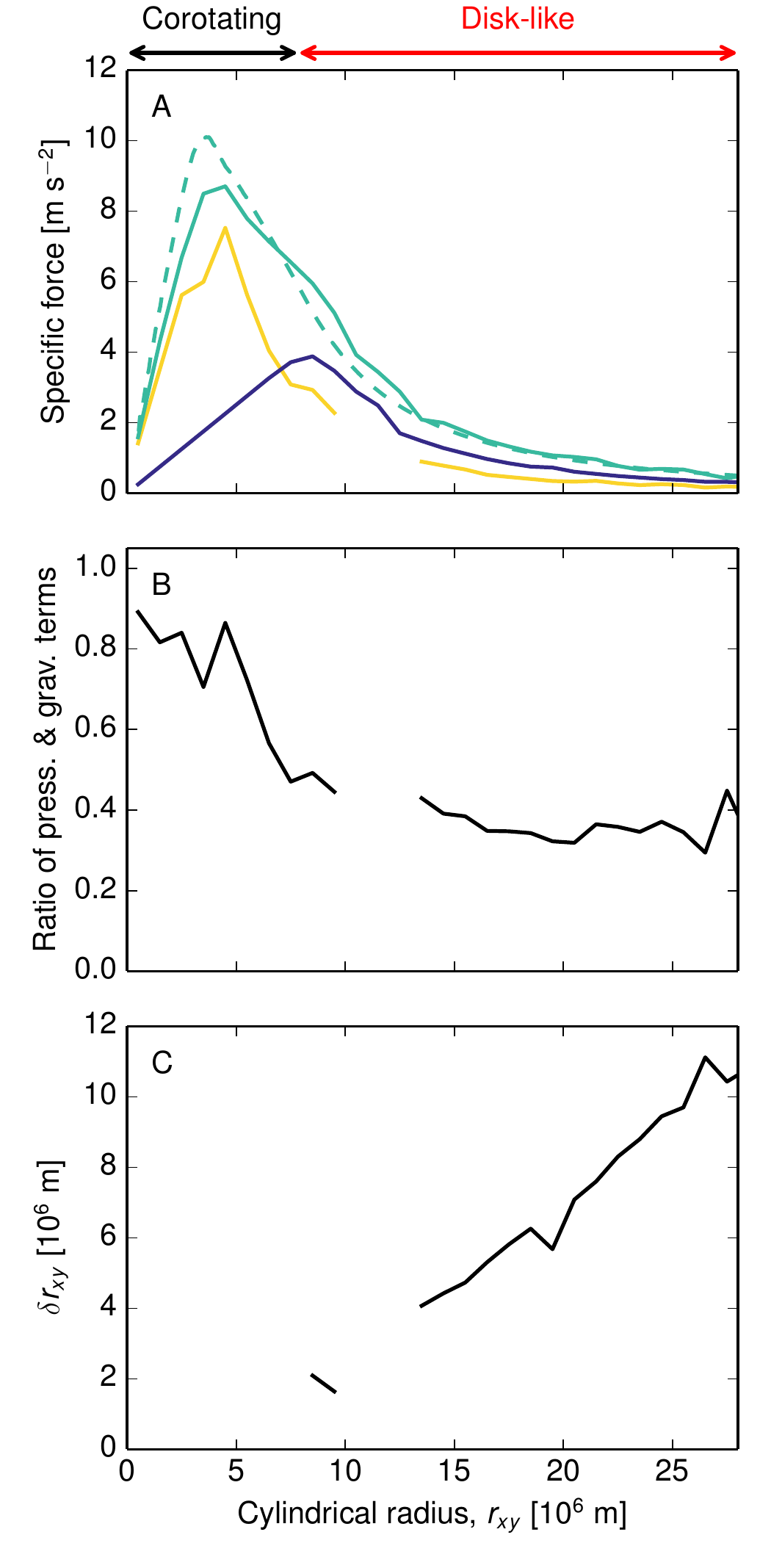}
\caption{Same information as presented in Figure~\ref{fig:pressure_support_main}, but for the initial processed, post-impact synestia shown in Figure~\ref{sup:fig:coolingB}.
}
\label{sup:fig:pressure_supportB}
\end{figure}

\begin{figure}
\centering
\includegraphics[scale=0.833333333]{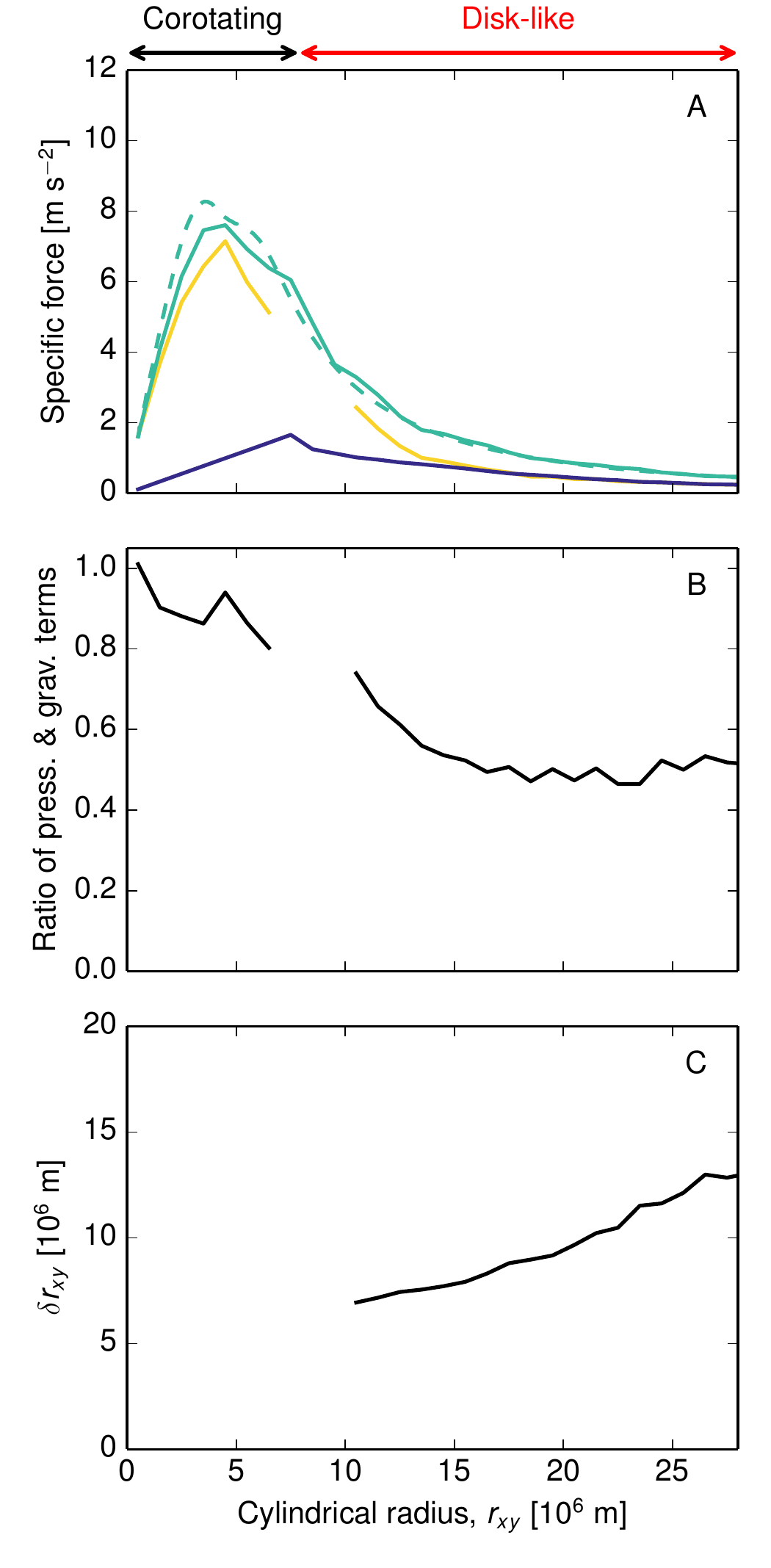}
\caption{Same information as presented in Figure~\ref{fig:pressure_support_main}, but for the initial processed, post-impact synestia shown in Figure~\ref{fig:SPH_cooling}. Note the change of the y-axis scale in panel C.
}
\label{sup:fig:pressure_supportp}
\end{figure}

\begin{sidewaysfigure*}
\centering
\includegraphics[scale=0.8333333333333]{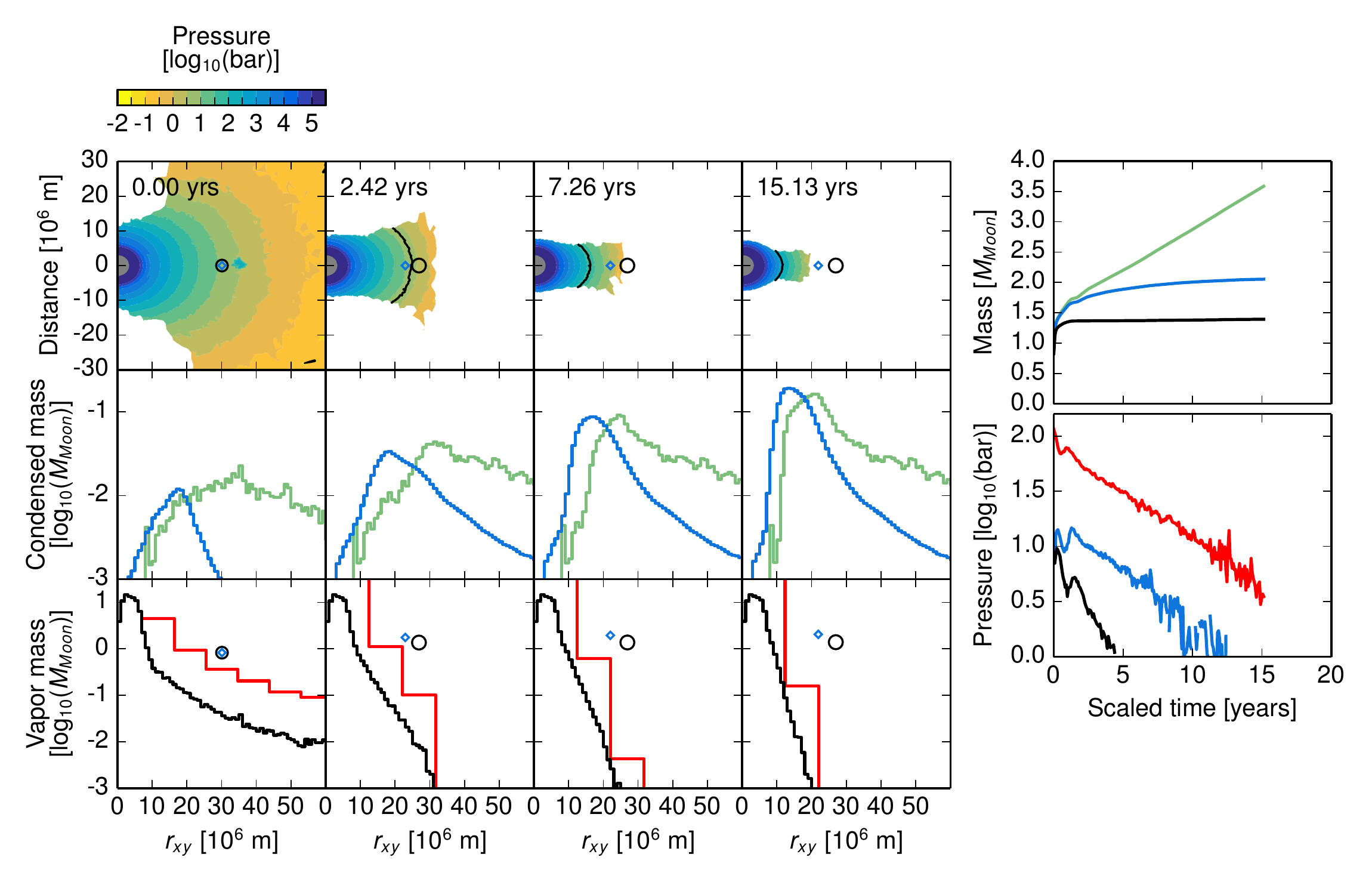}
\caption{An additional example of the cooling of a post-impact synestia and formation of a moon. The information presented is the same as in Figure~\ref{fig:SPH_cooling} (left) and Figure~\ref{fig:SPH_cooling_time} (right). Columns show different time steps. Top row: pressure contours of the vapor structure, where the black line denotes the boundary between the isentropic and vapor-dome regions; middle row: cumulative histograms of the radii at which mass condensed (green) and the locations to which falling condensate was redistributed (blue); bottom row: histograms (red, black) of the instantaneous mass distribution in the synestia. Red histograms binned by Hill diameter of moon A at that time step; other histograms binned by $1$~Mm. The black circles represents moon A, an estimate based on the total mass of condensing material beyond the Roche limit. In the top row, the size of moon A is shown to scale assuming a bulk density of 3000~kg~m$^{-3}$. Blue diamonds represent moon B, an estimate which includes some falling condensate, not shown to scale. Moons A and B are plotted at the radius of a circular Keplerian orbit corresponding to the integrated angular momentum of their constituent mass.
On the right, the mass (A) and vapor pressures (B) at moon A (black lines) and B (blue lines) are shown. The green line is the total condensed mass (corresponding to the green histograms on the left). The red line is the midplane vapor pressure at the Roche limit.
In this example, the initial synestia was formed by a $0.3M_{\rm Earth}$ body striking a $0.75M_{\rm Earth}$ body at 11.3 km~s$^{-1}$ and an impact parameter of 0.6. The seed of the moon in the first time step has a mass of $0.829M_{\rm Moon}$. The total mass of falling condensates over the time period shown was 6.8~$M_{\rm Moon}$.}
\label{sup:fig:coolingA}
\end{sidewaysfigure*}

\begin{sidewaysfigure*}
\centering
\includegraphics[scale=0.8333333333333]{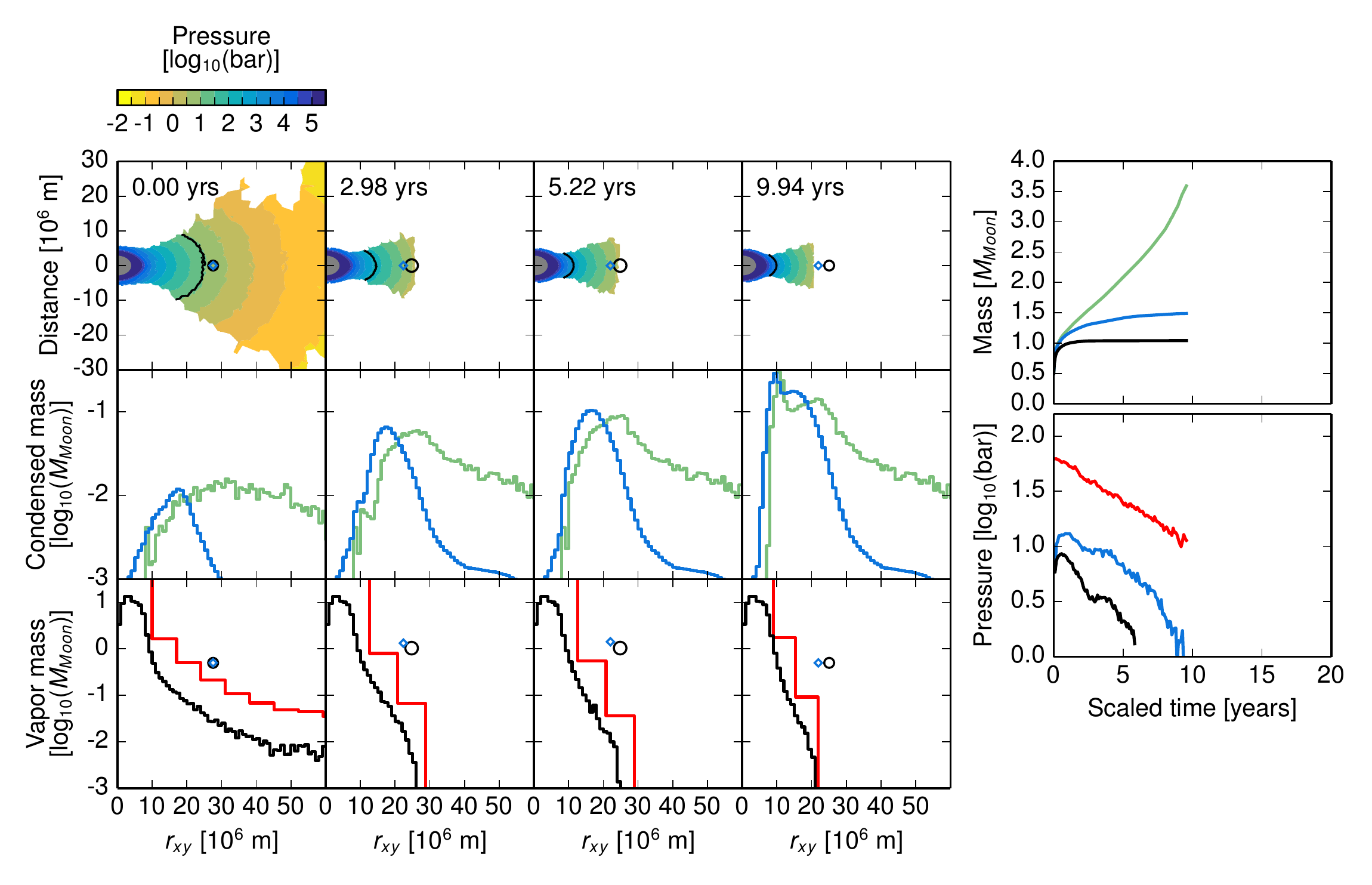}
\caption{An additional example of the cooling of a post-impact synestia and formation of a moon.  The information presented is the same as in Figure~\ref{fig:SPH_cooling} (left) and Figure~\ref{fig:SPH_cooling_time} (right), see Figure~\ref{sup:fig:coolingA} for details. In this example, the initial synestia was formed by a $0.1M_{\rm Earth}$ body striking a $0.99M_{\rm Earth}$ body spinning with a 2.3~hr period (an AM of $3L_{\rm EM}$) at 15~km~s$^{-1}$ and an impact parameter of 0.4. The seed of the moon in the first time step has a mass of $0.501M_{\rm Moon}$. The total mass of falling condensates over the time period shown was 7.5~$M_{\rm Moon}$.}
\label{sup:fig:coolingB}
\end{sidewaysfigure*}

\begin{sidewaysfigure*}
\centering
\includegraphics[scale=0.833333333333]{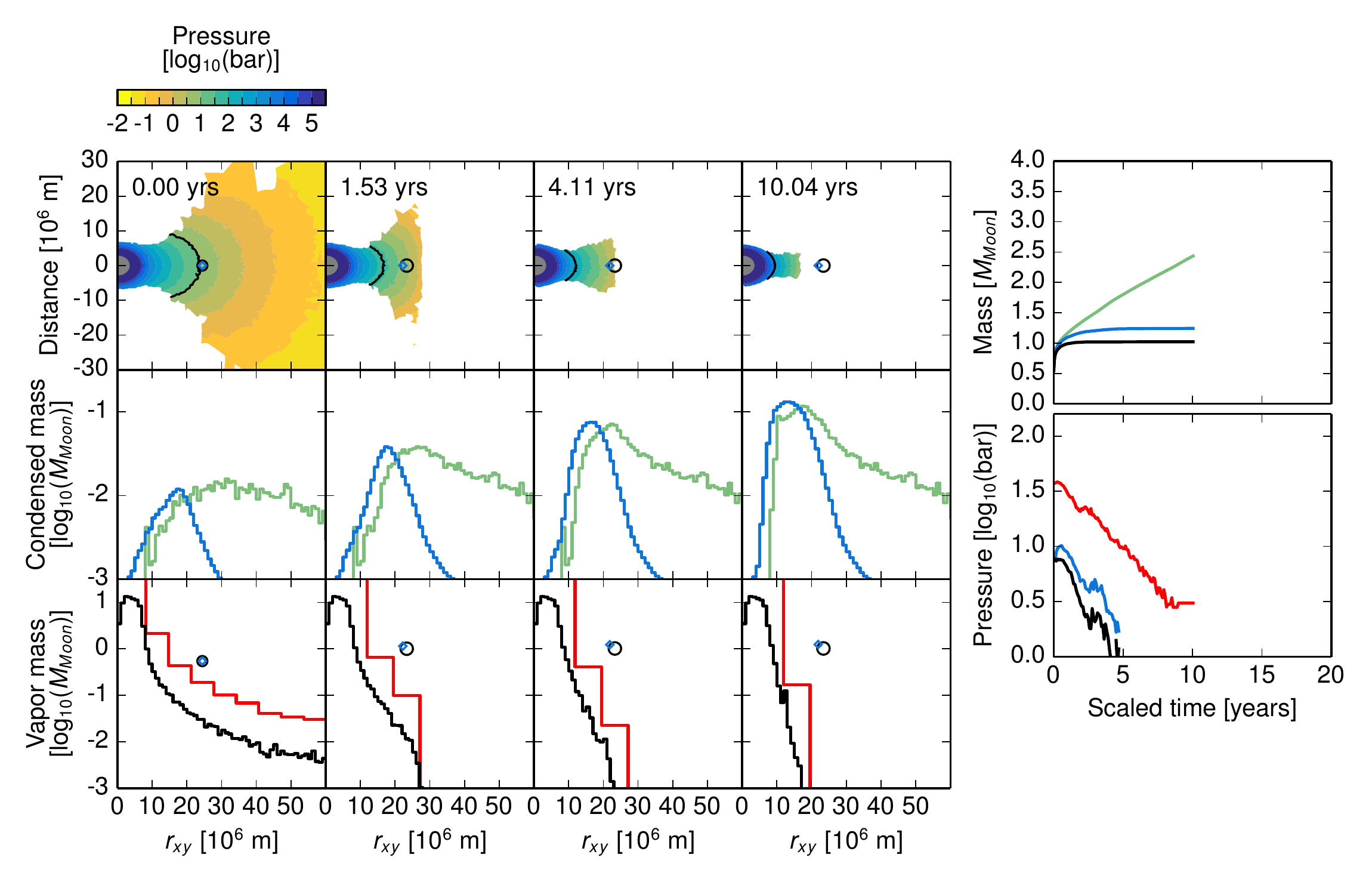}
\caption{An additional example of the cooling of a post-impact synestia and formation of a moon.  The information presented is the same as in Figure~\ref{fig:SPH_cooling} (left) and Figure~\ref{fig:SPH_cooling_time} (right), see Figure~\ref{sup:fig:coolingA} for details. In this example, the initial synestia was formed by a $0.468M_{\rm Earth}$ body striking a $0.572M_{\rm Earth}$ body at 9.7 km~s$^{-1}$ and an impact parameter of 0.55. The seed of the moon in the first time step has a mass of $0.546M_{\rm Moon}$. The total mass of falling condensates over the time period shown was 5.0~$M_{\rm Moon}$.
}
\label{sup:fig:coolingC}
\end{sidewaysfigure*}

We did not conduct a comprehensive study of satellite formation from impact-generated synestias, which we leave for future work. Using our current database of giant impacts [LS17], we found potential Moon-forming events based on the results of the cooling calculation described above. In this work, we focused on demonstrating that different giant impact configurations could generate potential Moon-forming synestias, where a lunar mass of material may accrete within 10~bars or more of vapor. Here we present additional examples of synestias that could form a body similar to our Moon, generated by a variety of different impacts. The degree of pressure support in example synestias are shown in Figures~\ref{sup:fig:pressure_supportA}, \ref{sup:fig:pressure_supportB}, and \ref{sup:fig:pressure_supportp}. Figure~\ref{sup:fig:pressure_supportp} corresponds to the structure shown in Figure~\ref{fig:SPH_cooling}. Examples of the time evolution of cooling synestias are shown in Figures \ref{sup:fig:coolingA}, \ref{sup:fig:coolingB}, and \ref{sup:fig:coolingC}.

We note that, after many impacts in our database, there is too little mass and AM beyond the Roche radius to form a satellite approaching a lunar mass during the initial period of post-impact cooling. Not every synestia may form a large satellite and the formation of smaller moons from synestias is likely to be common.

%xxxxxxxxxxxxxxxxxxxxxxxxxxxxxxxxxxxxxxxxxxxxxxxxxxxxxxxxxxxxxxxxxxxxxxxxxxxxxxxxxxxxxxxxxxxxxxxxxxxxxxxxxxxxxxxxxxxxxxxxxxxxxxxxxxxxxxxx
%xxxxxxxxxxxxxxxxxxxxxxxxxxxxxxxxxxxxxxxxxxxxxxxxxxxxxxxxxxxxxxxxxxxxxxxxxxxxxxxxxxxxxxxxxxxxxxxxxxxxxxxxxxxxxxxxxxxxxxxxxxxxxxxxxxxxxxxx
%xxxxxxxxxxxxxxxxxxxxxxxxxxxxxxxxxxxxxxxxxxxxxxxxxxxxxxxxxxxxxxxxxxxxxxxxxxxxxxxxxxxxxxxxxxxxxxxxxxxxxxxxxxxxxxxxxxxxxxxxxxxxxxxxxxxxxxxx
\clearpage
\section*{References}
\bibliographystyle{elsarticle-harv} 
\bibliography{References}

%%Supplemental

%This top part might not be allowed but they shouldn't care too much
%make all the sections labelled with an S for supp
\renewcommand{\thepage}{S\arabic{page}}  
\renewcommand{\thesection}{S\arabic{section}}   
\renewcommand{\thetable}{S\arabic{table}}   
\renewcommand{\thefigure}{S\arabic{figure}}
\renewcommand{\theequation}{S\arabic{equation}}

%reset page count and section count
\setcounter{page}{1}
\setcounter{section}{0}
\setcounter{figure}{0}
\setcounter{table}{0}
\setcounter{equation}{0}
\setcounter{tnote}{0}
\setcounter{fnote}{0}
\setcounter{footnote}{0}
\setcounter{cnote}{0}
\setcounter{author}{0}
\setcounter{affn}{0}

\renewenvironment{abstract}{\global\setbox\absbox=\vbox\bgroup
  \hsize=\textwidth\def\baselinestretch{1}%
  \noindent\unskip\textbf{Contents}
 \par\medskip\noindent\unskip\ignorespaces}
 {\egroup}
 %\renewenvironment{frontmatter}{}{}

%xxxxxxxxxxxxxxxxxxxxxxxxxxxxxxxxxxxxxxxxxxxxxxxxxxxxxxxxxxxxxxxxxxxxxxxxxxxxxxxxxxxxxxxxxxxxxxxxxxxxxxxxxxxxxxxxxxxxxxxxxxxxxxxxxxxxxxxx
%xxxxxxxxxxxxxxxxxxxxxxxxxxxxxxxxxxxxxxxxxxxxxxxxxxxxxxxxxxxxxxxxxxxxxxxxxxxxxxxxxxxxxxxxxxxxxxxxxxxxxxxxxxxxxxxxxxxxxxxxxxxxxxxxxxxxxxxx
%xxxxxxxxxxxxxxxxxxxxxxxxxxxxxxxxxxxxxxxxxxxxxxxxxxxxxxxxxxxxxxxxxxxxxxxxxxxxxxxxxxxxxxxxxxxxxxxxxxxxxxxxxxxxxxxxxxxxxxxxxxxxxxxxxxxxxxxx

\begin{frontmatter}

\title{Supplementary materials for ''The origin of the Moon within a terrestrial synestia"}

\begin{abstract}
%%%Remove or add items as needed%%%
\begin{enumerate}
\item Text S1 to S6
\item Figures S1 to S5
\item Tables S1 to S3
\end{enumerate}

\section*{Additional supporting information (files uploaded separately)}

\begin{enumerate}
\item GADGET-2 cooling routine
\item Condensate orbital evolution model
\item GRAINS code and corresponding thermodynamic data
\end{enumerate}

\end{abstract}
\end{frontmatter}

%xxxxxxxxxxxxxxxxxxxxxxxxxxxxxxxxxxxxxxxxxxxxxxxxxxxxxxxxxxxxxxxxxxxxxxxxxxxxxxxxxxxxxxxxxxxxxxxxxxxxxxxxxxxxxxxxxxxxxxxxxxxxxxxxxxxxxxxx
%xxxxxxxxxxxxxxxxxxxxxxxxxxxxxxxxxxxxxxxxxxxxxxxxxxxxxxxxxxxxxxxxxxxxxxxxxxxxxxxxxxxxxxxxxxxxxxxxxxxxxxxxxxxxxxxxxxxxxxxxxxxxxxxxxxxxxxxx
%xxxxxxxxxxxxxxxxxxxxxxxxxxxxxxxxxxxxxxxxxxxxxxxxxxxxxxxxxxxxxxxxxxxxxxxxxxxxxxxxxxxxxxxxxxxxxxxxxxxxxxxxxxxxxxxxxxxxxxxxxxxxxxxxxxxxxxxx
\section{The canonical giant impact model}
\label{sup:sec:canonical}

In this section, we support our statements in the introduction about the difficulty of forming a lunar mass satellite from a canonical Moon-forming giant impact. In previous giant impact studies \citep{Canup2001,Canup2004,Canup2008, Canup2012, Cuk2012, Reufer2012,Rufu2017}, the mass of the moon formed by a given impact has been approximated based on the mass and specific AM of the material injected into orbit using scaling laws derived from $N$-body lunar accretion studies. 
$N$-body simulations treat all the mass in orbit as condensed particles that interact by gravity and accrete to form a moon.
However, giant impact simulations have shown that the material injected into orbit in the impact is a mixture of liquid and vapor.

In condensate-dominated disks, like those considered in models of the canonical case, the condensate is thought to settle to the midplane and form a liquid layer overlain by a vapor atmosphere \citep{Thompson1988,Ward2012,Ward2014,Ward2017,Charnoz2015}.
Material in orbit beyond the Roche limit would likely condense rapidly and accrete in a manner similar to an $N$-body disk.
\citet{Salmon2012} used a hybrid code to simultaneously model a simplified one-dimensional, Roche-interior multiphase disk and an $N$-body Roche exterior disk.
They produced scaling laws relating the mass and specific AM of the disk to the mass of the final moon and showed that lunar accretion from a multiphase disk is much less efficient than from a pure $N$-body disk.
The growing Moon interacts with the multiphase disk, truncating the edge of the disk. Thus, the Moon must migrate and the disk viscously spread beyond the Roche limit before more mass can be accreted to the moon. These processes bottleneck lunar accretion.

\begin{figure}
\centering
\includegraphics[scale=0.83333333]{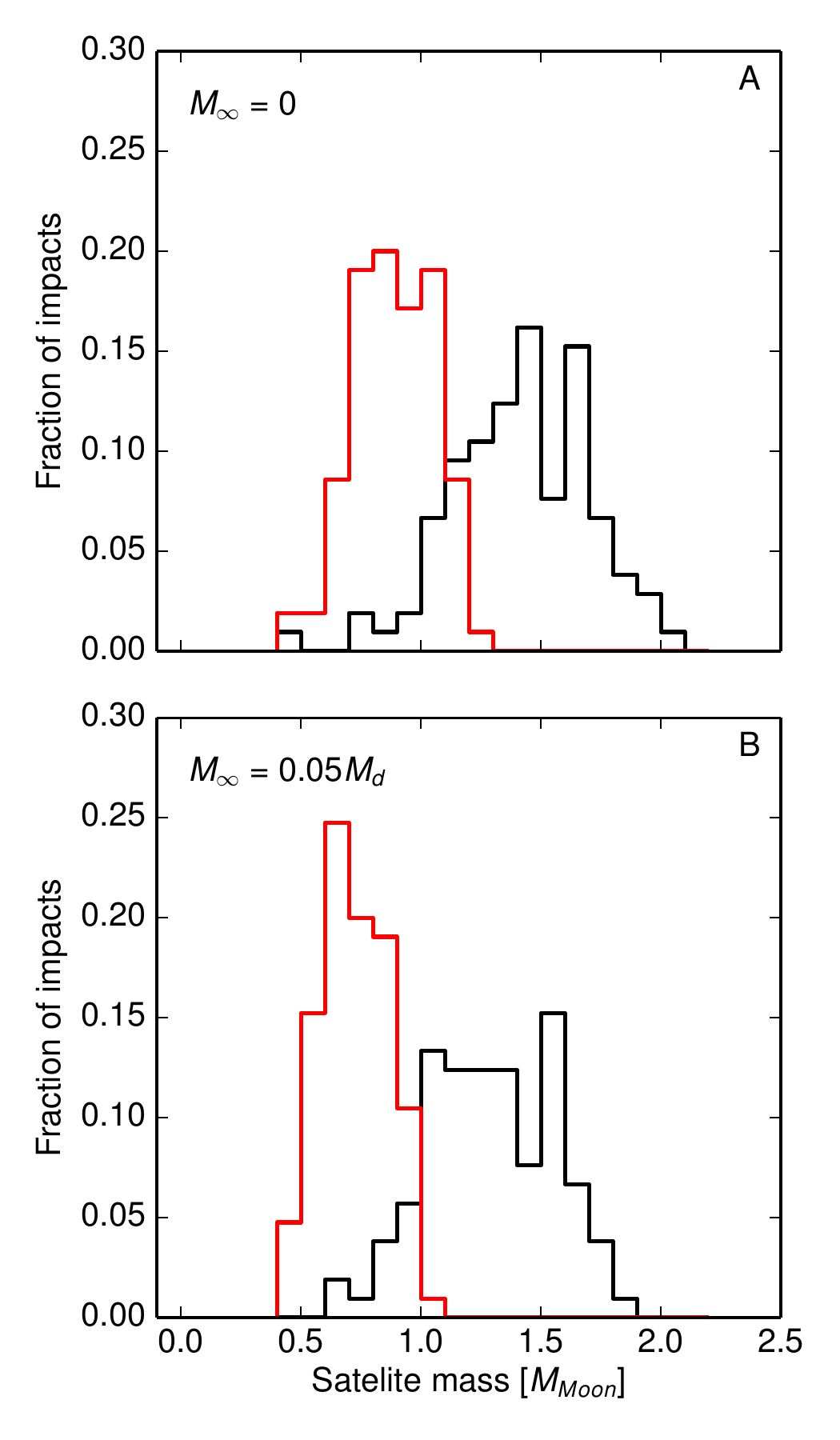}
\caption{Hybrid scaling laws for satellite accretion from a circumterrestrial disk predict the formation of less than a lunar mass moon for most published canonical impact events. Histograms show the predicted lunar mass from different scaling laws applied to published canonical impact simulations \citep{Canup2001,Canup2004,Canup2008a}. In both panels, satellite masses calculated using the $N$-body accretion scaling law \citep{Ida1997} are shown in black and masses calculated using the scaling law from the hybrid disk evolution models of \cite{Salmon2012} are in red. Ejected masses of $M_{\infty}=0$ (A) and $M_{\infty}=0.05M_{\rm d}$ (B) were assumed, where $M_{\rm d}$ is the initial disk mass.}
\label{sup:fig:can_hist}
\end{figure}

To date, a determination of the mass of the moon formed by canonical Moon-forming giant impacts using the more conservative hybrid scaling laws has not been published.
We use the results of 105 published canonical Moon-forming impacts \citep{Canup2001,Canup2004,Canup2008a} to consider the likelihood of forming a lunar mass moon from such impacts.
Figure~\ref{sup:fig:can_hist} shows the range of satellite masses predicted using both $N$-body \citep[][black]{Ida1997} and hybrid \citep[][red]{Salmon2012} scaling laws.
Both scaling laws require an assumption about the mass of material that is ejected from the system during satellite accretion, $M_{\infty}$.
A good fit between accretion simulations and the scaling laws is generally achieved for $0$~$<$~$M_{\infty}$~$\le$~$0.05 M_{\rm d}$, where $M_{\rm d}$  is the initial mass of the disk.
The two panels of Figure~\ref{sup:fig:can_hist} show the predicted satellite mass for loss of 0 and 5\% of the disk mass in A and B respectively. 
The number of impacts that are expected to form a greater than lunar mass moon is significantly lower when using the hybrid scaling laws.
Assuming $M_{\infty}=0$, only 29\% of the published impacts would be able to produce a lunar mass moon and only one impact formed a lunar mass moon using $M_{\infty}$~$=$~$0.05 M_{\rm d}$. 
Furthermore, the disks that do form greater than lunar mass moons using the hybrid scaling laws tend to have $L_{\rm d}/M_{\rm d}$~$>$~$\left( G M_{\rm Earth} a_{\rm R} \right )^{1/2}$, where $L_{\rm d}$ is the AM of the disk, $G$ is the gravitational constant, $M_{\rm Earth}$ is the mass of the Earth, and $a_{\rm R}$ is the radius of the Roche limit.
The accretion efficiency of such high specific AM disks are best captured by scaling laws with higher $M_{\infty}$ and lower accretion efficiencies \citep{Ida1997,Kokubo2000,Salmon2012,Salmon2014}.

Given the lower efficiency of moon formation in multiphase disks, it is uncertain whether canonical style impacts can inject enough mass and AM into orbit to form a lunar mass moon.
Recent work by \citet{Charnoz2015} has also shown that the spreading of the inner disk is slower that assumed by \citet{Salmon2012} and more mass is lost to the planet, further impeding mass addition to a moon from a Roche interior disk.
\citet{Charnoz2015} suggested that the issue of spreading mass beyond the Roche limit could be circumvented if the Moon accreted mostly from mass that was injected beyond the Roche limit in the impact.
For the published studies that reported the mass injected beyond the Roche limit \citep{Canup2001,Canup2004}, $\sim$65\% of impacts injected more than a lunar mass of material beyond the Roche limit.
The accretion efficiency of this material is uncertain as it depends on the surface density profile of the disk.
The simulations of \citet{Salmon2012} showed a wide range of accretion efficiencies (0 to 98\%) for mass initially outside the Roche limit for the idealized disks they initialized. 
If a moon did accrete from material emplaced beyond the Roche limit, it would still have to meet the other observational (chemical and isotopic) constraints. 
More work is needed to ascertain whether the multiphase disks produced by canonical giant impact simulations can form lunar mass moons, while satisfying the other observational constraints.

%xxxxxxxxxxxxxxxxxxxxxxxxxxxxxxxxxxxxxxxxxxxxxxxxxxxxxxxxxxxxxxxxxxxxxxxxxxxxxxxxxxxxxxxxxxxxxxxxxxxxxxxxxxxxxxxxxxxxxxxxxxxxxxxxxxxxxxxx
%xxxxxxxxxxxxxxxxxxxxxxxxxxxxxxxxxxxxxxxxxxxxxxxxxxxxxxxxxxxxxxxxxxxxxxxxxxxxxxxxxxxxxxxxxxxxxxxxxxxxxxxxxxxxxxxxxxxxxxxxxxxxxxxxxxxxxxxx
%xxxxxxxxxxxxxxxxxxxxxxxxxxxxxxxxxxxxxxxxxxxxxxxxxxxxxxxxxxxxxxxxxxxxxxxxxxxxxxxxxxxxxxxxxxxxxxxxxxxxxxxxxxxxxxxxxxxxxxxxxxxxxxxxxxxxxxxx
\section{Determination of the photic surface}
\label{sup:sec:photosphere}

Cooling of a synestia is controlled by radiation of energy from the effective radiating surface of the structure, where the material is optically thin. 
We refer to the layer from which the structure radiates as the photosphere. The thermal structure is determined in part by radiative transfer through a gas-condensate mixture. A full combined thermal and radiative calculation is beyond the scope of this paper.
Here, we approximate the photospheric pressure, and hence temperature, using a simple calculation of radiative transfer through a fixed hydrostatic structure.

As we show in \S\ref{sup:sec:adiabats}, adiabats in a synestia are mostly vapor in the high-pressure midplane of the structure but condense a few tens of percent of condensate at lower pressures.
The radius at which the structure becomes optically thin therefore depends both on the absorption of the vapor and condensate. 
The probability of a photon being absorbed as it traverses over a distance $\mathrm{d}r$ is given by
\begin{linenomath*}
\begin{equation}
  \mathcal{P}(r)=\alpha(r) \mathrm{d}r = \frac{\mathrm{d}r}{L_{\rm MFP}(r)} + \alpha_{\rm vap}(r) \mathrm{d}r \, ,
\end{equation}
\end{linenomath*}
where $\alpha (r)$ is the average absorption coefficient at radius $r$, $ L_{\rm MFP}$ is the mean free path of a photon traveling through a droplet suspension, and $\alpha_{\rm vap} (r)$ is the absorption coefficient of the vapor which is also a function of $r$.
The absorption of silicate vapor in the conditions relevant for post-impact states is poorly known.
Here we use the semiconductor type Drude model for the absorption of silica vapor constructed by \cite{Kraus2012} and constrained by them using first principles molecular dynamics simulations.
In our calculations, the droplet absorption dominates over the vapor absorption so errors in $\alpha_{\rm vap}$ will have only a small effect on our conclusions.
$ L_{\rm MFP}$ for a photon passing through a cloud of spherical condensates of diameter, $D_0$, is
\begin{linenomath*}
\begin{equation}
 L_{\rm MFP} = \frac{4 D_0}{6} \frac{V_{\rm avg}}{V_{\rm cond}} \, ,
\end{equation}
\end{linenomath*}
where $V_{\rm liq}/V_{\rm avg}$ is the volume fraction of condensate. 
The inner edge of the photosphere is defined as when, integrating from the outside of the structure inwards, the optical depth is unity,
\begin{linenomath*}
\begin{equation}
 \int_{\infty}^{r_{\rm rad}} \! \alpha (r) \mathrm{d}r = 1 \, ,
\end{equation}
\end{linenomath*}
where $r_{\rm rad}$ is the radius of the photic surface. 

We approximate the photosphere for post-impact and thermally equilibrated structures. The low pressure regions of the structure are not resolved in SPH. To overcome this, we integrate radially outwards along the rotation axis from the lowest resolved pressure along an isentrope to find the hydrostatic profile to low pressures. For simplicity, we used the single-phase M-ANEOS forsterite EOS \citep{Melosh2007,Canup2012}. We then integrate back along the same profile to find where the optical depth is unity. The location of the photosphere depends strongly on the mass fraction of condensate. In \S\ref{sup:sec:adiabats} we calculate the mass fraction of condensate that is present along adiabats in bulk BSE material. However, the true mass fraction of condensate at a point in the synestia depends on how efficiently condensates separate from the vapor, which is poorly constrained. To remove this complexity, we assume a constant mass fraction of condensate, $f_{\rm cond}$, in the mixed phase region when calculating both the hydrostatic profile and the optical depth. We now explore the range of possible photospheric pressures, and hence temperatures, varying $f_{\rm cond}$ and the diameter of condensates.

\begin{figure}[t]
\centering
\includegraphics[scale=0.833333333]{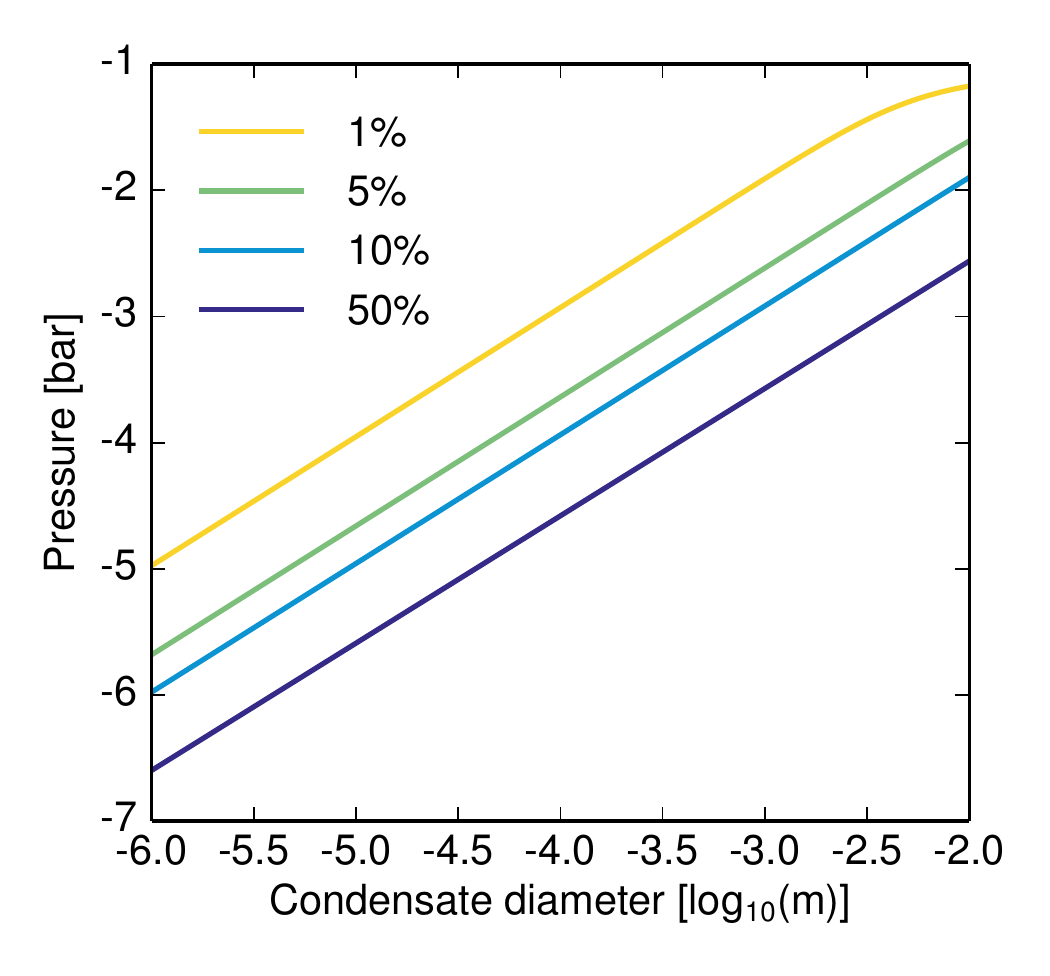}
\caption{The photospheric pressure is low but depends on the size of condensates. Shown is the photospheric pressure at the pole as a function of condensate diameter assuming varying mass fractions of condensate in the vapor (colored lines) calculated for the thermally equilibrated synestia shown in Figure~\ref{fig:contourstructures}E-G.}
\label{sup:fig:photo}
\end{figure}

Synestias typically become optically thin at low pressures. For example, Figure~\ref{sup:fig:photo} shows the photospheric pressure at the poles for the synestia shown in Figure \ref{fig:contourstructures}E-G calculated using different $f_{\rm cond}$ and $D_0$.
For $f_{\rm cond}=0.1$ the photospheric pressure ranges from 10$^{-6}$~-~10$^{-2}$~bar, assuming condensates of diameter 10$^{-6}$~-~10$^{-2}$~m. A few $10^{-2}$~m is the largest size for falling condensates that we approximate in \S\ref{sec:dynamics_cooling}. 
The corresponding temperatures on an adiabat starting at the dew point in the midplane (\S\ref{sup:sec:adiabats}) range from approximately 1900 to 2600~K. On a saturated adiabat the range is from 2200 to 2700~K.
Increasing the mass fraction of condensate decreases the pressure at the photosphere. The relatively narrow range of photospheric temperatures means that variation in condensate size has a limited effect on our calculations with a maximum change in radiative flux of a factor of two. Away from the poles, the effective gravity in the synestia can be substantially lower. A lower gravity, and associated larger scale height, results in a lower pressure photosphere. The effect of changing gravity is similar in magnitude to increasing the mass fraction of condensates.
The photospheric pressure, and hence temperature, we calculate remains relatively constant as a synestia cools.
For the calculations in this work, we use a photospheric temperature of 2300~K. This radiative temperature is similar to that used for studies of canonical post-impact states \citep[e.g.,][]{Thompson1988,Pahlevan2007}.

Because synestias are extended structures that radiate at a high temperature, a substantial amount of power is radiated from the photosphere.
The surface area of the photosphere of the initial post-impact state is on the order 10$^{16}$~-~10$^{17}$~m$^2$ which, assuming that the structure is radiating as a black body, corresponds to a radiated power of 10$^{22}$~-~10$^{23}$~W.
Most of the energy radiated at the photosphere goes into condensing silicates and the radiated power corresponds to an initial rate of condensation of about $1 M_{\rm Moon}$~yr$^{-1}$.
This is a significant mass of condensate providing a large driver for mass transport and mixing.
As the structure evolves, it contracts reducing the surface area of the photosphere and hence the rate of production of condensate. However, a significant mass of condensate is produced throughout the evolution of synestias (see green histograms in Figures~\ref{fig:SPH_cooling}, \ref{sup:fig:coolingA}, \ref{sup:fig:coolingB} and \ref{sup:fig:coolingC}).

%xxxxxxxxxxxxxxxxxxxxxxxxxxxxxxxxxxxxxxxxxxxxxxxxxxxxxxxxxxxxxxxxxxxxxxxxxxxxxxxxxxxxxxxxxxxxxxxxxxxxxxxxxxxxxxxxxxxxxxxxxxxxxxxxxxxxxxxx
%xxxxxxxxxxxxxxxxxxxxxxxxxxxxxxxxxxxxxxxxxxxxxxxxxxxxxxxxxxxxxxxxxxxxxxxxxxxxxxxxxxxxxxxxxxxxxxxxxxxxxxxxxxxxxxxxxxxxxxxxxxxxxxxxxxxxxxxx
%xxxxxxxxxxxxxxxxxxxxxxxxxxxxxxxxxxxxxxxxxxxxxxxxxxxxxxxxxxxxxxxxxxxxxxxxxxxxxxxxxxxxxxxxxxxxxxxxxxxxxxxxxxxxxxxxxxxxxxxxxxxxxxxxxxxxxxxx
\section{Mixing in a terrestrial synestia}
\label{sup:sec:mixing}
In \S\ref{sec:dynamics_cooling}, we argue that vertical convection would rapidly mix vertical columns of material in a terrestrial synestia. Here we present supporting calculations for this argument and address some of the finer points.

Convection within a synestia is unlike any system that has been studied to date.
The main component of a terrestrial synestia, silicate vapor, is condensible. Unlike most studies of planetary atmospheres, where condensible species make up a small mass fraction of the system, convection in a synestia is governed by the phase boundary.
Furthermore, unlike most planets and astrophysical bodies, convection is driven by cooling from the top rather than heating from below.
The thermal structure of post-impact states is highly stratified [LS17], and 
the densities and pressures in the structure also vary by many orders of magnitude. 
Hence, we are not able to make direct analogies to convection in other well studied systems.
Here we describe the basic aspects of convection in a synestia and approximate the mixing timescale.

Synestias rotate rapidly which has a significant effect on convection. 
Convection in rapidly rotating systems tends to organize into columns parallel to the rotation axis \citep[as in giant planets, see e.g.,][]{Vasavada2005,Kaspi2009}, as dictated by the Taylor-Proudman theorem.
In rotating systems with shear similar to the disk-like regions of synestias, such as astrophysical disks, similar columns are formed but they can be transient as they are destroyed by the shear stresses in the bulk flow \citep[see discussion in][]{Shariff2009}.
Given the small Rossby number of the system, we expect synestias to form similar columnar flow patterns in regions of the structure where there is a strong AM gradient with radius.
Such a flow pattern can significantly impede the ability of fluid convection to transport mass radially as this requires exchange of mass between columns, which is slow relative to the vertical convective velocities \citep{Kaspi2009}. In the outer regions of synestias, the specific AM gradient with radius can be small and radial fluid convection may be possible. Exploration of this possibility is left to future work.

Despite the uncertainty in the convective pattern, we wish to be able to estimate the timescale for vertical mixing in a synestia.
To do this, we used mixing length theory (MLT), a technique that has been used
extensively in stellar astrophysics \citep[see e.g.,][]{Kippenhahn2012}. MLT
has also been applied to magma oceans on terrestrial planets 
\citep[e.g.,][]{Solomatov2000} and previously to the canonical lunar disk \citep{Pahlevan2007}.
MLT considers the movement of convecting parcels of material that are able to travel a mixing length (or mean free path), $\ell_{\rm m}$, before becoming indistinguishable from their surroundings.
From MLT, it is possible to estimate an average convective velocity,
\begin{linenomath*}
\begin{equation}
  v_{\rm conv} \sim \left ( \frac{F_{\rm conv} \ell_{\rm m}}{\rho H} \right )^{\frac{1}{3}} \, ,
  \label{sup:eqn:MLT}
\end{equation}
\end{linenomath*}
where $F_{\rm conv}$ is the convective flux, $\rho$ is the density of the fluid, and $H$ is the scale height \citep[see e.g.,][]{Priestley1959,Kraichnan1962,Stevenson1979}. 
In equilibrium, the convective flux and the radiative flux are equal, $F_{\rm conv} $~$\sim$~$\sigma T_{\rm rad}^4 $, and $v_{\rm conv}$ is determined by the mixing length parameter, $\Lambda_{\rm m}  = \ell_{\rm m} / H $.
In astrophysics, $\Lambda_{\rm m}$ is typically taken to be of order unity \citep[e.g.][]{Spiegel1971,Kippenhahn2012}, but in the post-impact structure, rapid rotation could shorten the convective length scale due to the Coriolis effect.
In MLT, rotation is often compensated for by simply using a smaller value for the
mixing length parameter. Here we use $\Lambda_{\rm m} =0.1$.
Alternatively, \cite{Stevenson1979} considered the effect of rotation on convection in the rapidly rotating limit and found that the convective velocity scaled as
\begin{linenomath*}
\begin{equation}
  v_{\rm conv} \sim 1.5 \left ( 2 \Omega \ell_{\rm m} \right )^{-\frac{1}{5}} \left (v_{\rm conv}^0 \right )^{\frac{2}{5}} \, ,
  \label{sup:eqn:MLT_stevenson}
\end{equation}
\end{linenomath*}
where $\Omega$ is the rotational angular velocity, and $v_{\rm conv}^0$ is the convective velocity in the absence of rotation as defined by Equation \ref{sup:eqn:MLT}.
The description of \cite{Stevenson1979} has been shown to match well the results of numerical simulations of thermal convection between fixed temperature plates in a rotating system \citep{Barker2014}.
A third approach to account for rotation was suggested by \citet{Solomatov2000}, based on experimental results.
With rotation, the mixing length scales as 
\begin{linenomath*}
\begin{equation}
\ell_{\rm m} \sim  \frac{v_{\rm conv}}{\Omega} \, .
\end{equation}
\end{linenomath*}
For an adiabatic scale height, $H= c_{p} / \alpha_{p} g$,
the convective velocity is then
\begin{linenomath*}
\begin{equation}
  v_{\rm conv} \sim \left (  \frac{\alpha_{p} g F_{\rm conv} }{\rho c_{p} \Omega} \right )^{\frac{1}{2}} \, ,
\end{equation}
\end{linenomath*}
where $\alpha_{p}$ is the coefficient of thermal expansion, $g$ is the gravitational acceleration, and $c_{p}$ is the specific heat capacity at constant pressure.
These three estimates of the convective velocity were used to calculate a convective mixing timescale $\tau_{\rm mix} $~$\sim$~$ L / v_{\rm conv} $, where $L$ is the length scale over which mixing occurs.

MLT has been shown to work well for purely thermal convection \citep[e.g.,][]{Barker2014} but its applicability to thermochemical convection has not been demonstrated.
However, in the limit where the condensates do not separate from the vapor, the buoyancy forcing of the system is similar for a purely thermal and purely condensation driven convection. 
The change in density of a parcel of material due to thermal contraction alone is given by
\begin{linenomath*}
\begin{equation}
\Delta \rho_{\rm therm} = \rho_{\rm vap} \alpha_{p} \Delta T \, ,
\end{equation}
\end{linenomath*}
where $\rho_{\rm vap}$, $\alpha_{p}$ and $\Delta T$ are the density of the vapor, the thermal expansivity at constant pressure, and the change in temperature of the parcel respectively.
If we assume that the temperature change of the parcel is due to an energy change $\Delta E$, we can rewrite the change in density as 
\begin{linenomath*}
\begin{equation}
\Delta \rho_{\rm therm} = \rho_{\rm vap} \alpha_{p} \left ( \frac{\Delta E}{m c_{p}} \right ) \, ,
\end{equation}
\end{linenomath*}
where $m$ is the mass of the parcel.
For comparison, we calculate the density change of a parcel if we instead assume that $\Delta E$ was doing work, not to cool the parcel, but to condense silicate vapor.
The density of a mixed phase with mass fraction $f$ of condensates is
\begin{linenomath*}
\begin{equation}
\rho = \frac{\rho_{\rm vap} \rho_{\rm cond}}{f \rho_{\rm vap} + (1-f)\rho_{\rm cond}} \, ,
\end{equation}
\end{linenomath*}
where $\rho_{\rm cond}$ is the density of the condensate.
We assume that the parcel is initially entirely vapor and calculate the change in density with a change in mass fraction of condensate $\Delta f$.
To first order in $\Delta f$,
\begin{linenomath*}
\begin{equation}
\Delta \rho_{\rm cond} \sim \Delta f \rho_{\rm vap} \left ( 1- \frac{\rho_{\rm vap}}{\rho_{\rm cond}} \right ) \, .
\end{equation}
\end{linenomath*}
Since $\rho_{\rm vap} \ll \rho_{\rm cond}$, we approximate this as $\Delta \rho_{\rm cond} \sim \Delta f \rho_{\rm vap}$.
$\Delta f$ is determined by the energy lost and the latent heat of vaporization, $\ell$, where
\begin{linenomath*}
\begin{equation}
\Delta \rho_{\rm cond} \sim \rho_{\rm vap}  \left ( \frac{\Delta E}{m l} \right ) \, .
\end{equation}
\end{linenomath*}
The ratio of the negative buoyancy produced by purely thermal and purely condensation-driven convection is therefore given by
\begin{linenomath*}
\begin{equation}
\frac{\Delta \rho_{\rm therm}}{\Delta \rho_{\rm cond}} \sim \frac{\alpha_{p} l }{c_{p}}.
\end{equation}
\end{linenomath*}
For silicates, the latent heat of vaporization is $\sim$$10^{7}$~J~kg$^{-1}$, and we assume that the vapor is an ideal gas; $\alpha_{p} \sim 10^{-4}$ and $c_{p} \sim 10^{3}$ (see \S\ref{sup:sec:adiabats}).
The ratio of the buoyancies for the purely thermal and purely condensate end members is then of order unity.
The work that would need to be done to heat up a downwelling parcel and re-equilibrate it thermally with its surroundings would also be similar in each case.
We suggest that in the limiting case of perfect condensate-vapor coupling that the system will behave similarly to thermally driven convection, and so we can use MLT to approximate the mixing timescale in a synestia.
Our estimates based on MLT do not include the effect of falling condensates transferring mass radially and only give an estimate of the mixing velocity and timescale parallel to the rotation axis.

Due to the range of material properties in the post-impact structure, to calculate the mixing timescale, we consider separately convective mixing in the low density and high density regions of a synestia.
For the high density regions, we use $g $~$\sim $~$5$-$10$~m~s$^{-2}$, $\Omega $~$\sim$~$10^{-4}$~rad~s$^{-1}$, $L$~$\sim$~$10^7$~m and assume the fluid has properties comparable to a silicate liquid, where $\alpha_{p} $~$\sim$~$ 10^{-5}$~K$^{-1}$, $\rho $~$\sim$~$ 10^3$~kg~m$^{-3}$ and $c_{p} $~$\sim$~$ 10^{3}$~J~K$^{-1}$~kg$^{-1}$ \citep{Lange1987,Rivers1987}.
For the low-density regions of the structure, we use $g $~$\sim$~$ 0.1$-$5$~m~s$^{-2}$, $\Omega $~$\sim$~$ 10^{-4}$~rad~s$^{-1}$, $\alpha_{p} $~$\sim$~$ 10^{-4}$~K$^{-1}$ (comparable to that of an ideal gas), $c_{p} $~$\sim$~$ 10^{3}$~J~K$^{-1}$~kg$^{-1}$, and $L$~$\sim$~$10^7$~m. We considered two different densities for the midplane ($\rho \sim 10$~kg~m$^{-3}$) and the photosphere ($\rho \sim 10^{-3}$~kg~m$^{-3}$).

In the high density regions, the convective velocity is about $10$~m~s$^{-1}$ without rotation, and the timescales for mixing are on order a week.  Including the effect of rotation decreases the convective velocity to a few meters per second and increases the
mixing time to weeks. The low density outer regions of a structure can mix faster. The convective velocities in the midplane are on the order of tens of meters per second, accounting for rotation. The corresponding mixing times are on the order of days.  At the photosphere, the convective velocities are hundreds of meters per second and the mixing timescale is less than a day. The three different methods for including the effect of rotation give mixing timescales within the same order of magnitude for the regime considered here with the formulation of \cite{Solomatov2000} giving slower convective velocities and longer mixing times compared to the other two methods. Thus, we expect that the vertical mixing in synestias is efficient.

%xxxxxxxxxxxxxxxxxxxxxxxxxxxxxxxxxxxxxxxxxxxxxxxxxxxxxxxxxxxxxxxxxxxxxxxxxxxxxxxxxxxxxxxxxxxxxxxxxxxxxxxxxxxxxxxxxxxxxxxxxxxxxxxxxxxxxxxx
%xxxxxxxxxxxxxxxxxxxxxxxxxxxxxxxxxxxxxxxxxxxxxxxxxxxxxxxxxxxxxxxxxxxxxxxxxxxxxxxxxxxxxxxxxxxxxxxxxxxxxxxxxxxxxxxxxxxxxxxxxxxxxxxxxxxxxxxx
%xxxxxxxxxxxxxxxxxxxxxxxxxxxxxxxxxxxxxxxxxxxxxxxxxxxxxxxxxxxxxxxxxxxxxxxxxxxxxxxxxxxxxxxxxxxxxxxxxxxxxxxxxxxxxxxxxxxxxxxxxxxxxxxxxxxxxxxx
\section{Synestia cooling calculations}
\label{sup:sec:SPH_cooling}

Here, we provide details of the implementation of the calculation summarized in \S\ref{sec:cooling_methods}.

\subsection{Cooling method}
We developed a simple model to estimate the shortest timescale possible for lunar accretion and separation from an impact-generated terrestrial synestia. We focused on the process of condensation of the silicate vapor by radiative cooling and neglect internal heating by viscous dissipation. We assumed that the quasi-isentropic vapor region of the synestia is well-mixed with a constant specific entropy down to pressures at which the isentrope intersects the vapor dome. Then, at lower pressures, the specific entropy follows the vapor side of the vapor dome. The size and shape of a synestia were estimated by removing the condensate fraction in the disk-like region and only calculating the pressure field for the remaining vapor and high-pressure fluid. The mass and orbit of a single moon was estimated assuming perfect accretion of Roche-exterior condensates. Based on the estimated circular orbit of the growing moon, we determined the vapor pressure at that location in the synestia. We consider this a minimum vapor pressure around the moon because it does not take into account the gravitational field of the moon.

First, giant impact simulations using the GADGET-2 code were calculated for 24 to 48 hours of simulation time, when most structures were nearly axisymmetric and had reached a quasi-hydrostatic equilibrium. Escaping particles were removed and the system was truncated at a radius of $1.5\times10^8$~m for the cooling calculation. Any iron particles remaining in the disk-like region were removed. In a few cases, clumps of self-gravitating pure liquid silicate particles (small moonlets) were present in the inner disk-like region and in the process of falling into the corotating region; these parcels were also removed.

\begin{sidewaysfigure*}
\centering
\includegraphics[scale=0.833333333333]{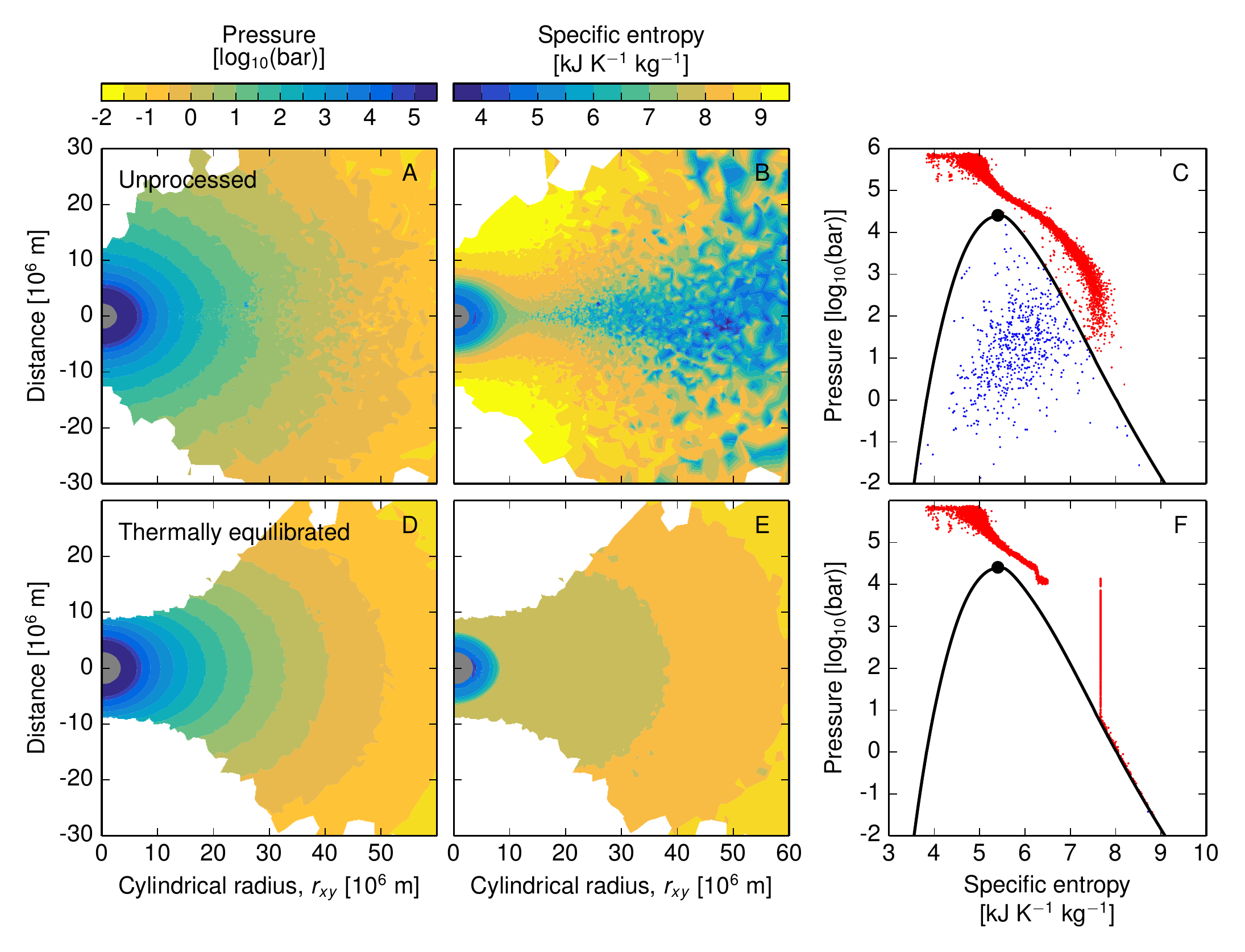}
\caption{
Example of thermal equilibration of a impact-generated synestia, showing contours of pressure (A,D) and specific entropy (B,E), and the distribution of silicate particles in the midplane in pressure-entropy space (C,F with the vapor dome shown as a black line). The post-impact structure (top row) is thermally stratified with partial condensates in the midplane. During thermal equilibration, condensates are removed from the Roche-exterior region and the middle region is mixed to a single isentrope. Then, the structure gravitationally re-equilibrates. The thermally equilibrated structure (bottom row) approximates the pressure and entropy structure of the fluid portion of the post-impact synestia, assuming that the outer regions are well-mixed and that the gravitational influence of condensates is negligible. Note the change in shape of the structure before and after thermal equilibration. This SPH calculation does not resolve closing of pressure and entropy contours at the surface of the flared structure. In this example, which is the same as used for the cooling calculation in the main paper (Figures 
\ref{fig:SPH_cooling} and
\ref{fig:SPH_cooling_time}), the mass of condensate removed during thermal equilibration was $0.466M_{\rm Moon}$. In this impact, a $0.468M_{\rm Earth}$ body struck a $0.572M_{\rm Earth}$ body at 12.33 km~s$^{-1}$ and an impact parameter of 0.4. 
}
\label{sup:fig:thermalequil}
\end{sidewaysfigure*}

Second, the post-impact structure was thermally equilibrated, and the pressure field was recalculated. Most post-impact structures had approximately constant entropy particles in the inner disk-like region, as shown in Figure~\ref{sup:fig:thermalequil}C. In this step, the masses of the SPH particles that were a mixture of liquid and vapor were modified to remove the condensed mass fraction from the particle. The specific entropy and density of the particle was then set to the value for vapor on the vapor dome at the same pressure (Figure~\ref{sup:fig:thermalequil}F). The particles retained the same specific angular momentum. Based on the unique entropy distribution for each post-impact structure, we chose a value for the specific entropy, $S_{\rm inner}$, that divided the structure between a thermally stratified inner region and a quasi-isentropic outer region. All the fully vapor particles in the quasi-isentropic region were averaged to a constant specific entropy. This modified structure was evolved in GADGET-2 to attain quasi-hydrostatic equilibrium. As the pressure field equilibrated, any particles that partially condensed had the condensed mass fraction removed and the remaining mass was set to pure vapor at the same pressure. Most of the condensate was removed in the first few time steps, and the structure attained quasi-hydrostatic equilibrium on dynamical timescales (hours). The total mass and AM vector of condensate with specific AM greater than the value for a circular orbit at the Roche radius, $j_{\rm Roche}$, was recorded. We assume that this material quickly accretes into a single moonlet that we refer to as the seed of the moon. In the high-energy post-impact structures considered here, the midplane of the Roche-interior region was fully vaporized. Condensates with specific AM below $j_{\rm Roche}$ were redistributed into the vapor structure in the same manner as described in step 3 below. The structure becomes less flared upon thermal equilibration because high entropy particles above and below the midplane are averaged to an intermediate value (Figure~\ref{sup:fig:thermalequil}). Note that the overall pressure field is similar before and after thermal equilibration and thus this step does not affect our inference of the vapor pressure around the growing moon. However, in the initial unmodified structure, the concentration of condensates in the midplane reduces the pressure compared to the pressure above and below the midplane because of the local reduction in vapor pressure support. Therefore, in the direct output of SPH calculations, the midplane pressures in mixed phase disk-like regions are biased to lower pressures. At the end of thermal equilibration, any escaping particles were removed.

Third, the thermally equilibrated synestia was evolved in GADGET-2 using a simple treatment of radiative cooling. The steps included in this process are illustrated in Figure~\ref{fig:methods_cartoon}. Because the time steps in an SPH code are set by the Courant criterion, a direct calculation of cooling time is not possible. Thus, we implement an artificially large effective radiative temperature, $T_{\rm eff}$, and scale the simulation time by the factor $\left(T_{\rm eff}/T_{\rm rad}\right)^4$, where $T_{\rm rad}$ is the true photospheric temperature. The expected photospheric temperature is about 2300~K (\S\ref{sec:thermo}). We typically used $T_{\rm eff}=$15000~K or 20000~K. For each full time step:
\begin{enumerate}
\item The structure is centered on the iron core.
\item Each silicate particle is assigned to a group: inner, isentropic, and vapor-dome. The mass-weighted average specific entropy of the isentropic group particles is calculated and assigned to all particles in the isentropic region. 
\item For each radial bin $k$ with annulus area $A_{\rm k}$, the radiative energy loss is $dQ_{k}=2A_{\rm k}\sigma T_{\rm eff}^4dt$, where $\sigma$ is the Stefan-Boltzmann constant and $dt$ is the time step. The factor of 2 accounts for radiation from the top and bottom of the structure. Radiative cooling is accommodated by reducing the total enthalpy of the structure under the assumption that the material in vertical columns in the cooling regions is well mixed. For each radial bin, the specific entropy of each isentropic or vapor-dome group particle $i$ is reduced by $dS_{k}$ such that $\sum_{i}^{} m_i T_i dS_k = dQ_{k}$.  
\item After reducing the enthalpy of the system, each silicate particle is re-assigned to a group. For particles that partially condense, the mass fraction of condensate is removed from the particle and the specific entropy of the remaining mass is set to that of pure vapor. The mass-weighted average specific entropy of the isentropic region is recalculated, and all particles in the isentropic group are assigned the new mean value. The initial location and AM of all condensate is recorded. For condensate with specific AM that exceeds $j_{\rm Roche}$, the mass is removed from the system and the mass and AM vector is recorded to estimate the mass and location of the growing moon. For condensate with specific AM less than $j_{\rm Roche}$, the mass is evenly distributed in radial bins between the initial location of the condensate and the radius corresponding to the circular Keplerian orbit for the specific AM. This redistribution of mass is a simple function to assess the influence of falling condensates on the evolving structure. 
\item The component of falling condensate that is redistributed into the Roche-interior region is added only to the isentropic group particles in each bin. For the total condensate added to each radial bin, the total enthalpy of the isentropic particles in that bin is reduced by the corresponding latent heat of vaporization. The density of the particle is then updated for the change in specific entropy at the same pressure. The mass of each particle in the isentropic group is incremented to accommodate the additional mass, increasing the total AM of the bin. The redistribution of mass from this simple cooling model typically led to a 1 to 2\% error (reduction) in tracking total AM over the duration of the cooling calculation.
\item In order to estimate the fastest cooling rate for the structure, if falling condensate is redistributed to a Roche-interior radial bin that is only occupied by particles in the vapor-dome group, the falling condensate mass is removed from the system. This simple calculation does not attempt to model the dynamics of Roche-interior condensates and the potential accretion of Roche-interior material onto a growing satellite. 
\end{enumerate}

Typically, the cooling simulations were run until the edge of the structure in the midplane receded to the Roche radius. In most cases, the structure was still comprised of an inner, isentropic and vapor-dome region. In some cases, the isentropic region cooled to the value of $S_{\rm inner}$ and the simulation was stopped. 

The disk-like regions of thermally equilibrated synestias were vertically hydrostatic and radially expanding due to viscous spreading. To estimate the magnitude of the effect of viscous spreading, we calculated the spreading of synthetic, constant-entropy synestias without cooling, i.e., the specific entropy of the system remained constant. We found that, over the scaled duration of our cooling calculations, the effect of artificial viscosity in GADGET-2 is comparable to that due to viscosity values previously used for strong thermal turbulence (e.g., $\alpha$ of $10^{-3}$ to $10^{-4}$, as suggested by \cite{Pahlevan2007}). We estimate that viscous spreading contributed a small fraction of material to the Roche-exterior region (e.g., less than to about $0.1M_{\rm Moon}$) during thermal equilibration. Because the outer regions of the synestia condense on a timescale faster than viscous spreading, the initial mass of Roche-exterior condensate is not substantially affected by viscous spreading from artificial viscosity. At later times, there is some variation in the estimated mass of the satellite depending on the details of viscosity in the outer regions of the synestia. In this work, we focus on the initial growth of the moon, and this uncertainty in the late evolution of the synestia is left to future refinements on our proposed lunar origin model.

Here, we have neglected some physical processes that were emphasized in previous studies of lunar accretion. We do not include viscous heating of the Roche interior region because the initial rapid period of condensation dominates satellite accretion in this simple cooling calculation. As a result, we cannot investigate the late stages of cooling of the structure. We also neglect dynamical resonances between the synestia and the growing moon, such as Lindblad resonances. Prior studies of circumterrestrial disks \citep{Salmon2012,Salmon2014} implemented Lindblad torques applicable to cool, thin, near-Keplerian disks \citep{Goldreich1979,Goldreich1980}; however the structure of the synestia is unlike any structure for which Lindblad torques have been calculated. Impact-generated synestias have thermal velocities of the same order as orbital velocities, large scale heights compared to the distance to the center of mass, and strong radial pressure support. In some cases, the primary inner Lindblad resonance is located near the boundary between the corotating and disk-like regions of the synestia, calling into question the development of an inner Lindblad density wave. We do not consider the effect of gas drag on the orbit of the growing moon, which will be investigated in future work. Finally, we neglect tides between the synestia and growing satellite because tidal migration is minimal for the short duration of these calculations. In addition, we expect the tidal quality factor of the terrestrial synestia to be very large, e.g., closer to present-day gas giant planets than fully condensed bodies.

%xxxxxxxxxxxxxxxxxxxxxxxxxxxxxxxxxxxxx
\subsection{Estimating the satellite mass}
Based on our cooling calculations, we estimated the mass of materials that could accrete to form a moon. We provide two estimates for potential satellite masses, one that incorporates only material with sufficient AM to remain beyond the Roche radius (moon A) and one that incorporates some condensates that could fall within the Roche radius (moon B). 

The Roche-exterior mass and total AM of condensate removed in the thermal equilibration step is assumed to quickly accrete into a seed body for the moon. This assumption is motivated by previous $N$-body calculations of efficient Roche-exterior accretion \citep{Ida1997,Kokubo2000,Salmon2012,Salmon2014}; however, most $N$-body studies began with very compact disks. For the range of giant impact simulations considered here, the radial extent of the Roche-exterior condensates varies widely. For simplicity we processed all the post-impact structures in the same way, and future work will revisit the accretion of a wider variety of Roche-exterior mass distributions. 

During the cooling calculation, the mass and AM of condensates with specific AM that exceed $j_{\rm Roche}$ are assumed to be added to the seed. Under the assumption of perfect accretion, the seed and Roche-exterior condensates form our first estimate (moon A) for the mass and orbital location of the satellite formed from a particular synestia. Moon A is shown by the black circles and black lines in Figure~\ref{fig:SPH_cooling}, \ref{fig:SPH_cooling_time}, \ref{sup:fig:coolingA}, \ref{sup:fig:coolingB}, and \ref{sup:fig:coolingC}.

The pressure-support in the synestia leads to the generation of condensates that originate beyond the Roche radius but do not have specific AM exceeding $j_{\rm Roche}$. In most cases, the location of the equivalent circular Keplerian orbit of this material is slightly within the Roche radius. In our model, we assumed that this material would be deposited (either as condensate or revaporized) over the volume between its point of origin and the equivalent Keplerian orbit, and the mass was redistributed evenly between the encompassed radial bins. The portion of the mass, and its corresponding AM, that was distributed beyond the Roche radius was included in moon B, under the assumption that falling condensate is likely to encounter and accrete to the moon. The falling condensate increased the total mass and decreased the specific AM of moon B compared to moon A. Moon B is shown by the blue diamond and blue lines in Figures \ref{fig:SPH_cooling}, \ref{fig:SPH_cooling_time}, \ref{sup:fig:coolingA}, \ref{sup:fig:coolingB}, and \ref{sup:fig:coolingC}.

%xxxxxxxxxxxxxxxxxxxxxxxxxxxxxxxxxxxxxxxxxxxxxxxxxxxxxxxxxxxxxxxxxxxxxxxxxxxxxxxxxxxxxxxxxxxxxxxxxxxxxxxxxxxxxxxxxxxxxxxxxxxxxxxxxxxxxxxx
%xxxxxxxxxxxxxxxxxxxxxxxxxxxxxxxxxxxxxxxxxxxxxxxxxxxxxxxxxxxxxxxxxxxxxxxxxxxxxxxxxxxxxxxxxxxxxxxxxxxxxxxxxxxxxxxxxxxxxxxxxxxxxxxxxxxxxxxx
%xxxxxxxxxxxxxxxxxxxxxxxxxxxxxxxxxxxxxxxxxxxxxxxxxxxxxxxxxxxxxxxxxxxxxxxxxxxxxxxxxxxxxxxxxxxxxxxxxxxxxxxxxxxxxxxxxxxxxxxxxxxxxxxxxxxxxxxx
\section{Condensation calculations}
\label{sup:sec:cond}

Here, we provide supporting information about the physiochemical calculations presented in \S\ref{sec:thermo} and our comparisons to lunar data (\S\ref{sec:Moon_comp}).

%%%%%%%%%%%%%%%%%%%%%%%%%%%%%%%%%%%%%%%%%%%%%%%%%%%%%%%%%%%%%%%%%%%%%%
\subsection{Bulk composition of the Moon}
\label{sup:sec:BSM}

There are a wide range of estimates for the bulk composition of the Moon due to the difficulties of inferring a bulk composition from a limited number of surface samples, and seismic and gravity data.
Different estimates of the bulk Moon (BM) are compared with the bulk silicate Earth (BSE) in Figure~\ref{fig:lunar_comp} for the major and minor elements. For Earth, we chose the widely used BSE composition of \cite{McDonough1995}. The gray band in Figure~\ref{fig:lunar_comp} shows a range of estimates of the bulk Moon composition normalized to bulk silicate Earth. 

Our range of estimates of the bulk Moon composition is based on estimates and discussions provided by \cite{Ringwood1977,Waenke1977,Morgan1978,Ringwood1979,Taylor1982,Wanke1982,Ringwood1987,Warren2005,Longhi2006,Taylor2009,Taylor2014}, and \cite{Hauri2015}. All BM estimates are depleted in volatile elements (Na, K, Mn) relative to BSE, but there are substantial differences in Fe enrichment. The estimate of \cite{Taylor1982} reflects an early view that the Moon, in addition to volatile element depletion, is enriched in refractory elements and Fe (Mg\# of 84). Other early estimates, such as \cite{Waenke1977,Ringwood1979}, and \cite{Ringwood1987} predict similar enrichment in Fe, but not in refractory elements. Because olivines with an Mg\# of 87.5 have been found in two old lunar rocks (troctolite 76535 and dunite 72415), it is difficult to understand how the bulk silicate Moon could have an Mg\# as low as 84. More recently, \cite{Longhi2006,Warren2005}, and \cite{Hauri2015} proposed BM compositions with Mg\#'s of 87~-~90, similar to Earth. Re-evaluation of lunar seismic data \citep{Lognonne2003,Weber2011,Garcia2011}, and constraints from the recent GRAIL mission \citep{Wieczorek2013}, now suggest a lunar crustal thickness of 30 to 40~km, half the thickness of Apollo-era estimates. Therefore, \cite{Taylor2014} no longer supports refractory element (Ca, Ti, Al) enrichment in the Moon. 

Compositions plotted in Figure~\ref{fig:lunar_comp} are normalized to the refractory element Al because Ca, Al, Ti and other refractory elements are believed to have close to chondritic ratios in both Earth and the Moon \citep[c.f.][]{Taylor2014}; i.e. the enrichment factors of the gray band for the selected elements, E, are calculated as (E/Al)$_{\rm BM}$/(E/Al)$_{\rm BSE}$ ratios. The refractory elements have a very small range as they reflect the uncertainty in various estimates of their chondritic ratios. The main planet-building elements Mg and Si have larger uncertainties due to the lack of samples that directly represent the lunar mantle composition. Fe is tied to Mg through the possible range of Mg\#'s. The Fe/Mn ratio of the BSE is about 60 while that of the Moon is about 75. The moderately volatile elements K and Na have depletion factors in the BM of 5 to 10, with K being particularly well established based on K/U ratios \citep[c.f.][]{Taylor2014}. This is a major constraint on the model discussed in this paper. The elements Cr, Co and Ni may also provide further constraints if their bulk lunar compositions can be better established than their current uncertain values.

Recent work has emphasized the observation of volatile species \citep[e.g., water, see][and references
therein]{Hauri2015} in the Moon. Thus, a complete model for lunar origin must also address the origin of the most volatile components, which we do not consider in this work.

%%%%%%%%%%%%%%%%%%%%%%%%%%%%%%%%%%%%%%%%%%%%%%%%%%%%%%%%%%%%%%%%%%%%%%
\subsection{Effect of varying the silicate vaporization temperature buffer}
\label{sup:sec:lunar_si_buffer}

\begin{figure}[t]
\centering
\includegraphics[scale=0.8333333]{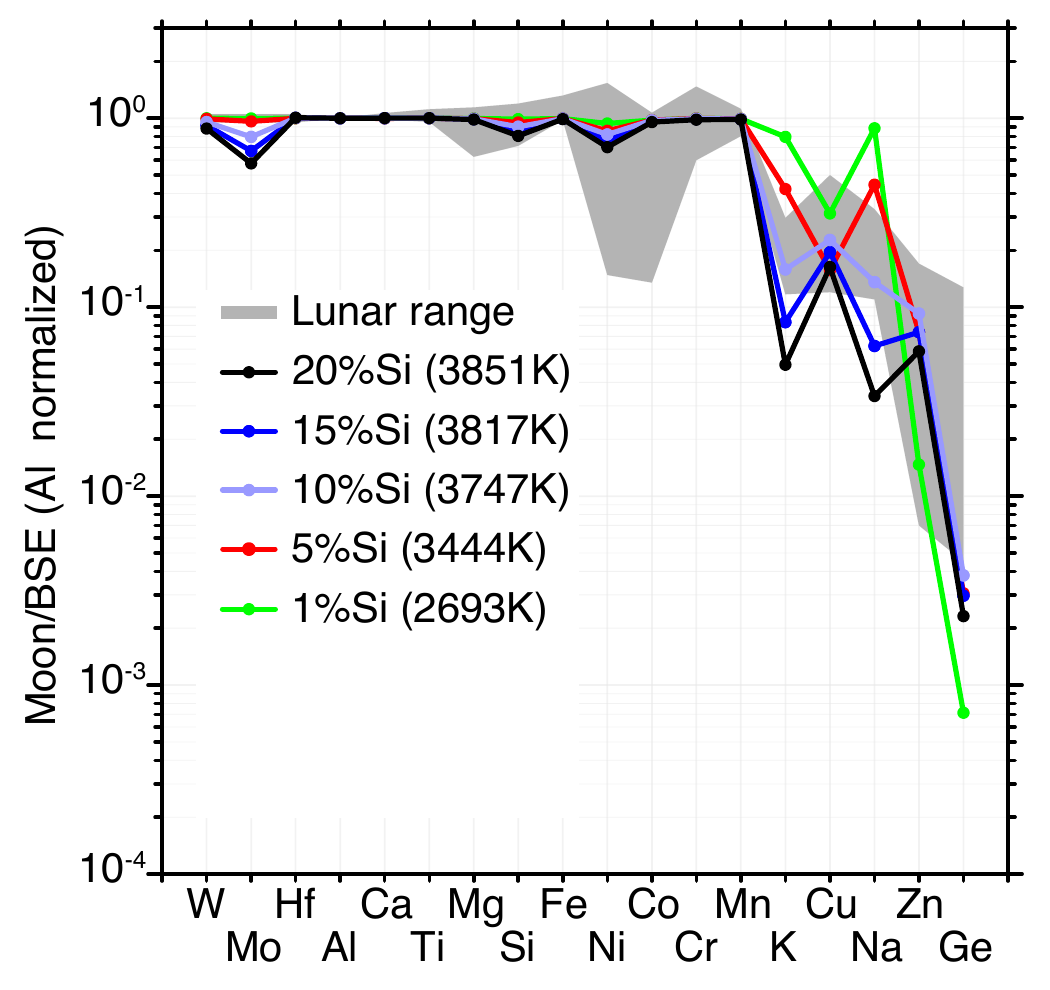}
\caption{The composition of the moon formed within a terrestrial synestia depends on the exact position of the silicate vaporization temperature buffer. Colored lines present the composition of the condensate at 25~bar and different equilibration temperatures, corresponding to the indicated mass fraction of Si in the vapor. The gray band presents the observed estimate and uncertainties on the bulk Moon composition.}
\label{sup:fig:lunar_comp_buffer}
\end{figure}

In our model, the temperature of equilibration of moonlets is controlled by the onset of silicate vaporization.
In \S\ref{sec:buffer}, we argue that the equilibration temperature is buffered near the point where $\sim$10~wt\% of the total Si is vaporized.
However, it is possible that the buffer varies somewhat due to either highly efficient or highly inefficient vaporization of moonlets.
Varying the amount of Si in the gas at which the moonlets equilibrate can significantly affect the composition of the resulting moon (Figure~\ref{sup:fig:lunar_comp_buffer}).
However, as noted in \S\ref{sec:Moon_comp}, the moderately volatile element composition of the liquid at slightly lower and higher temperatures are somewhat complimentary.
A better understanding of the dynamics and thermodynamics of boundary layers around moonlets will be needed to determine the exact temperature of equilibration for moonlets in a synestia and hence the composition of the moon formed.

%%%%%%%%%%%%%%%%%%%%%%%%%%%%%%%%%%%%%%%%%%%%%%%%%%%%%%%%%%%%%%%%%%%%%%
\subsection{Pressure constraints on lunar origin}
\label{sup:sec:lunar_comp_press}

In this work, we present a physical model for the formation of our Moon that predicts a range of vapor pressures around the growing moon. However, our condensation calculations can place an independent, empirical constraint on the pressure of lunar origin assuming the Moon formed by equilibrium condensation from BSE vapor.

\begin{figure}[t]
\centering
\includegraphics[scale=0.8333333333]{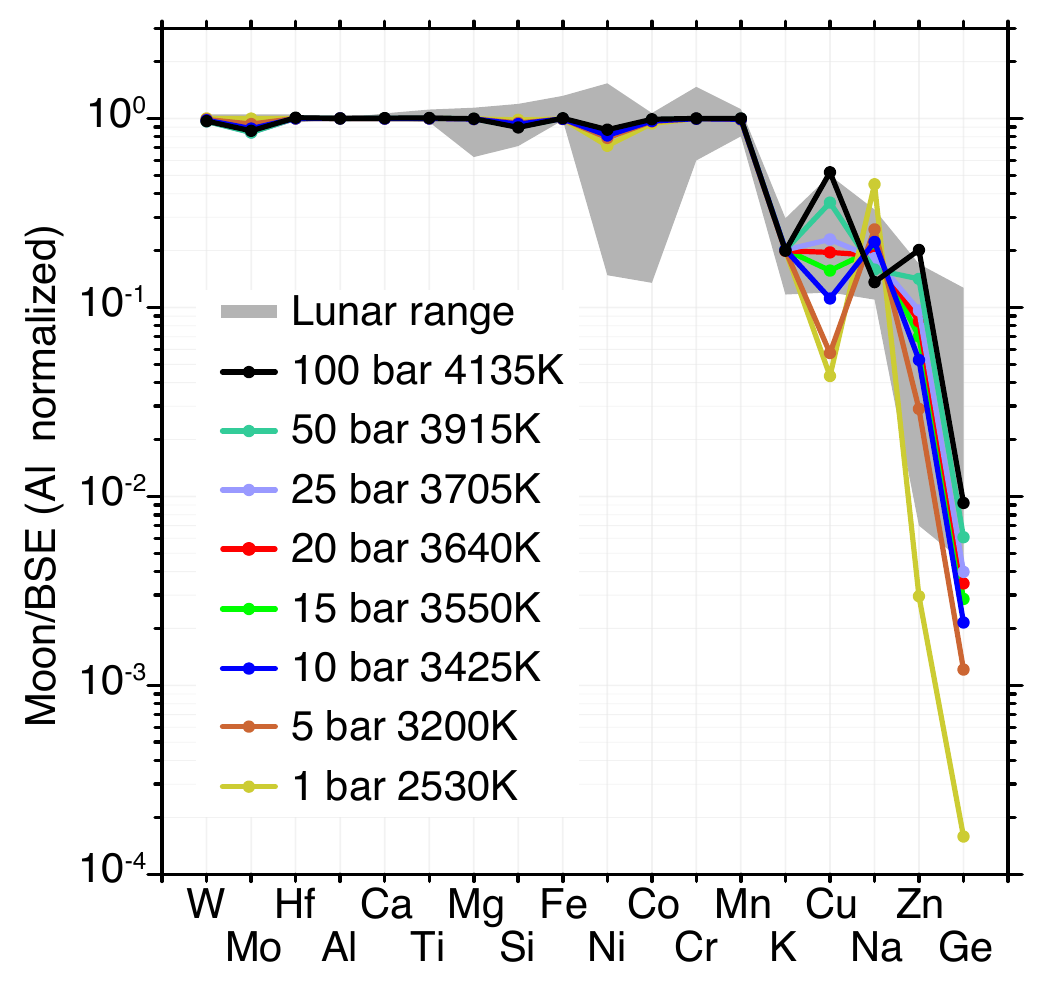}
\caption{The composition of the bulk Moon can constrain the pressure of lunar origin. We used potassium depletion as a proxy for accretion temperature and calculated the composition of condensate at different pressures and temperatures set by requiring a depletion factor of five for potassium. The range of published estimates for the bulk Moon composition are shown by the gray band. The lunar composition can be matched by equilibration at tens of bar. Equilibration at all pressures can match the depletion of potassium but other moderately volatile elements, in particular copper and sodium, are not satisfied simultaneously .}
\label{sup:fig:lunar_comp_press}
\end{figure}

The lunar depletion in potassium is probably the best known of all the moderately volatile element depletions.
Figure~\ref{sup:fig:lunar_comp_press} shows the composition of the condensate at different pressures where the temperature is set by requiring a depletion in potassium of a factor of 5 compared to BSE. 
For pressures that are too low ($<10$~bar) or too high ($>\sim50$~bar), the composition of the condensate cannot simultaneously satisfy the potassium depletion and the depletion of other moderately volatile elements.
As we saw in Figure~\ref{fig:lunar_comp}, copper is a particularly good pressure indicator.
Therefore, any lunar origin model that relies on equilibrium condensation, or equilibration between terrestrial and lunar material, requires modest vapor pressures (tens of bar) to be able to produce the observed bulk lunar composition.

%xxxxxxxxxxxxxxxxxxxxxxxxxxxxxxxxxxxxxxxxxxxxxxxxxxxxxxxxxxxxxxxxxxxxxxxxxxxxxxxxxxxxxxxxxxxxxxxxxxxxxxxxxxxxxxxxxxxxxxxxxxxxxxxxxxxxxxxx
%xxxxxxxxxxxxxxxxxxxxxxxxxxxxxxxxxxxxxxxxxxxxxxxxxxxxxxxxxxxxxxxxxxxxxxxxxxxxxxxxxxxxxxxxxxxxxxxxxxxxxxxxxxxxxxxxxxxxxxxxxxxxxxxxxxxxxxxx
%xxxxxxxxxxxxxxxxxxxxxxxxxxxxxxxxxxxxxxxxxxxxxxxxxxxxxxxxxxxxxxxxxxxxxxxxxxxxxxxxxxxxxxxxxxxxxxxxxxxxxxxxxxxxxxxxxxxxxxxxxxxxxxxxxxxxxxxx
\section{Calculation of adiabats}
\label{sup:sec:adiabats}

In order to determine the composition of parcels of material traveling through different pressure-temperature paths within a terrestrial synestia (\S\ref{sec:paths}), we calculated adiabats incorporating the effect of condensation.  Entropy
cannot be calculated directly from our condensation calculations but an adiabat can be approximated by treating the vapor and
condensate as two phases with a single latent heat of vaporization,
$l$.  This approach is commonly used in atmospheric physics to consider the
effect of water vapor on moving parcels of air \citep{Chamberlain1987};
however, various assumptions are made that are not suitable for our
purposes.  Therefore, we derived the adiabats for a parcel of material based on our calculated BSE phase diagram and the assumption that two phases are in equilibrium and these is no dynamical phase separation.

The differential of specific entropy assuming $S(p,T,w)$, can be written
\begin{linenomath*}
\begin{equation}
TdS = T \frac{\partial S}{\partial T} \bigg |_{p,w} dT + T \frac{\partial S}{\partial p} \bigg |_{T,w} dp +T \frac{\partial S}{\partial w} \bigg |_{p,T} dw \, ,
\label{sup:eqn:dS}
\end{equation}
\end{linenomath*}
where $S$ is specific entropy, $T$ is temperature, $p$ is pressure, and $w$ is the mass fraction of vapor.
The first term can be simplified using the definition for specific heat capacity at constant pressure (and $w$),
\begin{linenomath*}
\begin{equation}
c_{p,w} = T \frac{\partial S}{\partial T} \bigg |_{p,w} \, .
\end{equation}
\end{linenomath*}
The pre-factor to the second term is converted using a Maxwell relation to
\begin{linenomath*}
\begin{equation}
T \frac{\partial S}{\partial p} \bigg |_{T,w}  = - T \frac{\partial }{\partial T} \left ( \frac{1}{\rho} \right)\bigg |_{p,w} \, ,
\end{equation}
\end{linenomath*}
which we write in terms of the coefficient of thermal expansion at
constant pressure and $w$,
\begin{linenomath*}
\begin{equation}
T \frac{\partial S}{\partial p} \bigg |_{T,w}  = - \frac{T \alpha_{p,w}}{\rho} \, .
\end{equation}
\end{linenomath*}
The final term can be simplified to $l dw$ assuming that the amount of latent heat released during condensation is linear with the mass condensed.
Combining these expressions into Equation \ref{sup:eqn:dS} gives
\begin{linenomath*}
\begin{equation}
TdS = c_{p,w} dT - \frac{T \alpha_{p,w}}{\rho} dp +l dw \, .
\label{sup:eqn:dS2}
\end{equation}
\end{linenomath*}
To determine the adiabat in $p$-$T$ space, we set $dS=0$ and expand $dw$,
\begin{linenomath*}
\begin{equation}
0 = c_{p,w} dT - \frac{T \alpha_{p,w}}{\rho} dp +l \left (\frac{\partial w}{\partial T} \bigg |_{p} dT +\frac{\partial w}{\partial p} \bigg |_{T} dp \right ) \, .
\end{equation}
\end{linenomath*}
Collecting terms and dividing through,
\begin{linenomath*}
\begin{equation}
dp= \left (c_{p,w}  +l \frac{\partial w}{\partial T} \bigg |_{p} \right ) \left ( \frac{T \alpha_{p,w}}{\rho} - l \frac{\partial w}{\partial p} \bigg |_{T} \right )^{-1} dT \, .
\end{equation}
\end{linenomath*}
%acquire
Therefore,
\begin{linenomath*}
\begin{equation}
\frac{\partial p}{\partial T} \bigg |_{S}= \left (c_{p,w}  +l \frac{\partial w}{\partial T} \bigg |_{p} \right ) \left ( \frac{T \alpha_{p,w}}{\rho} - l \frac{\partial w}{\partial p} \bigg |_{T} \right )^{-1} \, .
\label{sup:eqn:adiabat}
\end{equation}
\end{linenomath*}
This equation can be integrated to find an adiabat.

The behavior of Equation \ref{sup:eqn:adiabat} is as expected in two commonly used limits.
First, if the pressure dependence of $w$ is negated and an ideal gas is assumed (i.e. $\alpha_{p,w}^{\rm vap}=1/T$), Equation \ref{sup:eqn:adiabat} becomes
\begin{linenomath*}
\begin{equation}
\frac{\partial p}{\partial T} \bigg |_{S}=\rho \left (c_{p,w}  +l \frac{\partial w}{\partial T} \bigg |_{p} \right ) \, ,
\label{sup:eqn:wet_lapse}
\end{equation}
\end{linenomath*}
which is the expression commonly used in atmospheric physics for moist convection \citep[e.g.,][]{Chamberlain1987}.
Secondly, in the limit of $l=0$, the equation for a standard adiabat is recovered,
\begin{linenomath*}
\begin{equation}
\frac{\partial p}{\partial T} \bigg |_{S}=\frac{\rho c_{p}}{T \alpha_p}  \, .
\end{equation}
\end{linenomath*}
This expression has been widely used in geophysics \citep[e.g.,][]{Solomatov2000}.

Before we can use Equation \ref{sup:eqn:adiabat} to calculate adiabats, we must make some further assumptions. 
The vapor is treated as a monotonic ideal gas. Therefore, the molar heat capacity is $C_{p,w}^{\rm vap}=5R/2$ and the coefficient of thermal expansion is $\alpha_{p,w}^{\rm vap}=1/T$. $R$ is the universal gas constant.
The molar heat capacity of the condensate is given by the high temperature limit of the Debye model: $C_{p,w}^{\rm cond}$~$\sim$~$ C_{V,w}^{\rm cond} = 3 R$.
Note that we have assumed that the heat capacities are
identical and we will discuss the validity of this assumption shortly.
The coefficient of thermal expansion for the condensate is taken to be a constant $ \alpha_{p,w}^{\rm cond}= 5 \times 10^{-5}$~K$^{-1}$, consistent with the properties of silicate melts \citep{Lange1987,Rivers1987}.
The density for the condensate and vapor were taken from the M-ANEOS forsterite vapor dome \citep{Melosh2007} at the relevant pressure.
The bulk density is then given by
\begin{linenomath*}
\begin{equation}
\frac{1}{\rho}=\frac{w}{\rho_{\rm vap}}+\frac{1-w}{\rho_{\rm cond}} \, ,
\label{sup:eqn:density}
\end{equation}
\end{linenomath*}
where $\rho_{\rm vap}$ and $\rho_{\rm cond}$ are the density of the vapor and condensate, respectively.
The latent heat is also taken from the M-ANEOS forsterite vapor dome \citep{Melosh2007}.
The latent heat is calculated as
\begin{linenomath*}
\begin{equation}
l= T \Delta S
\label{sup:eqn:latent_heat} \, ,
\end{equation}
\end{linenomath*}
where $\Delta S$ is the entropy difference between liquid and vapor at the required pressure.
The latent heat calculated from M-ANEOS at the pressures of our calculations are similar to that widely used in studies of the canonical lunar disk of $l = 1.7 \times 10^{7}$~J~kg$^{-1}$ \citep[e.g.,][]{Thompson1988,Ward2012}.

In order to integrate Equation \ref{sup:eqn:adiabat}, we need to combine the molar heat capacities to give a specific heat capacity of the mixed phase.
The specific heat capacity is defined as
 \begin{linenomath*}
\begin{equation}
c_{p,w}=\frac{\partial U }{\partial T} \bigg |_{p,w} \, ,
\label{sup:eqn:comb_cp1}
\end{equation}
\end{linenomath*}
where $U$ is the internal energy. 
For the bulk
 \begin{linenomath*}
\begin{equation}
 U=n_{\rm vap} U_{\rm vap} + n_{\rm cond}U_{\rm cond} \, ,
 \label{sup:eqn:U_expanded}
 \end{equation}
 \end{linenomath*}
where $U_{\rm vap}$ and $U_{\rm cond}$ are the internal energy per mol of the vapor and condensate, respectively. $n_{\rm vap}$ and $n_{\rm cond}$ are the number of moles of the vapor and condensate in a unit mass of bulk material.
Substituting Equation \ref{sup:eqn:U_expanded} in Equation \ref{sup:eqn:comb_cp1} gives
 \begin{linenomath*}
\begin{equation}
c_{p,w}= n_{\rm vap} \frac{\partial U_{\rm vap} }{\partial T} \bigg |_{p,w}+ n_{\rm cond} \frac{\partial U_{\rm cond} }{\partial T} \bigg |_{p,w} \, ,
\label{sup:eqn:comb_cpw}
\end{equation}
\end{linenomath*}
since the fraction of vapor, $w$, and therefore $n_{\rm i}$, is held constant. 
Simplifying, 
 \begin{linenomath*}
\begin{equation}
 U=n_{\rm vap} C_{p,w}^{\rm vap} + n_{\rm cond} C_{p,w}^{\rm cond} \, ,
 \end{equation}
 \end{linenomath*}
where $C_{p,w}^{\rm vap}$ and $C_{p,w}^{\rm cond}$ are the molar heat capacities for the vapor and condensate.
The number of moles of vapor and condensate per unit mass of bulk are given by
 \begin{linenomath*}
 \begin{equation}
n_{\rm vap} = \frac{w}{\bar{m}_{\rm a}^{\rm vap}}   \, \,\,\,\,\,\, \& \,\,\,\,\,\,\, n_{\rm cond} = \frac{1-w}{\bar{m}_{\rm a}^{\rm cond}} \, ,
\end{equation}
\end{linenomath*}
where $\bar{m}_{\rm a}$ is the average atomic weight of either the vapor or condensate which is taken from our condensation calculations.
Note the inherent assumption that the vapor and condensate are both monotonic; relaxing this assumption has minimal effect on our conclusions.

In a similar fashion, we combine the coefficients of thermal expansion, $\alpha$, to find the relevant value for the bulk.
The thermal expansion coefficient is defined as
 \begin{linenomath*}
\begin{equation}
\alpha_{p,w}=\rho \frac{\partial  }{\partial T} \left (\frac{1}{\rho}\right ) \bigg |_{p,w} \, .
\label{sup:eqn:comb_alpha1}
\end{equation}
\end{linenomath*}
Substituting in Equation \ref{sup:eqn:density} and noting the the differential is at constant $w$,
 \begin{linenomath*}
\begin{equation}
\alpha_{p,w}=\rho \left \{ \frac{w \alpha_{p,w}^{\rm vap} }{\rho_{\rm vap}}+\frac{(1-w) \alpha_{p,w}^{\rm cond}}{\rho_{\rm cond}}  \right \} \, .
%\label{sup:eqn:comb_alpha1}
\end{equation}
\end{linenomath*}

We now return to consider the effect of assuming that $C_{p} $~$\sim$~$ C_{V}$ for the condensate.
The general expression relating the specific heat capacities at constant temperature and pressure is
\begin{linenomath*}
\begin{equation}
c_{p} = c_{V} + \frac{T \alpha_{p}^2}{\rho \beta_{T}} \, ,
\end{equation}
\end{linenomath*}
where $\beta_{T}$ is the isothermal compressibility. 
Given that $T $~$\sim$~$ 10^3$~K, $\alpha_{p} $~$\sim$~$ 10^{-5}$~K$^{-1}$ \citep{Lange1987,Rivers1987}, $\rho $~$\sim$~$ 10^3$~kg~m$^{-3}$ and $\beta_{T} $~$\sim$~$ 10^{-11}$~Pa$^{-1}$ \citep{Lange1987,Rivers1987}, the difference between the specific heat capacities is of order 10~J~K$^{-1}$~kg$^{-1}$. 
The specific heat capacities for typical condensates in our
calculations are $\sim$~$ 10^3$~J~K$^{-1}$~kg$^{-1}$; thus, the assumption of $C_{p,w}^{\rm cond}$~$\sim$~$ C_{V,w}^{\rm cond} $ has only a small effect on our results.

Using this formulation, we calculated adiabats for material of BSE composition. 
For material that is initially entirely vapor, adiabats cool rapidly with decreasing pressure until they encounter the dew point of the BSE system. 
Adiabats then stay close to the dew point as pressure continues to decrease with a slowly increasing condensed mass fraction.
For adiabats that begin close to the dew point at the pressures in the midplane of our example Moon-forming structures (tens of bar) the fraction of condensate is small but at lower pressures (tenths of a bar to a bar) the condensed fraction can be on the order of tens of percent. 
Similarly, adiabats that begin within the stability field of condensate follow lines of approximately constant condensed fraction through the phase diagram.
Adiabats that begin with a large condensed fraction tend to have less change in the mass fraction of condensate with pressure. 
The adiabats of chemically and thermally isolated condensate over the pressure range in the disk-like regions of post-impact structures are
isothermal as previously pointed out by \citet{Machida2004}.

%xxxxxxxxxxxxxxxxxxxxxxxxxxxxxxxxxxxxxxxxxxxxxxxxxxxxxxxxxxxxxxxxxxxxxxxxxxxxxxxxxxxxxxxxxxxxxxxxxxxxxxxxxxxxxxxxxxxxxxxxxxxxxxxxxxxxxxxx
%xxxxxxxxxxxxxxxxxxxxxxxxxxxxxxxxxxxxxxxxxxxxxxxxxxxxxxxxxxxxxxxxxxxxxxxxxxxxxxxxxxxxxxxxxxxxxxxxxxxxxxxxxxxxxxxxxxxxxxxxxxxxxxxxxxxxxxxx
%xxxxxxxxxxxxxxxxxxxxxxxxxxxxxxxxxxxxxxxxxxxxxxxxxxxxxxxxxxxxxxxxxxxxxxxxxxxxxxxxxxxxxxxxxxxxxxxxxxxxxxxxxxxxxxxxxxxxxxxxxxxxxxxxxxxxxxxx
%additional tables

%xxxxxxxxxxxxxxxxxxxxxxxxxxxxxxxxxxxxxxxxxxxxxxxxxxxxxxxxxxx
%table of added species

%xxxxxxxxxxxxxxxxxxxxxxxxxxxxxxxxxxxxxxxxxxxxxxxxxxxxxxxxxxx
%table of added species
\clearpage
\onecolumn
\begin{center}
\LTcapwidth=0.9\textwidth

\begin{longtable}{l l} 
\caption[]{Species added to the GRAINS code \citep{Petaev2009} for this study. Letters indicate the phase; solid (c), liquid (l) or gas (g).}  
\label{sup:tab:added_species} \\

Species & Phase(s)\\
\toprule
\endfirsthead

\multicolumn{2}{c}{{\tablename\ \thetable{} -- continued from previous page}} \\
\multicolumn{2}{l}{} \\
Species & Phase(s)\\
\toprule
\endhead

\multicolumn{2}{l}{} \\
\multicolumn{2}{r}{{continued on next page}} \\
\endfoot

\endlastfoot

\begin{filecontents*}{Tables/Added_species_25_1_17.csv}
GeO$_2$,c l g
Zn,c l g
ZnO,c l g
ZnCl$_2$,c l g
Ga$_2$O$_3$,c l g
\end{filecontents*}
\csvreader[
    column count=2,	
    table head=\hline,	
  late after line=\\ ]{Tables/Added_species_25_1_17.csv}{
  1=\one, 2=\two}
{\one & \two}

\end{longtable}
\end{center}
%\clearpage
%\twocolumn

%xxxxxxxxxxxxxxxxxxxxxxxxxxxxxxxxxxxxxxxxxxxxxxxxxxxxxxxxxxx
%table of vapor species
%\clearpage
%\onecolumn
\begin{center}
\LTcapwidth=0.9\textwidth

\begin{longtable}{l l l l l l} 
\caption[]{Vapor species used in condensation calculations whose thermodynamic data was extended to high temperature as described in text.}  
\label{sup:tab:vap_species} \\

\endfirsthead

\multicolumn{6}{c}{{\tablename\ \thetable{} -- continued from previous page}} \\
\multicolumn{6}{l}{} \\

\endhead

\multicolumn{6}{l}{} \\
\multicolumn{6}{r}{{continued on next page}} \\
\endfoot

\endlastfoot

\begin{filecontents*}{Tables/HighT_vapor_species_25_1_17.csv}
H$_2$,AlOH,MgS,KOH, ClO,MoO$_3$
Mg,OAlOH,SO,PH,PCl,Re
H$_2$O , AlO,S$_2$O,PH$_2$, POCl$_3$,W
 SiO,HCO,SO$_2$,NH$_3$, PCl$_3$,WO
Si,CH$_4$,Na, PH$_3$,PSCl$_3$,WO$_2$
O$_2$, CO, NaO,KO,PCl$_5$,WO$_3$
SiO$_2$, CO$_2$, Na$_2$,K$_2$, Cl,GeO$_2$
O,Ca, NaH,NO,NiO,Ga$_2$O 
MgO , CaOH,NaOH, PN, MnO,GaO
Mg$_2$,CaO,Cr,PO, Co, GeO
Mg(OH)$_2$ ,Fe,Mn,TiO,CoO,IrO
H, FeO,Ni,PO$_2$,He,ReO
SiH,S,N$_2$,TiO$_2$,Cu,Hf
MgH,SiS, K,(P$_2$O$_3$)$_2$,CuO,HfO
MgOH,S$_2$,P,P$_2$O$_5$ ,Ga, Zn
OH,COS,Ti,PS,Ge,ZnO
Al,CS,N, Cl$_2$,Ir,ZnCl$_2$
Al$_2$O,CaS, CrO,HCl,Mo,
 AlH,SH,CrO$_2$,KCl,MoO,
OAlH,H$_2$S,KH, NaCl,MoO$_2$,
\end{filecontents*}
\csvreader[
    column count=6,	
    table head=\hline,	
  late after line=\\ ]{Tables/HighT_vapor_species_25_1_17.csv}{
  1=\one, 2=\two, 3=\three, 4=\four,
  5=\five, 6=\six}
{\one & \two & \three & \four & \five & \six}

\end{longtable}
\end{center}
\clearpage
\twocolumn

%xxxxxxxxxxxxxxxxxxxxxxxxxxxxxxxxxxxxxxxxxxxxxxxxxxxxxxxxxxx
\clearpage
\onecolumn
\begin{center}
\LTcapwidth=0.9\textwidth

\begin{longtable}{l l} 
\caption[]{Condensed species used in condensation calculations whose thermodynamic data were extended to high temperature as described in text. Letters indicate the phase; solid (c) or liquid (l).}  
\label{sup:tab:cond_species} \\
Species & Phase(s)\\
\toprule
\endfirsthead

\multicolumn{2}{c}{{\tablename\ \thetable{} -- continued from previous page}} \\
\multicolumn{2}{l}{} \\
Species & Phase(s)\\
\toprule
\endhead

\multicolumn{2}{l}{} \\
\multicolumn{2}{r}{{continued on next page}} \\
\endfoot

\endlastfoot

\begin{filecontents*}{Tables/HighT_condensed_species_25_1_17.csv}
FeS,c l
CaS,s l
Fe,l
Si,l
Na$_2$SiO$_3$,l
Al$_2$O$_3$,l
CaO,l
FeO,l
MgO,l
Na$_2$O,l
SiO$_2$,l
Ni,l
P,l \& white
P$_2$O$_5$,c l
Mg$_3$(PO$_4$)$_2$,c l
TiO$_2$,l
K$_2$O,l
K$_2$SiO$_3$,l
Cr$_2$O$_3$,l
MnO,c l
NiO,c l
Ca$_3$(PO$_4$)$_2$,c l
CrO,l
Co,l
CoO,l
Cu,c l
CuO,c
Cu$_2$O,c l
Ga,c l
Ir,c l
Mo,c l
MoO$_2$,c l
Re,c l
W,c l
WO$_2$,c l
WO$_3$,c l
Hf,c l
HfO$_2$,c l
GeO$_2$,c l
Zn,l
ZnO,c l
ZnCl$_2$,l
Ge,l
Ga$_2$O$_3$,c l
\end{filecontents*}
\csvreader[column count=2,	
    table head=\hline,	
  late after line=\\]{Tables/HighT_condensed_species_25_1_17.csv}{
  1=\one, 2=\two}
{\one & \two }

\end{longtable}
\end{center}

\clearpage
\twocolumn

%xxxxxxxxxxxxxxxxxxxxxxxxxxxxxxxxxxxxxxxxxxxxxxxxxxxxxxxxxxxxxxxxxxxxxxxxxxxxxxxxxxxxxxxxxxxxxxxxxxxxxxxxxxxxxxxxxxxxxxxxxxxxxxxxxxxxxxxx
%xxxxxxxxxxxxxxxxxxxxxxxxxxxxxxxxxxxxxxxxxxxxxxxxxxxxxxxxxxxxxxxxxxxxxxxxxxxxxxxxxxxxxxxxxxxxxxxxxxxxxxxxxxxxxxxxxxxxxxxxxxxxxxxxxxxxxxxx
%xxxxxxxxxxxxxxxxxxxxxxxxxxxxxxxxxxxxxxxxxxxxxxxxxxxxxxxxxxxxxxxxxxxxxxxxxxxxxxxxxxxxxxxxxxxxxxxxxxxxxxxxxxxxxxxxxxxxxxxxxxxxxxxxxxxxxxxx
\section*{Supplementary References}
\bibliographystyle{elsarticle-harv} 
%IF NEEDED USE A SEPERATE BIB FILE
\bibliography{References}

\end{document}